\documentclass[a4paper,12pt]{article}
\pdfoutput=1
\usepackage{setspace}
\usepackage[USenglish]{babel}
\usepackage[utf8]{inputenc}
\usepackage{csquotes}
\usepackage{amsmath}

\usepackage{subfiles}
\usepackage{nameref}
\usepackage{array,multirow,graphicx}
\usepackage{authblk}
\usepackage{stmaryrd}

\usepackage{subfloat}
\usepackage{subcaption}
\usepackage[labelfont=bf,justification=justified,singlelinecheck=true]{caption}
\usepackage{mathtools}
\usepackage{float}
\usepackage{graphicx} 
\graphicspath{{Figures/}}

\usepackage[square,numbers,sort&compress]{natbib}
\usepackage[hyperindex,breaklinks]{hyperref}

\usepackage{xcolor}

\newcommand{\norm}[1]{\left\lVert#1\right\rVert}

\title{A computationally efficient and mechanically compatible multi-phase-field model applied to coherently stressed three-phase solids}
\author{\small{Sourav Chatterjee$^{a,b\star}$, Daniel Schwen$^c$, Nele Moelans$^a$}}
\date{%
   \small{$^a$Department of Materials Engineering, KU Leuven, Kasteelpark Arenberg 44, Leuven BE-3001, Belgium\\%
       \small{$^b${Department of Materials Science and Engineering, University of Florida, Gainesville, 32611, FL,  USA}}\\%
   $^c$Computational Mechanics and Materials Department, Idaho National Laboratory, Idaho Falls, ID 83415, United States}\\%
    \small{$^*$Corresponding author. 
    E-mail addresses:}\\
    \small{chatterjee.s@ufl.edu (S. Chatterjee),  daniel.schwen@inl.gov (D. Schwen), nele.moelans@kuleuven.be (N. Moelans)}
}

\begin{document}
\maketitle
\clearpage
\section*{\center Abstract}
Engineering alloys generally exhibit multi-phase microstructures. For simulating their microstructure evolution during solid-state phase transformation, CALPHAD-guided multi-phase-field models coupled with micro-mechanics have proven to be a reliable simulation tool. Nevertheless, their efficiency and accuracy still depend on the homogenization scheme used to interpolate the elastic properties in the interfacial regions. In this paper, we present a phase-field model for multi-phase and multi-component solids using a partial rank-one homogenization scheme that enforces static and kinematic compatibilities in the interfacial regions. To this end, we first extend the rank-one homogenization scheme to multi-phase systems. Moreover, for computational efficiency, we analytically solve the static compatibility equations for linear elastic three-phase solids. For quantitative accuracy, a coupling technique is used to extract the prerequisite thermodynamic and kinetic properties from CALPHAD databases. The model is solved numerically in an open source finite-element framework. As numerical applications, the microstructure of two elastically stressed intermetallic-containing three-phase alloys: Ni-Al and Al-Cr-Ni, are simulated. The accuracy of the model is verified against analytically obtained solutions for planar and concentric ring interfaces. We show that the simulation results remain unaltered with varying interface width. Except for one simulation, all cases show better or nearly equal convergence using the partial rank-one scheme compared to the Voigt-Taylor scheme.\\
\textit{Keywords:} chemo-mechanical processes; microstructure; phase transformation; inhomogeneous material; homogenization

\section{Introduction}
Engineering alloys, such as Ni-base superalloys, steels, etc., generally comprise multiple chemical constituents and phases. Their physical and mechanical properties are strongly related to the microstructure formed during interdiffusion processes at elevated temperatures. However, predicting the kinetics of microstructure evolution during diffusive transformations, especially when elastic stresses are included, is difficult since this requires solving a free-boundary problem, which is seldom analytically soluble \cite{Fried1999, Gurtin1993, NProvatas2011}. Thus, reliable and efficient computational approaches are often needed to gain a quantitative understanding of microstructure evolution in elastically stressed multiphase and multicomponent alloys. 

 The phase-field method has emerged as a useful tool to predict microstructure evolution in engineering alloys \cite{Chen1996, LQChen2002,Moelans2008,Steinbach2009,Nestler2011}. Its well-known advantage is that the interface or interphase separating either the grains or phases is implicitly represented by a phase-field variable that varies smoothly across a finite region of thickness, referred to as the interface width. Further, for simulations to be well-resolved, the interface width has to be at least five times the grid spacing \cite{Moelans2011}. Therefore, simulations using this method are particularly difficult when the desired microstructural length scale is in the micro to millimeter range due to a stringent limit on interface width \cite{Plapp2011}, \cite{Karma2016}. In addition, this limit typically varies with the bulk alloy properties \cite{Kim1998}.  To overcome this limitation, it is thus essential that the interface width in a phase-field model can be independently controlled without affecting the accuracy of the simulation.
    
This requirement has led to the development of alloy phase-field models in which the interface width is treated as a simulation parameter that can be selected depending on numerical convenience  \cite{Kim1999}, \cite{Plapp2011}. This is because the bulk and interfacial properties in such models are independent, even when the interface width is artificially enlarged \cite{Plapp2011}, \cite{Plapp2015}.  Nevertheless, the generalization of such alloy phase-field models to problems that require coupling with mechanics is not straightforward due to the dependence of bulk properties on elastic fields. More precisely, in a mechanically coupled phase-field model, the scheme of interpolation or homogenization of elastic fields in the interfacial regions may affect this desired separation of bulk from interfacial properties due to an interfacial excess elastic energy contribution that depends on the homogenization scheme \cite{DurgaA2013}, \cite{Schneider2015}.  

So far, two types of mechanically uncoupled phase-field models for alloys have been proposed that allow the interface width to be selected arbitrarily \cite{Plapp2011}. As pointed out by Plapp \cite{Plapp2011}, the first derives the evolution equations starting from a Helmholtz functional \cite{Kim1999}, while the second derives it from a grand-potential functional \cite{Plapp2011}, \cite{Choudhary2012}. Moreover, the former approach requires thermochemical properties as functions of composition(s), while the latter requires them as functions of diffusion potential(s) \cite{Plapp2011}. Although both models are equivalent \cite{Plapp2011}, the latter offers possible computational gains as it requires solving $(n-1)(p-1)$ less equations for a $n$-component and $p$-phase alloy system \cite{Chatterjee2021}. It is worth noting that this is strictly true assuming either non-dilute or non-ideal or non-quadratic free energies since only then the equal diffusion potential or ``quasiequilibrium" conditions have to be numerically solved at each grid point and time step \cite{Eiken2006} \cite{Kim2021}. Further, to decrease the computational costs in such simulations, some studies have developed simplified approaches to solving these conditions \cite{Eiken2006}, \cite{Bottger2015}, \cite{Jiang2019}. Nevertheless, the appropriate homogenization approach for coupling these alloy phase-field models with mechanics is still debatable.
 
Specifically, the coupling of the above-mentioned models with small-strain elasticity theory has been considered by many workers;  either based on a Helmholtz functional, e.g., \cite{Zhu2004}, \cite{Ammar2009}, \cite{Steinbach2007Acta}, \cite{DurgaA2013}, \cite{Durga2015}  or a grand-potential functional \cite{Mushongera2015, Tschukin2017a, Simon2020, SC2}. Nevertheless, the accuracy of such coupled models still depends on the homogenization assumptions with regard to the elastic fields \cite{Ammar2009}, \cite{DurgaA2013}, \cite{Mosler2014}, \cite{Schneider2015}, \cite{Svendsen2018}. To be precise, depending on the scheme of homogenization, these mechanically coupled models can be subdivided into two categories. The models in the first category follow those homogenization schemes that are either statically or kinematically compatible. For instance, Khachaturyan \cite{Zhu2004}, \cite{Mushongera2015},  Reuss/Sachs \cite{Steinbach2006}, \cite{Steinbach2007Acta}, and Voigt/Taylor  \cite{Ammar2009}, \cite{Simon2020}. On the other hand, models in the second category follow those schemes that enforce both static and kinematic compatibilities: by either using a mixed scheme that is a combination of Reuss/Sachs and Voigt/Taylor, e.g.,  \cite{DurgaA2013},  \cite{Durga2015}, \cite{Tschukin2017a}, or a partial rank-one scheme \cite{SC2}. Moreover, it has been argued that models in the first category are less accurate compared to models in the second category due to an interfacial excess energy contribution coming from the interpolated elastic strain energy \cite{DurgaA2013}, \cite{Durga2015}, \cite{Schneider2015}. 

Nevertheless, from the standpoint of computational efficiency, the particular scheme used for enforcing the static and kinematic compatibilities is also a topic of relevance. For example, the mixed scheme that is a combination of Reuss/Sachs and Voigt/Taylor proposed in the works of Durga \textit{et al.} \cite{DurgaA2013}, \cite{Durga2020} and Schneider \textit{et al.} \cite{Schneider2015}, \cite{Schneider2018} requires a coordinate transformation of elastic fields in order to formulate the interfacial elastic driving force contribution as a function of only \textit{continuous} elastic fields. As shown in \cite{DurgaA2013}, \cite{Schneider2015}, this is needed because then this interfacial excess contribution vanishes in the model. Consequently, their approaches are computationally intensive \cite{Tschukin2017}. Naturally, this limits the application of their model to simple systems. Specifically, Durga's model has been so far applied to simulate an elastically anisotropic four-phase Cu-Sn alloy having only planar interfaces \cite{Durga2020}, while Schneider's model has been limited to elastically isotropic two-phase \cite{Tschukin2017a} and multi-phase \cite{Amos2018} binary alloys. These works, however, assume only small-strain deformations. For sake of completeness, it is worth mentioning that Schneider \textit{et al.} and Hermann \textit{et al.} have also proposed a numerical approach to enforce the static compatibility equations for multiphase solids undergoing finite-strain and small-strain inelastic deformations, respectively. Svendsen \textit{et al.} \cite{Svendsen2018} independently proposed a more unified framework that extends Helmholtz-based models, such as \cite{DurgaA2013}, to multiphase multicomponent solids undergoing finite-strain and inelastic deformations. 

Contrary to the mixed scheme, Mosler \textit{et al.} \cite{Mosler2014} proposed a partial rank-one homogenization scheme to enforce static and kinematic compatibilities for two-phase solids undergoing finite deformations. The advantage of this scheme over the mixed scheme is that it does not require coordinate transformation. Keifer \textit{et al.} showed improved convergence using this scheme compared to schemes that ensure either static or kinematic compatibility for two-phase solids undergoing small-strain deformations. Subsequently, Bartels \textit{et al.} \cite{Bartel2021} applied this scheme to couple mechanics with a WBM (Wheeler-Boettinger-McFadden) type chemical model \cite{Wheeler1992}. Unfortunately, unlike the previously discussed mechanically uncoupled models, the interface width in this model cannot be controlled due to an interfacial excess energy contribution coming from bulk chemical free energies \cite{Plapp2011}, \cite{Choudhary2012}. Later, Bartel's model was improved by the present authors by combining a grand-potential model with the partial rank-one scheme for two-phase solids undergoing small-strain deformations \cite{SC2}. Using this model, we also found that the rank-one scheme offered improved numerical convergence compared to either static or kinematically compatible schemes \cite{SC2}. 

Despite these advantages, the partial rank-one scheme has so far not been extended to multiphase and multicomponent solids undergoing linear elastic deformations. To our knowledge, the only published work that extends the rank-one scheme to multi-phase solids is by Sarhil \textit{et al.} \cite{Sarhil2021}. However, there are two limitations to this model. The first limitation is that it has not been coupled with diffusion equations and hence cannot be applied to simulate diffusive transformations. The second limitation is that it takes an interpolation function that is equal to the phase-field variable, i.e., $h_{\theta}(\boldsymbol{\phi})= \phi_{\theta}$, to interpolate elastic properties. As noted by Moelans \cite{Moelans2011}, this assumption may shift the local minima of the free energies and may cause inaccuracies. Hence, this paper aims to fill these gaps by formulating a multi-phase-field model based on a partial rank-one homogenization scheme starting from a grand-potential functional, thereby ensuring that the interfacial excess contribution due to bulk properties vanishes. To this end, we present an analytical approach to solving the static compatibility equations for a three-phase linear elastic solid. For quantitative accuracy, we use a coupling method developed in \cite{Chatterjee2021} that allows incorporating thermodynamic and kinetic properties obtained from CALPHAD databases into a grand-potential-based model. 

The paper is organized as follows. The phase-field formulation with the rank-one homogenization scheme is introduced in Section 2. In Section 3, the prerequisite chemical properties and the elastic properties for two—a Ni-Al and an Al-Ni-Cr—three-phase alloys are given. To demonstrate the application of our model, four numerical simulations are performed, and the results are discussed in Section 4. The accuracy of our numerical results is tested by comparing the phase-field simulations with analytically obtained solutions.  Finally, the conclusions of the paper are discussed in Section 5.
	
\section{Formulation}

\subsection{Notations}
In this paper, we assume an isothermal system consisting of $n$ diffusing component and $p$ phases. We denote a set of scalar fields with boldface letters. For example, the set of $(n-1)$ independent diffusion potential fields is denoted as ${\boldsymbol{\tilde{\mu}}}=\left\lbrace \tilde{\mu}_{k=1\hdots (n-1)}\right\rbrace$. Similarly, the set of $p$ phase-field variables is shown as $\boldsymbol{\phi} = \left\lbrace\phi_{\theta=\alpha \hdots p} \right\rbrace$. Vector and tensors are also represented with boldface letters, e.g., the displacement is written as $\boldsymbol{u} = u_{i}\boldsymbol{e}_{i}$, where $u_{i=1,2,3}$ are the components of $\boldsymbol{u}$ relative to a chosen orthonormal basis $\left\{\boldsymbol{e}_{i}\right\}$. Following standard notations, the Einstein summation convention is used throughout the paper to indicate summation over spatial dimensions. The dot, outer and inner products between two vectors, say $\boldsymbol{a}$ and $\boldsymbol{b}$, are written as $\boldsymbol{a}\cdot\boldsymbol{b}$, $\boldsymbol{a}\otimes \boldsymbol{b}$ and $\boldsymbol{a}:\boldsymbol{b}$, respectively. The norm, divergence, gradient and laplacian of a physical quantity, say $\boldsymbol{\Phi}$, are written as $\norm{\Phi}$, $\text{div}\,\boldsymbol{\Phi}$, $\text{grad}\,\boldsymbol{\Phi}$, and $\Delta \Phi$, respectively.

\subsection{Definitions of field variables and jump}
As mentioned before, since the diffusion potentials are the independent variables in a grand-potential-based model, any prerequisite property in the model should be expressed as functions of diffusion potentials. Precisely, the diffusion potential of a diffusing component, say $k$, is defined as the difference between its chemical potential and the chemical potential of the dependent component, i.e., $\tilde{\mu}_{k} = (\mu_{k} - \mu_{n})$, and it has units of $\text{J/mol}$. Further, an arbitrary phase in the system, say $\theta$, at any given spatial point $\boldsymbol{x}$ and time $t$ is indicated by the phase-field variable, $\phi_{\theta}(\boldsymbol{x},t)$, such that the bulk regions occupied by this phase are when $\phi_{\theta}=1$. Moreover, the jump of a field or property, $\llbracket \Phi\rrbracket^{\alpha\beta} = \Phi^{\alpha} - \Phi^{\beta}$, at an interface, say $\alpha/\beta$,  is defined as the difference between its bulk values within the two phases.
 
\subsection{Partial rank-one scheme for multi-phase systems}
Starting from the two-phase approach of Mosler \textit{et al.} \cite{Mosler2014}, we assume that the total strain, $\boldsymbol{\epsilon}(\boldsymbol{u})$, in the interfacial regions is a smooth function of the \textit{phase} strains assigned to each phase in the system. Precisely,
\begin{align}
\epsilon _{ij}(\boldsymbol{u}) &= \sum_{\theta=1}^{p}\epsilon_{ij}^{\theta}h_{\theta}(\boldsymbol{\phi}),
\label{Eqn1}
\end{align}
where $\boldsymbol{\epsilon}(\boldsymbol{u)}$ is the total strain as a function of the displacement $\boldsymbol{u}$, $\boldsymbol{\epsilon}^{\theta}$ and $h_{\theta}(\boldsymbol{\phi})$ are the (total) \textit{phase} strain and interpolation function attributed to phase $\theta$. Moreover, the total strain at a point is calculated using the linear strain-displacement relations
\begin{align}
\epsilon _{ij}(\boldsymbol{u})  = \left(1/2\right) [\text{grad}\,\boldsymbol{u} + \left(\text{grad}\,\boldsymbol{u}\right)^{T}].
\label{Extra0}
\end{align}
The choice of the interpolation function, $h(\boldsymbol{\phi})$, is such that in the bulk regions: $h_{\theta}= 1$ for $(\phi_{\theta} = 1,\phi_{\sigma\neq\theta} =0)$ and $h_{\theta} =0 $ for $(\phi_{\theta}=0, \phi_{\sigma\neq\theta} =1)$, while in the interfacial regions: $0<h_{\theta}<1$ for $0<\phi_{\theta}<1$. Further, as noted in \cite{Folch2003}, \cite{Moelans2011}, the function $h(\boldsymbol{\phi})$ must satisfy two additional requirements: i) $\sum_{\theta=1}^{p}h_{\theta}=1$, and ii) $dh_{\theta}/d\phi_{\theta}\left[\phi_{\theta}=1, \phi_{\sigma\neq\theta} =0\right]= 0$. Thus, similar to the function proposed by Moelans \cite{Moelans2011}, three different interpolation functions that fulfill these requirements have been formulated by Schneider and co-workers \cite{Schneider2017, Schneider2018, Herrmann2018}. However, for the sake of convenience, in this work we chose the function first proposed by Moelans \cite{Moelans2011}:
\begin{align}
h_{\theta}(\boldsymbol{\phi}) = \frac{\phi_{\theta}^{2}}{\sum_{\theta=1}^{p}\phi_{\theta}^{2}}\quad \text{for} \quad \theta=\lbrace \alpha, \beta \hdots, p \rbrace.
\label{Extra2}
\end{align}
As discussed in the Introduction section, it should be noted that Sarhil \textit{et al.} \cite{Sarhil2021} proposed $h_{\theta}(\boldsymbol{\phi}) = \phi_{\theta}$ which does not satisfy the above-mentioned requirements and may lead to inaccuracies. Another noteworthy difference between our model and the models proposed by Sarhil \textit{et al.}  \cite{Sarhil2021}  and Schneider  \textit{et al.} \cite{Schneider2017, Schneider2018, Herrmann2018} is that our model does not require a constraint that the sum of phase-fields should add up to $1$, i.e., $\sum_{\theta=1}^{p}\phi_{\theta} = 1$. 

It is worth noting that the phase strains introduced in Eq. (\ref{Eqn1}) are physically meaningful only in the bulk regions of a phase $(h_{\theta}= 1)$ but not in the interfacial regions. It is because, in the bulk regions, they become equal to the total strain, which is a physically measurable quantity that depends on the stiffness tensor and boundary conditions. But in the interfacial regions, the variation of phase strains depends on the interface width; a numerical parameter selected arbitrarily. In section \ref{three_phase}, we will show that the phase strains also depend on the homogenization scheme in the interfacial regions. Similar to \textit{phase} concentrations introduced in solidification studies \cite{Steinbach2012}, their primary purpose is to separate the bulk and interfacial contributions in the total energy for artificially enlarged interfaces.

Following the two-phase approach \cite{Mosler2014}, to ensure kinematic compatibility, the \textit{phase} strains must satisfy the Hadamard jump conditions. Consequently, the $p$ unknown \textit{phase} strains, $\{\boldsymbol{\epsilon}^{\alpha}, \boldsymbol{\epsilon}^{\beta},   \boldsymbol{\epsilon}^{\gamma} \hdots \boldsymbol{\epsilon}^{p}\}$, in Eq. (\ref{Eqn1}) must satisfy the following $(p-1)$ Hadamard jump conditions
\begin{align}
\begin{split}
\llbracket\epsilon_{ij}\rrbracket^{\alpha\beta} &= \epsilon^{\alpha}_{ij} - \epsilon^{\beta}_{ij} = \text{sym}  (a^{\alpha \beta}_{i}n^{\alpha \beta}_{j}),\\
\llbracket\epsilon_{ij}\rrbracket^{\beta\gamma} &=  \epsilon^{\beta}_{ij} - \epsilon^{\gamma}_{ij} = \text{sym} ( a^{\beta\gamma}_{i}n^{\beta\gamma}_{j}),\\
&\vdots\\
\llbracket\epsilon_{ij}\rrbracket^{(p-1),p} &=  \epsilon^{p-1}_{ij} - \epsilon^{p}_{ij} = \text{sym}\left[ a^{(p-1),p}_{i}n^{(p-1),p}_{j}\right],\\
\end{split}
\label{Eqn3}
\end{align}
where $\lbrace\boldsymbol{a}^{\alpha\beta}$, $\boldsymbol{a}^{\beta\gamma}$, $\hdots,$ $\boldsymbol{a}^{(p-1),p}\rbrace$ and $\lbrace\boldsymbol{n}^{\alpha\beta}$, $\boldsymbol{n}^{\beta\gamma},$ $\hdots,$ and $\boldsymbol{n}^{(p-1),p}\rbrace$ are the  jump vectors and unit normals at the $\lbrace \alpha/\beta, \beta/\gamma, \hdots, (p-1)/p\rbrace$ interfaces, respectively. Here, the notation $\llbracket\boldsymbol{\epsilon}\rrbracket^{(p-1),p}$ denotes the strain jump at the interface between phases $(p-1)$ and $p$, and is equal to the outer product between the jump vector and unit normal at that interface. It should be noted that the jump vector, $\boldsymbol{a}^{\alpha\beta}$, at the $\alpha/\beta$ interface is symmetric with respect to the superscripts $\alpha\beta$ \cite{Schneider2017}, i.e., $\boldsymbol{a}^{\alpha\beta} = \boldsymbol{a}^{\beta\alpha}$, since by definition $\llbracket \boldsymbol{\epsilon}\rrbracket^{\alpha\beta} =- \llbracket \boldsymbol{\epsilon}\rrbracket^{\beta\alpha}$ and $\boldsymbol{n}^{\alpha\beta} = -\boldsymbol{n}^{\beta\alpha}$.

Moreover, following our previous work \cite{SC2}, we define the unit normal at an interface as \cite{Kazaryan2001}, \cite{Schneider2017}
\begin{align}
\boldsymbol{n}^{\theta\sigma, \sigma\neq\theta} = -\text{grad}\, \phi_{\theta}/\norm{\text{grad}\,\phi_{\theta}},\quad  \theta=\{\alpha,\beta\hdots (p-1)\}\, \&\,\sigma = \lbrace\beta,\gamma, \hdots, p\rbrace
\label{Extra0.1},
\end{align}
where $\norm{\nabla \phi_{\theta}}$ is the norm of the gradient of the phase-field variable $\phi_{\theta}$.
We note that although a multi-phase-field version of Eq. (\ref{Extra0.1}) exists (see \cite{Nestler2005} \& \cite{Moelans2008PRB}), we have not used that definition in this paper for the sake of simplicity. Moreover, Schneider \textit{et al.} \cite{Schneider2017} has noted that the solution of elastic fields is not significantly dependent on the definition of the unit normal vector. 

Eqs. (\ref{Eqn1}) and (\ref{Eqn3}) form a system of $p$ equations that can be analytically solved to explicitly determine the $p$-\textit{phase} strains: $\{\boldsymbol{\epsilon}^{\alpha}, \boldsymbol{\epsilon}^{\beta},   \boldsymbol{\epsilon}^{\gamma} \hdots \boldsymbol{\epsilon}^{p}\}$, as functions of the total strain $\boldsymbol{\epsilon}(\boldsymbol{u})$, $p$ interpolation functions $h_{\theta=\alpha,\beta\hdots p}(\boldsymbol{\phi})$ and $(p-1)$ strain jumps: $\lbrace\llbracket\boldsymbol{\epsilon}\rrbracket^{\alpha\beta}, \llbracket\boldsymbol{\epsilon}\rrbracket^{\beta\gamma}, \hdots,\llbracket \boldsymbol{\epsilon}\rrbracket^{(p-1),p}\rbrace$. In Appendix A, we show how to analytically calculate the \textit{phase} strains for a multiphase system as functions of the total strain, interpolation functions and strain jumps.

Next, we calculate the unknown jump vectors: $\{\boldsymbol{a}^{\alpha\beta}, \boldsymbol{a}^{\beta\gamma}, \hdots,\boldsymbol{a}^{(p-1),p}\}$ in order to determine the $(p-1)$ strain jumps. Similar to previous works, e.g., \cite{Mosler2014}, \cite{SC2}, \cite{Schneider2017}, we also calculate the jump vector at an interface by solving the static compatibility equation. To be precise, the following $(p-1)$ static compatibility equations must be solved to determine the same number of unknown jump vectors
\begin{align}
\begin{split}
\llbracket\sigma_{ij}\rrbracket^{\alpha\beta}n_{j}^{\alpha\beta} &= \left(\sigma_{ij}^{\alpha} - \sigma_{ij}^{\beta}\right)n_{j}^{\alpha\beta} = 0_{i} ,\\
\llbracket\sigma_{ij}\rrbracket^{\beta\gamma}n_{j}^{\beta\gamma} &= \left(\sigma_{ij}^{\beta} - \sigma_{ij}^{\gamma}\right)n_{j}^{\beta\gamma} = 0_{i},\\
&\vdots\\
\llbracket\sigma_{ij}\rrbracket^{(p-1),p}n_{j}^{(p-1),p} &=\left(\sigma_{ij}^{p-1} - \sigma_{ij}^{p}\right)n_{j}^{(p-1),p} = 0_{i},
\end{split}
\label{Extra3}
\end{align}
where $\boldsymbol{\sigma}^{\theta}$ is the elastic stress associated with phase $\theta$. In this paper, we have introduced static compatibility equations as a means to calculate the jump vectors. Alternatively, these equations can be derived by minimizing the total elastic strain energy with respect to the jump vectors, as pointed out by Mosler \textit{et al.} \cite{Mosler2014} and Sarhil \textit{et al.} \cite{Sarhil2021}  .

Using linear elastic theory, the elastic \textit{phase} stresses, $\{\boldsymbol{\sigma}^{\alpha}, \boldsymbol{\sigma}^{\beta},   \boldsymbol{\sigma}^{\gamma} \hdots \boldsymbol{\sigma}^{p}\}$, are related to the \textit{phase} strains by the generalized Hooke's law
\begin{align}
\sigma_{ij}^{\theta} = \mathcal{C}_{ijkl}^{\theta}\left(\epsilon_{kl}^{\theta} - \epsilon_{kl}^{\star\theta}\right) \quad \text{for} \quad \theta=\lbrace \alpha, \beta \hdots, p \rbrace,
\label {Extra4}
\end{align}
where $\boldsymbol{\mathcal{C}}^{\theta}$ and $\boldsymbol{\epsilon}^{\star\theta}$ denote the fourth-rank stiffness tensor and eigenstrain belonging to a particular phase $\theta$, respectively.

It follows from the set of Eqs. (\ref{Eqn3}) \& (\ref{Extra3}) that the rank-one scheme ensures both kinematic and static compatibilities at $(p-1)$ interfacial regions of a $p$-phase system. It is worth pointing out that a system consisting of $p$-phases may have ${p(p-1)/2}$ two-phase junctions. However, Eqs. (\ref{Eqn3}) \& (\ref{Extra3}) only ensure static and kinematic compatibilities at $(p-1)$ of these junctions. It can be shown that mechanical compatibilities at remaining $(p-1)(p-2)/2$ junctions are implicitly ensured. For instance, by adding Eqs. (\ref{Eqn3}) \& (\ref{Extra3}), we obtain the following set of compatibility equations
\begin{align}
\llbracket\epsilon_{ij}\rrbracket^{\alpha,p} &= \epsilon_{ij}^{\alpha} - \epsilon_{ij}^{p} = \lbrace a_{i}^{\alpha\beta}n_{j}^{\alpha\beta} + a_{i}^{\beta\gamma}n_{j}^{\beta\gamma}+ \hdots+ a_{i}^{(p-1),p}\rbrace\label{ExtraN8},\\
\llbracket\sigma_{ij}\rrbracket^{\alpha,p} n_{j}^{\alpha,p}&= \left(\sigma_{ij}^{\alpha} - \sigma_{ij}^{p}\right)n_{j}^{\alpha,p} = 0_{i}.\label{ExtraN9}
\end{align} 

From Eqs. (\ref{ExtraN8}) and (\ref{ExtraN9}), it follows that both static and kinematic compatibilities are ensured at the interface between phases $\alpha$ and $p$.

Finally, it should be emphasized that, depending on the constitutive equations, the jump vectors can be solved either analytically or numerically. As discussed in the Introduction section, for non-linear elastic solids, the jump vectors can be obtained only by numerically solving the set of static compatibility equations at each grid point and time step (see \cite{Schneider2017},\cite{Herrmann2018}), i.e., Eqs. (\ref{Extra3}). However, for linear elastic solids, the jump vectors can be determined either analytically or numerically. Although restricted to two phases, the analytical approach was followed in \cite{Kiefer2017} \& \cite{SC2}, while a Newton-Raphson scheme was used in \cite{Hansen2021}. Nevertheless, to our knowledge, analytical expressions for the jump vectors in a multi-phase-field setting do not exist. Since analytical approaches may offer computational gains over numerical solutions, particularly when linear constitutive equations are assumed, we follow the former approach.

However, deriving a general analytical expression for the jump vectors for a $p$-phase system is not straightforward as it requires explicit analytical expressions for \textit{phase} strains. Since the analytical expressions for the \textit{phase} strains become increasingly complicated as the number of phases increases (see Appendix A), we therefore take a special case to illustrate how to derive the jump vectors in a multi-phase-field context. For the sake of convenience, we chose a three-phase system for this derivation.

\subsection{Partial rank-one scheme for three-phase systems}
\label{three_phase}
For a three-phase system, the \textit{phase} strains belonging to phases, say $\alpha$, $\beta$ and $\gamma$, may be written as (see Appendix A)
\begin{align}
\epsilon^{\alpha}_{ij}\left(\boldsymbol{\epsilon}, \boldsymbol{\phi}, \llbracket\boldsymbol{\epsilon}\rrbracket^{\alpha\beta}, \llbracket\boldsymbol{\epsilon}\rrbracket^{\beta\gamma}\right) &= \epsilon_{ij}(\boldsymbol{u}) + \left[h_{\beta}(\boldsymbol{\phi})  + h_{\gamma}(\boldsymbol{\phi})\right]\llbracket \epsilon_{ij} \rrbracket^{\alpha\beta} + h_{\gamma}(\boldsymbol{\phi})\llbracket \epsilon_{ij}\rrbracket^{\beta\gamma},\label{Eqn4}\\
\epsilon^{\beta}_{ij}\left(\boldsymbol{\epsilon}, \boldsymbol{\phi}, \llbracket\boldsymbol{\epsilon}\rrbracket^{\alpha\beta}, \llbracket\boldsymbol{\epsilon}\rrbracket^{\beta\gamma}\right)  &= \epsilon_{ij}(\boldsymbol{u})  - h_{\alpha}(\boldsymbol{\phi})\llbracket \epsilon_{ij} \rrbracket^{\alpha\beta}  + h_{\gamma}(\boldsymbol{\phi})\llbracket \epsilon_{ij}\rrbracket^{\beta\gamma},\label{Eqn5}\\
\epsilon_{ij}^{\gamma}\left(\boldsymbol{\epsilon}, \boldsymbol{\phi}, \llbracket\boldsymbol{\epsilon}\rrbracket^{\alpha\beta}, \llbracket\boldsymbol{\epsilon}\rrbracket^{\beta\gamma}\right)  &= \epsilon_{ij}(\boldsymbol{u})  - h_{\alpha}(\boldsymbol{\phi})\llbracket \epsilon_{ij}\rrbracket^{\alpha\beta} - \left[h_{\beta}(\boldsymbol{\phi}) + h_{\alpha}(\boldsymbol{\phi})\right]\llbracket \epsilon_{ij}\rrbracket^{\beta\gamma}\label{Eqn6}.
\end{align}
It follows from Eqs. (\ref{Eqn4})-(\ref{Eqn6}) that the \textit{phase} strains are always equal to the total strain $\boldsymbol{\epsilon}(\boldsymbol{u})$ in the bulk regions of the phases. However, in the interfacial regions, they differ depending on the definition of strain jumps, which in turn depends on the homogenization assumption. Concretely, the strain jumps vanishes, i.e., $\llbracket\boldsymbol{\epsilon}\rrbracket^{\alpha\beta} = \llbracket\boldsymbol{\epsilon}\rrbracket^{\beta\gamma} = \boldsymbol{0}$, for the case of Voigt-Taylor homogenization scheme (henceforth referred to as the VT scheme) or the Khachaturyan scheme \cite{DurgaA2013}, while for the partial rank-one scheme  (henceforth referred to as PR scheme) the strain jumps are given by (see Eqs. \ref{Eqn3})
\begin{align}
\llbracket \boldsymbol{\epsilon}\rrbracket^{\alpha\beta} &= \boldsymbol{a}^{\alpha\beta}\otimes \boldsymbol{n}^{\alpha\beta},\label{Eqn6a1}\\
\llbracket \boldsymbol{\epsilon}\rrbracket^{\beta\gamma} &= \boldsymbol{a}^{\beta\gamma}\otimes \boldsymbol{n}^{\beta\gamma}.
\label{Eqn6a2}
\end{align}

As previously discussed, the jump vectors $\boldsymbol{a}^{\alpha\beta}$ and $\boldsymbol{a}^{\beta\gamma}$ in Eqs. (\ref{Eqn6a1}) \& (\ref{Eqn6a2}) are obtained by solving the static compatibility equations. Precisely, the set of Eqs. (\ref{Extra3}) for a three-phase system reduced to
\begin{align}
\left(\sigma_{ij}^{\alpha} - \sigma_{ij}^{\beta}\right)n_{j}^{\alpha\beta} = 0_{i} \label{Eqn10},\\
\left(\sigma_{ij}^{\beta} - \sigma_{ij}^{\gamma}\right)n_{j}^{\beta\gamma} = 0_{i}\label {Eqn11}.
\end{align}
Next, it follows from Eq. (\ref{Extra4}) that the elastic \textit{phase} stresses in Eqs. (\ref{Eqn10}) \& (\ref{Eqn11}) are related to the \textit{phase} strains by
\begin{align}
\sigma_{ij}^{\alpha} &= \mathcal{C}_{ijkl}^{\alpha} \left[\epsilon_{kl}^{\alpha} - \epsilon_{kl}^{\star\alpha}\right]\label{Eqn7},\\
\sigma_{ij}^{\beta} &= \mathcal{C}_{ijkl}^{\beta}\left[\epsilon_{kl}^{\beta} - \epsilon_{kl}^{\star\beta}\right]\label{Eqn8},\\
\sigma_{ij}^{\gamma} &= \mathcal{C}_{ijkl}^{\gamma}\left[\epsilon_{kl}^{\gamma} -  \epsilon_{kl}^{\star\gamma}\right].
\label{Eqn9}
\end{align}
Now, substituting Eqs. (\ref{Eqn7})-(\ref{Eqn9}) in Eqs. (\ref{Eqn10})-(\ref{Eqn11}) yields
\begin{align}
\left[\left(\mathcal{C}_{ijkl}^{\alpha} - \mathcal{C}_{ijkl}^{\beta}\right)\epsilon_{kl} + \lambda^{1}_{ijkl}\llbracket \epsilon_{kl}\rrbracket^{1}+ \lambda^{2}_{ijkl}\llbracket \epsilon_{kl}\rrbracket^{2}\right]n^{1}_{j} &= \mathcal{Z}_{i}^{1}\label{Eqn12},\\
\left[\left(\mathcal{C}_{ijkl}^{\beta} - \mathcal{C}_{ijkl}^{\gamma}\right)\epsilon_{kl} + \mathcal{M}^{1}_{ijkl}\llbracket \epsilon_{kl}\rrbracket^{1}+ \mathcal{M}^{2}_{ijkl}\llbracket \epsilon_{kl}\rrbracket^{2}\right]n_{j}^{2} &= \mathcal{Z}_{i}^{2}\label{Eqn13},
\end{align}
where, we have denoted the superscripts $\alpha\beta$ and $\beta\gamma$ by 1 \& 2, respectively, and
\begin{align}
\begin{split}
\lambda^{1}_{ijkl}(\boldsymbol{\phi}) &= h_{\beta}(\boldsymbol{\phi})\mathcal{C}_{ijkl}^{\alpha}+ h_{\alpha}(\boldsymbol{\phi})\mathcal{C}_{ijkl}^{\beta} + h_{\gamma}(\boldsymbol{\phi})\mathcal{C}_{ijkl}^{\alpha} ,\\
\lambda^{2}_{ijkl}(\boldsymbol{\phi}) &= h_{\gamma}(\boldsymbol{\phi})\mathcal{C}_{ijkl}^{\alpha}- h_{\gamma}(\boldsymbol{\phi})\mathcal{C}_{ijkl}^{\beta},\\
\mathcal{M}_{ijkl}^{1}(\boldsymbol{\phi})  &= h_{\alpha}(\boldsymbol{\phi})\mathcal{C}^{\gamma}_{ijkl} - h_{\alpha}(\boldsymbol{\phi})\mathcal{C}_{ijkl}^{\beta},\\
\mathcal{M}_{ijkl}^{2}(\boldsymbol{\phi} )  &= h_{\gamma}(\boldsymbol{\phi})\mathcal{C}_{ijkl}^{\beta} + h_{\alpha}(\boldsymbol{\phi})\mathcal{C}_{ijkl}^{\gamma}  + h_{\beta}(\boldsymbol{\phi})\mathcal{C}_{ijkl}^{\gamma}, \\
\mathcal{Z}_{i}^{1}(\boldsymbol{n}^{1}) & = \left\{\mathcal{C}_{ijkl}^{\alpha}\epsilon_{kl}^{\star\alpha} - \mathcal{C}_{ijkl}^{\beta}\epsilon_{kl}^{\star\beta}\right\}n_{j}^{1},\\
\mathcal{Z}_{i}^{2}(\boldsymbol{n}^{2})    &=  \left\{\mathcal{C}_{ijkl}^{\beta}\epsilon_{kl}^{\star\beta} - \mathcal{C}_{ijkl}^{\gamma}\epsilon_{kl}^{\star\gamma}\right\}n_{j}^{2},
\end{split}
\end{align}
Then, substituting Eqs. (\ref{Eqn6a1}) \& (\ref{Eqn6a2}) in Eqs. (\ref{Eqn12}) \& (\ref{Eqn13}) yields
\begin{align}
\left(m_{i}^{\alpha1} - m_{i}^{\beta1} \right) + \mathcal{\lambda}_{ik}^{\#}a_{k}^{1} + \lambda_{ik}^{\star}a_{k}^{2} = \mathcal{Z}_{i}^{1},\label{Eq.14a}\\
\left(\psi_{i}^{\beta2} - \psi_{i}^{\gamma2} \right) + \mathcal{L}_{ik}^{\#}a_{k}^{1} + \mathcal{L}_{ik}^{\star}a_{k}^{2} = \mathcal{Z}_{i}^{2}\label{Eq.15a},
\end{align}
where
\begin{align}
\begin{split}
m_{i}^{\alpha1}\left(\boldsymbol{\epsilon}, \boldsymbol{n}^{1}\right) &= \mathcal{C}_{ijkl}^{\alpha}\epsilon_{kl}n_{j}^{1},\\
m_{i}^{\beta1}\left(\boldsymbol{\epsilon}, \boldsymbol{n}^{1}\right)  &= \mathcal{C}_{ijkl}^{\beta}\epsilon_{kl}n_{j}^{1},\\
\psi_{i}^{\beta2}\left(\boldsymbol{\epsilon}, \boldsymbol{n}^{2}\right) &= \mathcal{C}_{ijkl}^{\beta}\epsilon_{kl}n_{j}^{2},\\
\psi_{i}^{\gamma2}\left(\boldsymbol{\epsilon}, \boldsymbol{n}^{2}\right) &= \mathcal{C}_{ijkl}^{\gamma}\epsilon_{kl}n_{j}^{2},\\
\lambda_{ki}^{\#}(\boldsymbol{\phi}, \boldsymbol{n}^{1}) &= n_{l}^{1}\lambda_{lkij}^{1}(\boldsymbol{\phi})n_{j}^{1},\\
\lambda_{ki}^{\star}(\boldsymbol{\phi}, \boldsymbol{n}^{1}, \boldsymbol{n}^{2}) &= n_{l}^{2}\lambda_{lkij}^{2}(\boldsymbol{\phi})n_{j}^{1},\\
\mathcal{L}_{ki}^{\#}(\boldsymbol{\phi},\boldsymbol{n}^{1}, \boldsymbol{n}^{2}) &= n_{l}^{1}\mathcal{M}_{lkij}^{1}(\boldsymbol{\phi})n_{j}^{2},\\
\mathcal{L}_{ki}^{\star}(\boldsymbol{\phi}, \boldsymbol{n}^{2}) &= n_{l}^{2}\mathcal{M}_{lkij}^{2}(\boldsymbol{\phi})n_{j}^{2}.
\end{split}
\label{Extra20}
\end{align}
Rearranging Eq. (\ref{Eq.15a}) and solving for $\boldsymbol{a}^{2}$ yields
\begin{align}
a_{j}^{2}(\boldsymbol{\phi}, \boldsymbol{\epsilon}, \boldsymbol{n}^{1}, \boldsymbol{n}^{2}) = \mathcal{S}_{ji}(\boldsymbol{\phi}, , \boldsymbol{n}^{2})b_{i},
\label{Eqn16}
\end{align}
where $\mathcal{S}_{ji} = \left(\mathcal{L}_{ij}^{\star}\right)^{-1}$ and $b_{i} = \mathcal{Z}_{i}^{2}-\left(\psi_{i}^{\beta2} - \psi_{i}^{\gamma2} + \mathcal{L}_{ik}^{\#}a_{k}^{1}\right)$. 
Next, substituting $a_{j}^{2}$ in Eq. (\ref{Eq.14a}) yields
\begin{align}
\lambda^{\#}_{pq}a_{q}^{1} + \lambda_{pq}^{\star}\left(\mathcal{S}_{qi}b_{i}\right) = \mathcal{Z}_{i}^{1}-\left(m_{p}^{\alpha1} - m_{p}^{\beta1} \right)
\label{Eqn17}
\end{align}
Using the expression for $\boldsymbol{b}$ and then solving for $\boldsymbol{a}^{1}$ using Eq.(\ref{Eqn17}) finally yields
\begin{align}
a_{k}^{1}(\boldsymbol{\epsilon}, \boldsymbol{\phi},\boldsymbol{n}^{1}, \boldsymbol{n}^{2}) = \left(\mathcal{D}_{pk}\right)^{-1}\left[ \mathcal{Z}_{p}^{1}-\left(m_{p}^{\alpha1} - m_{p}^{\beta1} \right) - \lambda_{pr}^{\star}\mathcal{S}_{ri}\left\{\mathcal{Z}_{i}^{2} -\left(\psi_{i}^{\beta2} - \psi_{i}^{\gamma2}\right) \right\} \right]
\label{Eqn18}
\end{align}
where $\mathcal{D}_{pk} = \lambda_{pk}^{\#} - \lambda^{\star}_{pr}\mathcal{S}_{ri}\mathcal{L}_{ik}^{\#}$. For a three-phase-field model, Eqs. (\ref{Eqn16}) and (\ref{Eqn18}) are the most general expressions for the jump vectors $\boldsymbol{a}^{2}$ and $\boldsymbol{a}^{1}$, respectively. 

To further simplify these expressions, we assume that the two second-rank tensors, $\boldsymbol{\lambda}^{\star}$ and $\boldsymbol{\mathcal{L}}^{\#}$, are zero. Because from Eq. (\ref{Extra20}) we see that these tensors depend on both the unit vectors, $\boldsymbol{n}^{1}$ and $\boldsymbol{n}^{2}$, which are simultaneously non-zero only at the triple points. Perhaps not surprisingly, by making this assumption we have strictly restricted the definition of jump vectors to the two-phase regions. Stated differently, we have enforced static compatibility only at the two-phase junctions. Our assumption is justified since the set of Eqs. (\ref{Extra3}) is strictly valid at the two-phase junctions only where the unit normal vector to the interface is uniquely defined. As a consequence, Eqs. (\ref{Eqn16}) and (\ref{Eqn18}) simplifies to
\begin{align}
a_{j}^{2}  &= - \left(\mathcal{L}_{ij}^{\star}\right)^{-1}\left[ \left(\mathcal{C}^{\beta}_{ikpq} - \mathcal{C}^{\gamma}_{ikpq} \right)\epsilon_{pq} - \left(\mathcal{C}^{\beta}_{ikpq}\epsilon_{pq}^{\star\beta} - \mathcal{C}^{\gamma}_{ikpq}\epsilon_{pq}^{\star\gamma}\right)\right] n_{k}^{2},\label{Eqs29N}\\
a_{j}^{1}  &= - \left(\mathcal{\lambda}_{ij}^{\#}\right)^{-1}\left[ \left(\mathcal{C}^{\alpha}_{ikpq} - \mathcal{C}^{\beta}_{ikpq} \right)\epsilon_{pq} - \left(\mathcal{C}^{\alpha}_{ikpq}\epsilon_{pq}^{\star\alpha} - \mathcal{C}^{\beta}_{ikpq}\epsilon_{pq}^{\star\beta}\right)\right] n_{k}^{1},\label{Eqs30N}
\end{align}

where
\begin{align}
\mathcal{L}_{ij}^{\star}(\boldsymbol{\phi}, \boldsymbol{n}^{2}) &= n_{l}^{2}
\left[h_{\gamma}(\boldsymbol{\phi})\mathcal{C}_{lijr}^{\beta}  + h_{\alpha}(\boldsymbol{\phi})\mathcal{C}_{lijr}^{\gamma} + h_{\beta}(\boldsymbol{\phi})\mathcal{C}_{lijr}^{\gamma}\right]n_{r}^{2},\\
\mathcal{\lambda}_{ij}^{\#}(\boldsymbol{\phi}, \boldsymbol{n}^{2}) &= n_{l}^{1}
\left[h_{\beta}(\boldsymbol{\phi})\mathcal{C}_{lijr}^{\alpha}  + h_{\alpha}(\boldsymbol{\phi})\mathcal{C}_{lijr}^{\beta} + h_{\gamma}(\boldsymbol{\phi})\mathcal{C}_{lijr}^{\alpha}\right]n_{r}^{1}.
\end{align}

Expectedly, we see that the analytically derived expressions for jump vectors, i.e., Eqs. (\ref{Eqs29N}) \& (\ref{Eqs30N}), are similar to the expression of jump vector derived in a two-phase setting (cf. Eq. (9) in \cite{SC2}). As noted in a previous work \cite{SC2}, we find that the magnitude of the jump vector at an interface, say $\alpha/\beta$, is proportional to two elastic properties: i) the jump in stiffness tensors of the bulk phases, and ii) the eigenstrains in the bulk phases.

\subsection{Functional, overall molar density and elastic stresses} 
Here, starting from a grand-potential functional we derive expressions for the overall molar density of a diffusing component and elastic stresses. As discussed in the Introduction section, we follow the grand-potential approach \cite{Plapp2011}, \cite{Choudhary2012} in this work because we don't need to explicitly solve for the quasiequilibrium conditions \cite{Eiken2006}, that may lead to computational gains.

The grand-potential functional, $\Omega[\boldsymbol{\phi}, \tilde{\boldsymbol{\mu}}, \boldsymbol{u}]$, of the system for an elastically stressed multiphase multicomponent alloy is given by 
\begin{align}
\Omega\left[\boldsymbol{\phi}, \tilde{\boldsymbol{\mu}}, \boldsymbol{u}\right] = \int_{V}\left[\omega_{bulk}\left(\boldsymbol{\phi}, \tilde{\boldsymbol{\mu}}, \boldsymbol{\epsilon}\right) + \omega_{int}\left(\boldsymbol{\phi},\nabla \boldsymbol{\phi}\right)\right]dv,
\label{Eqn19}
\end{align}
where the bulk contribution to the total grand-potential density is denoted by $\omega_{bulk}\left(\boldsymbol{\phi}, \tilde{\boldsymbol{\mu}}, \boldsymbol{\epsilon}\right)$; the interfacial energy contribution to the total grand-potential density is indicated by $\omega_{int}(\boldsymbol{\phi}, \nabla \boldsymbol{\phi})$; and $V$ is the total volume. Further, the bulk contribution to the total functional, i.e., $\omega_{bulk}$ [J/m$^3$], is defined as
\begin{align}
\omega_{bulk}(\boldsymbol{\phi},\tilde{\boldsymbol{\mu}}, \boldsymbol{\epsilon}) &= \sum_{\theta=1}^{p}h_{\theta}(\boldsymbol{\phi})\omega^{\theta}_{bulk}\left(\tilde{\boldsymbol{\mu}}, \boldsymbol{\epsilon}^{\theta}\right),\label{Eqn20}
\end{align}
where $h_{\theta}(\boldsymbol{\phi})$ is the interpolation function, which is defined at Eq. (\ref{Extra2}) and $\omega_{bulk}^{\theta}$ is the grand-potential density of phase $\theta$ expressed as functions of diffusion potentials $\boldsymbol{\tilde{\mu}}$ and \textit{phase} strains $\boldsymbol{\epsilon}^{\theta}$. Under the assumption that each phase is represented by a single grain orientation, the interfacial energy contribution to the total energy may be written as \cite{Moelans2011}
 \begin{align}
\omega_{int}(\boldsymbol{\phi}, \nabla \boldsymbol{\phi}) &= \sum_{\theta=1}^{p}(1/2)\kappa\norm{\text{grad}\,\phi_{\theta}}^{2} + mg(\boldsymbol{\phi}),
\label{Eqn21}
\end{align}
where the two constant parameters $\kappa$ $[\text{J/m}]$ and $m$ $\left[\text{J/m}^{3}\right]$ are related to  the interfacial energy $\sigma_{\alpha\beta}$ and interface width $l_{\alpha\beta}$ by \cite{Moelans2011}
\begin{align}
\kappa = \left(3.0/4.0\right)\sigma_{\alpha\beta} l_{\alpha\beta},\quad m = 6.0\left(\sigma_{\alpha\beta}/l_{\alpha\beta}\right),
\label{Eqn22}
\end{align}
assuming uniform interface properties and a multi-well function $g(\boldsymbol{\phi})$ of the form
\cite{Moelans2011}
\begin{align}
\begin{split}
g(\boldsymbol{\phi}) &= \sum_{\theta=1}^{p}\left[\left(1/4\right)\phi_{\theta}^{4} - \left(1/2\right)\phi_{\theta}^{2} \right] + (3/4)\sum_{\theta=1}^{p}\sum_{\substack{\sigma=1\\
\sigma>\theta}}^{p}\phi_{\theta}^{2}\phi_{\sigma}^{2} + (1/4).
\end{split}
\label{Eqn23}
\end{align}
Following our previous work \cite{SC2}, the bulk grand-potential density $\omega^{\theta}_{bulk}(\tilde{\boldsymbol{\mu}}, \boldsymbol{\epsilon}^{\theta})$ in Eq. (\ref{Eqn20}) is written as
\begin{align}
\omega_{bulk}^{\theta}(\boldsymbol{\epsilon}^{\theta}, \boldsymbol{\tilde{\mu}}) = \omega_{chem}^{\theta}(\tilde{\boldsymbol{\mu}}) + (1/2)\mathcal{C}_{ijkl}^{\theta}(\boldsymbol{\tilde{\mu}})\left[\epsilon_{kl}^{\theta} - \epsilon_{kl}^{\star\theta}(\boldsymbol{\tilde{\mu}})\right]\left[\epsilon_{ij}^{\theta} - \epsilon_{ij}^{\star\theta}(\boldsymbol{\tilde{\mu}})\right].
\label{Eqn24}
\end{align}
The first and second terms in Eq. (\ref{Eqn24}) are the chemical and elastic energy contributions to the bulk grand-potential density of a phase $\theta$, respectively. Precisely, $\omega_{chem}^{\theta}$  is defined as $\Omega_{m}^{\theta}/V_{m}$, where $\Omega_{m}^{\theta}$ is the molar grand-potential and $V_{m}$ is the molar volume, which is assumed to be constant. Moreover, $\Omega_{m}^{\theta}$ can be analytically calculated by assuming either parabolic or dilute or ideal free energies \cite{Plapp2011}. This was the approach taken in our previous study \cite{SC2}. However, it is difficult to extend this approach to multi-phase and multi-component alloy systems. Thus, in this work we take a numerical approach to calculate the chemical grand-potential from CALPHAD databases using the method developed in \cite{Chatterjee2021}. It should be noted that, similar to our previous study \cite{SC2}, here we have assumed that the stiffness tensor and the eigenstrains are functions of diffusion potentials to account for composition-engendered stresses in the model.

Next, we derive an expression for the overall molar density. Thus, differentiating Eq. (\ref{Eqn20}) with respect to the diffusion potential gives
\begin{align}
c_{r}(\boldsymbol{\phi}, \boldsymbol{\tilde{\mu}},\boldsymbol{\epsilon}^{\theta}) = -\frac{\partial \omega_{bulk}}{\partial \mu_{r}} = \sum_{\theta=1}^{p}h_{\theta}(\boldsymbol{\phi})c_{r}^{\theta}\left(\boldsymbol{\tilde{\mu}},\boldsymbol{\epsilon}^{\theta}\right),
\label{Extra39N}
\end{align}
where $c_{r}$ and $c_{r}^{\theta}$ are the overall  and \textit{phase} molar densities of a diffusing component $r$, and have units of  mol/m$^3$. More precisely, using Eq. (\ref{Eqn24}) the \textit{phase} molar density may be explicitly written as \cite{SC2}
\begin{align}
\begin{split}
c_{r}^{\theta}\left(\boldsymbol{\epsilon}^{\theta}, \boldsymbol{\tilde{\mu}}\right) &= -\frac{\partial \omega_{bulk}^{\theta}}{\partial \mu_{r}} \\
&= \frac{X_{r}^{\theta}(\boldsymbol{\tilde{\mu}})}{V_{m}} - \frac{1}{2}\frac{\partial \mathcal{C}_{ijkl}^{\theta}}{\partial \tilde{\mu}_{r}}\left[\epsilon_{kl}^{\theta} -\epsilon_{kl}^{\star\theta}\right]\left[\epsilon_{ij}^{\theta} - \epsilon_{ij}^{\star\theta}\right] + \frac{\partial \epsilon_{ij}^{\star\theta}}{\partial \tilde{\mu}_{r}}\sigma_{ij}^{\theta} 
\label{Eqn25},
\end{split}
\end{align}
 where $X_{r}^{\theta}(\boldsymbol{\tilde{\mu}})=-\partial \Omega_{m}^{\theta}/\partial \mu_{r}$ \cite{Chatterjee2021} is the \textit{phase} mole fraction of component $r$ in phase $\theta$. It follows from Eq. (\ref{Eqn25}) that if the stiffness tensor and the eigenstrains are assumed to be uniform throughout the system, then Eq. (\ref{Eqn25}) simplifies to
 \begin{align}
 c_{r}^{\theta}\left(\boldsymbol{\tilde{\mu}}\right) &= \frac{X_{r}^{\theta}(\boldsymbol{\tilde{\mu}})}{V_{m}}\label{Eqn25N}.
 \end{align}
Since we will assume uniform elastic properties in this paper, we will use Eq. (\ref{Eqn25N}) to define the \textit{phase} molar densities. Moreover, the prerequisite \textit{phase} mole fractions, $X_{r}^{\theta}$, can be calculated either analytically assuming either parabolic or dilute or ideal free energies \cite{Plapp2011},  or can be directly obtained from CALPHAD databases  \cite{Chatterjee2021}. In this work, we will follow the latter approach since simplistic free energies may cause inaccuracies, particularly for multiphase and multicomponent alloys.
Finally, we derive an expression for the overall stress. Thus, differentiating Eq. (\ref{Eqn20}) with respect to total strain, it can be shown that (Appendix C)
 \begin{align}
 \sigma_{ij}(\boldsymbol{\phi}, \boldsymbol{\tilde{\mu}},\boldsymbol{\epsilon}^{\theta})   =\frac{\partial \omega_{bulk}}{\partial \epsilon_{ij}} =  \sum_{\theta=1}^{p}h_{\theta}(\boldsymbol{\phi})\sigma_{ij}^{\theta}(\boldsymbol{\tilde{\mu}},\boldsymbol{\epsilon}^{\theta}),
\label{Eqn25a}
 \end{align}
 where $\sigma_{ij}$ and $\sigma_{ij}^{\theta} = \partial \omega_{bulk}^{\theta}/\partial \epsilon_{ij}^{\theta}$ are the overall and \textit{phase} elastic stresses. The latter is defined at Eq. (\ref{Extra4})

\subsection{Governing equations}
Taking the first variation of Eq. (\ref{Eqn19}) and using Eqs. (\ref{Extra39N}) and (\ref{Eqn25a}) yields:
\begin{align}
\sum_{\theta=1}^{p}h_{\theta}(\boldsymbol{\phi})c_{k}^{\theta}(\tilde{\boldsymbol{\mu}}) - c_{k} &= 0\quad \label{Eqn26} \forall\, k=1\hdots (n-1), \\
\text{div}\left[\sum_{\theta=1}^{p}h_{\theta}(\boldsymbol{\phi})\sigma_{ij}^{\theta}\right] &= 0,\label{Eqn27}\\
\frac{\partial \phi_{\theta}}{\partial t} + L_{\phi}\left[ m\frac{\partial g\left(\boldsymbol{\phi}\right)}{\partial \phi_\theta} - \kappa\Delta\phi_{\theta} + \frac{\partial \omega_{bulk}}{\partial \phi_{\theta}}\right] &=0\quad  \label{Eqn28}\forall\, \theta=1\hdots p,
\end{align}
where $L_{\phi}$ is the Allen-Cahn mobility and is assumed to be uniform in this work. It should be noted that the standard diffusion equations do not naturally come out of the variational derivative in the case of grand-potential-based models. Thus, to ensure mass conservation, the evolution of overall molar density, $c_{k}$ in Eq. (\ref{Eqn26}), is given by \cite{Chatterjee2021}
\begin{align}
\frac{\partial c_{k}(\mathbf{x},t)}{\partial t} - \text{div}\left[\sum_{j=1}^{n-1}\frac{L_{kj}^{n}\left(\boldsymbol{\tilde{\mu}, \phi}\right)}{V_{m}} \text{grad}\,\tilde{\mu}_{j}\right] &= 0\quad \forall\, k=1\hdots (n-1),
\label{Eqn29}
\end{align}
where the components of the overall Onsager matrix $L_{kj}^{n}\left(\boldsymbol{\tilde{\mu}, \phi}\right)$ are interpolated as  \cite{Chatterjee2021}
\begin{align}
L_{kj}^{n}\left(\boldsymbol{\tilde{\mu}, \phi}\right) = \sum_{\theta=1}^{p}h_{\theta}\left(\boldsymbol{\phi}\right)L_{kj}^{n\theta}\left(\tilde{\boldsymbol{\mu}}\right).
\label{Eqn30}
\end{align}
Here, the notation $L_{kj}^{n\theta}\left(\tilde{\boldsymbol{\mu}}\right)$ represents the components of the Onsager matrix specific to a particular phase $\theta$ expressed as a function of diffusion potentials. Again, this term can also be either directly obtained as functions of diffusion potentials from CALPHAD databases \cite{Chatterjee2021} or can be assumed to be uniform.

It should be noted that we do not follow grand-potential-based models, e.g., \cite{Plapp2011}, \cite{Choudhary2012}, \cite{Hotzer2015, Cogswell2015, Mushongera2015, Amos2018, Aagensen2018}, that requires formulating a diffusion potential \textit{rate} equation by first taking the time derivative of Eq. (\ref{Eqn26}) and then substituting Eq. (\ref{Eqn29}). Instead, we calculate the diffusion potential by iteratively solving Eq. (\ref{Eqn26}). Consequently, this approach requires calculating a Jacobian matrix, that can be evaluated by differentiating Eq. (\ref{Eqn26}) with respect to the diffusion potential \cite{SC2}. This yields
\begin{align}
\sum_{\theta=1}^{p}h_{\theta}(\boldsymbol{\phi})\chi^{\theta}_{jr}\left(\boldsymbol{\tilde{\mu}}\right) -\frac{\partial c_{j}}{\partial\tilde{\mu}_{r}} = 0,
\label{Eqn_before31}
\end{align}
where $\chi_{jr}^{\theta}(\tilde{\boldsymbol{\mu}}) = \partial c_{j}^{\theta}/\partial \mu_{r}$ are the coefficients of the susceptibility matrix expressed as a function of diffusion potentials. Further, these coefficients can be determined either by analytical approaches assuming parabolic or dilute or ideal free energies \cite{Plapp2011} or numerically from CALPHAD databases \cite{Chatterjee2021}.

Moreover, as previously discussed, the scheme of homogenization may influence the independence of bulk and interfacial properties. More specifically, this independence is achieved provided that the last term in Eq. (\ref{Eqn28}) vanishes at equilibrium \cite{Plapp2011}. However, this may not be evident in mechanically coupled alloy phase-field models. Concretely, consider a three-phase system consisting of phases—$\alpha$, $\beta$ and $\gamma$; then the last term for a specific phase-field variable, say $\phi_\alpha$, may be explicitly written as (Appendix D)
\begin{align}
\begin{split}
\frac{\partial \omega_{bulk}}{\partial \phi_{\alpha}} =& -\frac{\partial h_{\beta}}{\partial \phi_{\alpha}}\left\{\left(\omega^{\alpha}_{bulk} - \omega^{\beta}_{bulk}\right) -\left(\sum_{\theta=1}^{p}h_{\theta}\sigma_{ij}^{\theta}\right)\llbracket \epsilon_{ij}\rrbracket^{\alpha\beta} \right\}\\
& -\frac{\partial h_{\gamma}}{\partial \phi_{\alpha}}\left\{ \left(\omega^{\alpha}_{bulk} - \omega^{\gamma}_{bulk}\right) - \left(\sum_{\theta=1}^{p}h_{\theta}\sigma_{ij}^{\theta}\right)\llbracket \epsilon_{ij} 
\rrbracket^{\alpha\gamma}\right\}.
\end{split}
\label{Eqn31}
\end{align}
It follows from Eq. (\ref{Eqn31}) that the terms within the large curly braces depend on the strain jumps, $\llbracket \boldsymbol{\epsilon}\rrbracket^{\alpha\beta}$ and $\llbracket \boldsymbol{\epsilon}\rrbracket^{\alpha\gamma}$, which are in turn dependent on the scheme of homogenization. For instance, if Voigt/Taylor or Khacturayan scheme is followed, then the strain jumps vanish and consequently these terms are proportional to the jump in the grand potentials, i.e., 
$\llbracket \omega_{bulk}\rrbracket^{\alpha\beta}$ \& $\llbracket \omega_{bulk}\rrbracket^{\alpha\gamma}$. Further, since the bulk grand-potentials are functions of both continuous and discontinuous (total) strain components (see Eq. (\ref{Eqn24})), these terms would not necessarily vanish at equilibrium. On the other hand, in the case of the partial rank-one scheme the strain jumps are non-zero and it can be shown that these terms reduce to the sharp interfacial chemical equilibrium conditions for coherently stressed two-phase solids (see Eq. (7.31) in \cite{Voorhees2004}). For sake of completeness, we have also provided the derivatives with respect to $\phi_\beta$ and $\phi_\gamma$ in Appendix D, which are similar to Eq. (\ref{Eqn31}).

Moreover, it must be noted that in writing Eq. (\ref{Eqn28}) we have tacitly assumed that the variational contribution to the driving force is negligible. Precisely, the variational term may be written as:
\begin{align}
\text{div}\left\{\frac{\partial \omega_{bulk}}{\partial \left( \text{grad}\,\phi_{\theta}\right)}\right\} = \text{div}\left\{\left(\sum_{\theta=1}^{p}h_{\theta}\sigma_{ij}^{\theta}\right)\frac{\partial \epsilon_{ij}^{\theta}}{\partial \left( \text{grad}\,\phi_{\theta}\right)}\right\}.
\label{Extra1}
\end{align}
Note that due to the dependence of \textit{phase} strains on the unit vectors: $\boldsymbol{n}^{\alpha\beta}$ and $\boldsymbol{n}^{\beta\gamma}$, the term within the curly braces in Eq. (\ref{Extra1}) is nonzero in case of the rank-one scheme. However, based on our previous study \cite{SC2}, we found that this term does not significantly affect the temporal variation of the interface for cases with a small difference in stiffness tensors \cite{SC2}. This is because the term is proportional to the magnitude of jump vectors, $\boldsymbol{a}^{\alpha\beta}$ and $\boldsymbol{a}^{\beta\gamma}$, and are consequently proportional to the difference in stiffness tensors (see Eqs. (\ref{Eqs29N}) \& (\ref{Eqs30N})). Thus, we have neglected this term in our calculations which renders our formulation non-variational.

Finally, the Allen-Cahn mobility is calculated using \cite{Moelans2011}
\begin{align}
L_{\phi}=4m/(3\kappa\zeta),
\label{Eqn51_new}
\end{align}
where $m$ and $\kappa$ are defined at Eq. (\ref{Eqn22}) and the parameter $\zeta = \sum_{k=1}^{n-1}(X_{k}^{\theta,eq} - X_{k}^{\sigma,eq})\sum_{j=1}^{n-1}\left(L_{kj}^{n\theta,eq}V_{m}\right)^{-1}(X_{j}^{\theta,eq} - X_{j}^{\sigma,eq})$, is obtained assuming infinite interface kinetics \cite{Kim2007}. This choice of $\zeta$  ensures that local equilibrium is maintained near the interface and the growth is diffusion-controlled \cite{Moelans2011}. 

\section{Coupling with CALPHAD databases}
As discussed before, our model requires thermodynamic properties and mobilities as functions of diffusion potentials. Specifically, four properties are needed for any given phase \cite{Chatterjee2021}. First, the molar grand-potential, $\Omega_{m}^{\theta}$, of an individual phase to calculate the chemical contribution to the bulk grand-potential density in Eq. (\ref{Eqn24}). Second, the \textit{phase} mole fractions to calculate the \textit{phase} molar densities using Eq. (\ref{Eqn25N}). Third, the susceptibility matrix to evaluate Eq. (\ref{Eqn_before31}). Finally, the Onsager matrix pertaining to each individual phase is also required to evaluate Eq. (\ref{Eqn30}). Moreover, for non-dilute and non-ideal solid solutions, these properties cannot be analytically expressed as functions of diffusion potentials. Thus, we numerically evaluated these properties using the MATLAB-ThermoCalc interface by minimising the prerequisite properties with respect to a discretized range of diffusion potential(s). 
This discretized range was predetermined based on the phase diagram \cite{Chatterjee2021}. 

\begin{figure}[!ht]
\begin{center}
\includegraphics[keepaspectratio,width=0.8\linewidth]{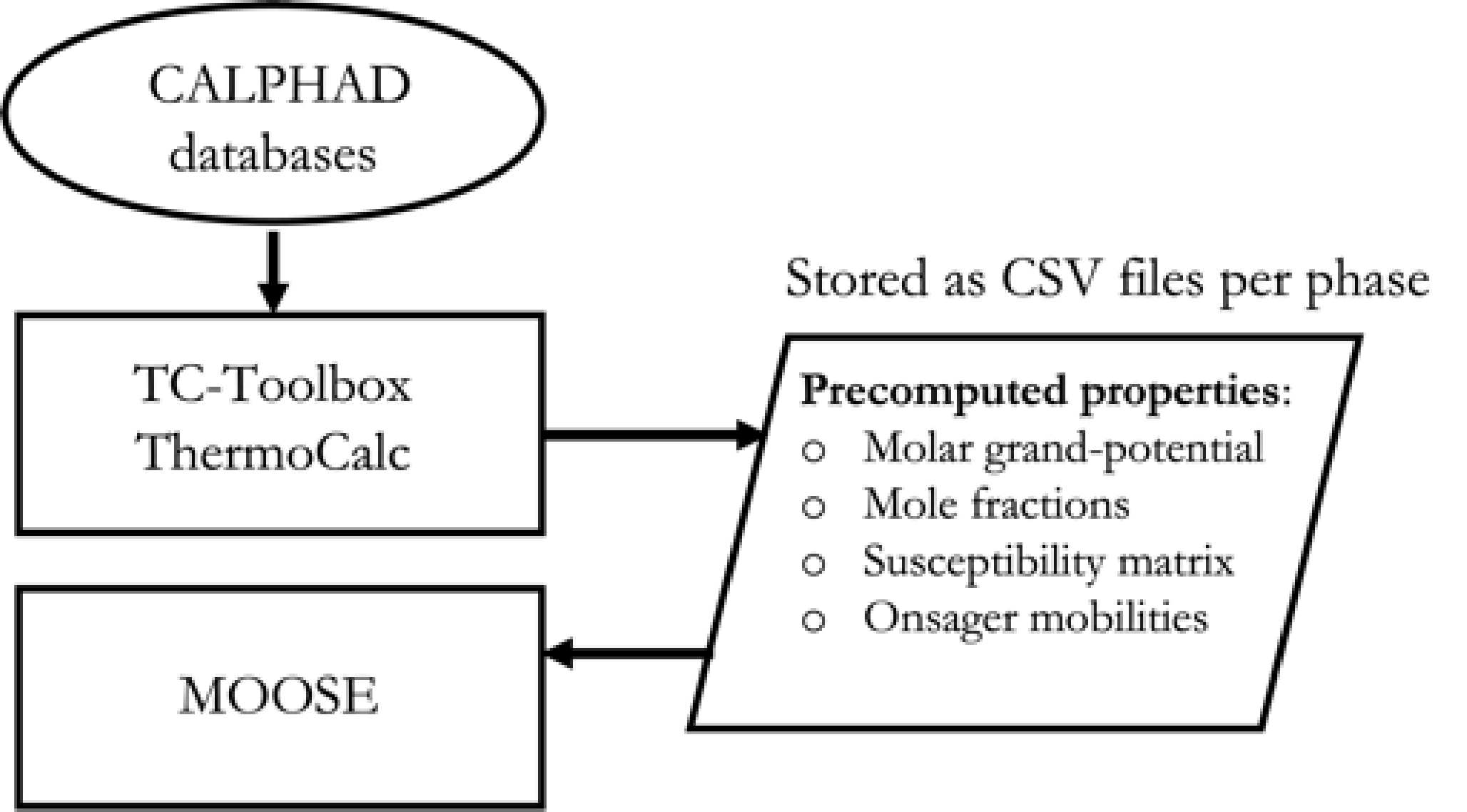}
\caption{Schematic showing the coupling procedure between CALPHAD databases and MOOSE in case of a grand-potential-based model.}
\label{coupling_schematic}
\end{center}
\end{figure}

Concretely, we chose two three-phase alloys: a binary Ni-Al and a ternary Ni-Al-Cr, to illustrate the coupling procedure. For all phases except the binary and ternary B$2$ phases, the above-mentioned properties were extracted as functions of diffusion potentials from the TCNi$8$ and MOBNi$4$ databases using the TC-Toolbox for MATLAB. Specifically, in the case of Ni-Al, we evaluated the thermodynamic properties and mobilities as discretized functions of Al diffusion potential in the interval of $[-2\mathrm{e}{5}, 2\mathrm{e}{5}]$ J/mol. Similarly, for the Al-Cr-Ni simulations, we obtained the discretized properties by varying the Cr and Al diffusion potentials from $-1\mathrm{e}{5}$ J/mol to $1\mathrm{e}{5}$ J/mol. These limits were selected to ensure that the Al and Cr mole fractions of an arbitrary phase are very close to the limits of $0$ and $1$ (see Appendix C in Ref. \cite{Chatterjee2021}, for details). Following this, the properties assigned to a given phase were non-dimensionalized and stored in a tabulated format and then supplied as an input to MOOSE (Multiphysics Object-Oriented Simulation Environment) \cite{Gaston2009} for phase-field simulations. Fig.\ref{coupling_schematic} shows the coupling procedure schematically.
The details of the non-dimenionalization are given in Appendix E.

For the binary and ternary B$2$ phases, we could obtain only the thermodynamic properties as discretized functions of diffusion potentials. The Onsager coefficients were assumed to be constants. Specifically, the mobilities were obtained from ThermoCalc at the equilibrium mole fractions. Following this, these mobilities were used to evaluate the $\zeta$ parameter in Eq. (\ref{Eqn51_new}), which is needed to calculate the Allen-Cahn mobility. 
The mobilities, the equilibrium mole fractions, the parameter $\zeta$, and the simulation temperatures are listed in Table \ref{Table1.1}. 

\begin{table}[!ht]
\begin{center}
\caption{Constant material parameters for the Ni-Al and Al-Cr-Ni alloy systems. The equilibrium mole fractions and the Onsager mobilities were obtained from ThermoCalc.}
\label{Table1.1}
\begin{tabular}{p{4cm}lll}
\hline
 & Ni-Al & Al-Cr-Ni  \\
\hline
$T$ $[\text{K}]$  &$1000$& $1473$  \\
$\sigma$ [J/m$^2$]  &$0.5$& $0.5$  \\
$V_{m}$ [m$^{3}$/mol] & $7.5\mathrm{e}{-5}$ & $7.5\mathrm{e}{-5}$\\
$X_{B}^{\alpha,eq}$ &$0.27457$& $0.2209$\\
$X_{B}^{\beta,eq}$ &$0.40646$& $0.2912$\\
$X_{C}^{\alpha,eq}$ &-&$0.07575$\\
$X_{C}^{\beta,eq}$ &-&$0.06756$\\
$L^{\beta,eq}$ [mol m$^{2}$/Js] &1.7534e-17&$\begin{bmatrix}
0.8238 & 0.0552\\
0.0552 & 0.2684
\end{bmatrix} \times 1\mathrm{e}{-17}$\\
$\zeta$ [Js/m$^{5}$] & $1.3228\mathrm{e}{19}$&$8.8423\mathrm{e}{18}$\\
\hline
\end{tabular}
\end{center}
\end{table}
\section{Results and discussion}
As previously discussed, we have considered two three-phase alloys, an Al-Ni alloy and an Al-Cr-Ni alloy, to demonstrate the application of our model. Further, we have considered two interface geometries per alloy system. Specifically, the first two cases assume planar interfaces, while the remaining two cases assume concentric ring interfaces. We have employed both the partial rank-one (hereafter referred to as PR) and the Voigt-Taylor (hereafter referred to as VT) homogenization schemes to simulate all four cases. As noted earlier, this was achieved by controlling the jump in phase strains, i.e., $\llbracket\boldsymbol{\epsilon}\rrbracket$, in Eqs. (\ref{Eqn4})-(\ref{Eqn6}).

For sake of clarity, Table \ref{table:bc} provides the mechanical boundary conditions and the eigenstrains for each considered case. From Table \ref{table:bc}, we note that the eigenstrains in the binary and ternary $\gamma^{\prime}$ phases are identical. Although in real alloys, the strength of the eigenstrain depends on the alloy composition, we made this simplifying assumption due to the lack of any experimental data in the literature. Moreover, the assumed elastic constants for each simulated case are listed in Table \ref{Table_ec}. Except for case II, we have assumed isotropic elastic constants for all considered cases (Table \ref{Table_ec}). Finally, to verify the accuracy of our model, we have compared the simulated elastic fields in each of these cases against the analytically obtained solution. The analytical solutions are provided in Appendix F. Here, it is worth emphasizing that the analytical solutions depend on the interface positions, which have been calculated numerically by tracking the phase-field variables ($\phi_{\theta=\alpha,\beta,\gamma}=0.5$). 

\begin{table}[ht]
\caption{Summary of eigenstrains and mechanical boundary conditions for all cases. The x- and y-components of displacement $\boldsymbol{u}$ are denoted by $u_{x}$ \& $u_{y}$, respectively. Here, $l_{c}=0.033\,\mu m$ denotes a characteristic length scale used for non-dimensionalization.} 
\centering
\begin{tabular}{l l l }
\hline\hline
Simulation & Eigenstrains $[\text{Phase}]$ &  Boundary conditions  \\  
\hline
\multirow{2}{*}{Planar Al-Ni}&\multirow{2}{*}{$\boldsymbol{\epsilon}^{\star}$  $\left[\gamma^{\prime} \right]= -0.3\%\boldsymbol{1}$}& $\boldsymbol{u}$ at left boundary = $\boldsymbol{0}$ \\
 \multirow{2}{*}{(Case I)}&&$\boldsymbol{u}$ at right boundary = $\boldsymbol{0}$\\
 &{$\boldsymbol{\epsilon}^{\star}$  $\left[\gamma \right]=  \boldsymbol{\epsilon}^{\star}  \left[\text{B}2 \right]= \boldsymbol{0}$}&\multirow{2}{*}{$\boldsymbol{u}$ is periodic along y-direction}\\
 &&\\
&& $\boldsymbol{u}$ at left boundary = $\boldsymbol{0}$ \\
 \multirow{2}{*}{Planar Al-Cr-Ni}&\multirow{2}{*}{$\boldsymbol{\epsilon}^{\star}$  $\left[\gamma^{\prime} \right]= -0.3\%\boldsymbol{1}$}&${u}_{x}/l_{c}$ at right boundary $= 5$\\
  \multirow{2}{*}{(Cases II)}&&${u}_{y}/l_{c}$ at right boundary $= -5$\\
 &{$\boldsymbol{\epsilon}^{\star}$  $\left[\gamma \right]=  \boldsymbol{\epsilon}^{\star}  \left[\text{B}2 \right] =\boldsymbol{0}$}&\multirow{1}{*}{$\boldsymbol{u}$ is periodic along y-direction}\\
 &&\\
\multirow{2}{*}{Non-planar Al-Ni}&\multirow{2}{*}{$\boldsymbol{\epsilon}^{\star}$  $\left[\gamma^{\prime} \right]= -0.3\%\boldsymbol{1}$}& $u_{x}$ at left boundary = 0 \\
 \multirow{2}{*}{(Case III)}&&$u_{y}$ at bottom boundary = 0\\
 &{$\boldsymbol{\epsilon}^{\star}$  $\left[\gamma \right]= \boldsymbol{0}$}&\multirow{2}{*}{traction is zero at outer boundary}\\
 &&\\
\multirow{2}{*}{Non-planar Al-Cr-Ni}&\multirow{2}{*}{$\boldsymbol{\epsilon}^{\star}$  $\left[\gamma^{\prime} \right]= \boldsymbol{0}$}& $u_{x}$ at left boundary = 0 \\
 \multirow{2}{*}{(Case IV)}&&$u_{y}$ at bottom boundary = 0\\
 &{$\boldsymbol{\epsilon}^{\star}$  $\left[\gamma \right]= \boldsymbol{\epsilon}^{\star}$ $\left[\text{B}2\right] = \boldsymbol{0}$}&\multirow{1}{*}{$u_{x}$ at outer boundary = $0.1\%x$}\\
 &&\multirow{1}{*}{$u_{y}$ at outer boundary = $0.1\%y$}\\
 \\
\hline
\end{tabular}
\label{table:bc}
\end{table}

\begin{table}[!ht]
\begin{center}
\caption[Elastic constants used for the simulations]{Summary of elastic constants for all simulated cases. Here, the left/inner label refers to the leftmost or the innermost phase in the simulations depending on the planar or concentric interface case. Likewise, the right/outer label refers to the rightmost or outermost phase, and the centre label refers to the intermediate phase.}
\label{Table_ec}
\begin{tabular}{l l l l l}
\hline
Simulation& Left/Inner & Centre &  Right/Outer & Refs. \\
\hline
&&&&\\
\multirow{2}{*}{Case I,} &$E = 158$ GPa&$E = 147$ GPa& $G=76.6$ GPa&\cite{Socrate1992}, \cite{Miracle1993}\\
&$\nu = 0.3$&$ \nu=0.3$&$\nu=0.3387$&\\
II \& III &&&&\\
&&&&\\
\multirow{3}{*}{Case II} &$C_{11} = 188.3$ GPa&$C_{11} = 194.37$ GPa& $G=76.6$ GPa&\cite{Ardell2014}, \cite{Miracle1993}\\
&$C_{12} = 143.54$ GPa &$C_{12} = 140.82$ GPa &$\nu=0.3387$&\\
&$C_{44} = 80.734$ GPa &$C_{44} = 84.04$ GPa & &\\
\hline
\end{tabular}
\end{center}
\end{table}

\subsection{Planar three-phase Ni-Al simulation}
\begin{figure}[!ht]
\begin{center}
\begin{subfigure}{0.8\textwidth}
\includegraphics[trim=0 0 0 0, clip, keepaspectratio,width=\linewidth]{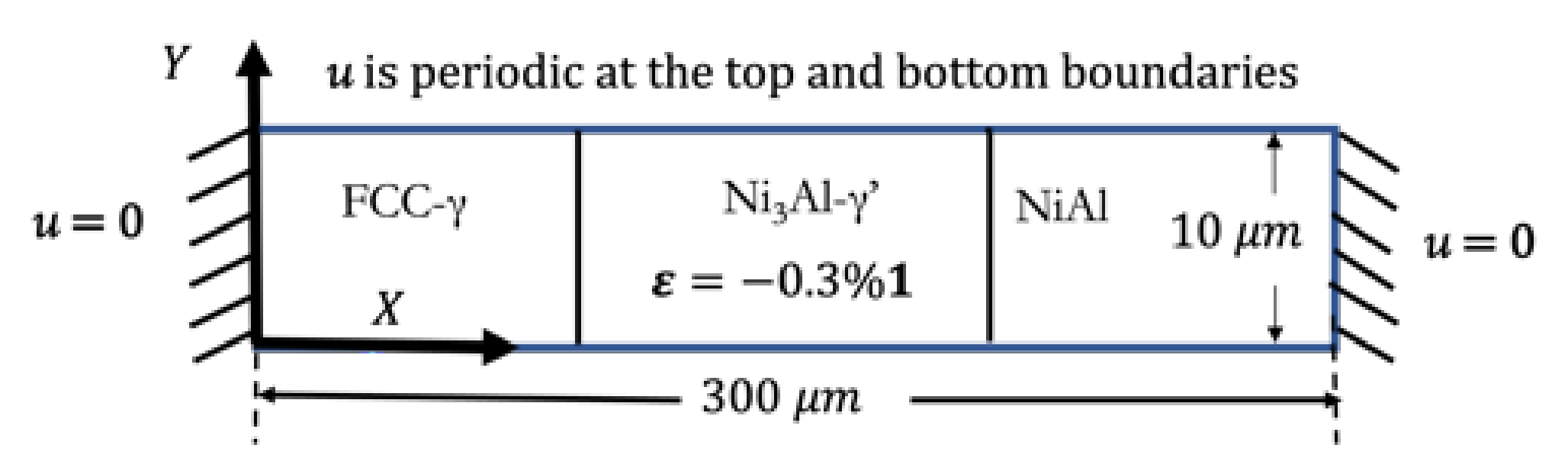}
\caption{}
\label{Fig1New}
\end{subfigure}
\end{center}
\begin{center}
\begin{subfigure}{1.0\textwidth}
\includegraphics[trim=0 10 95 0, clip, keepaspectratio,width=\linewidth]{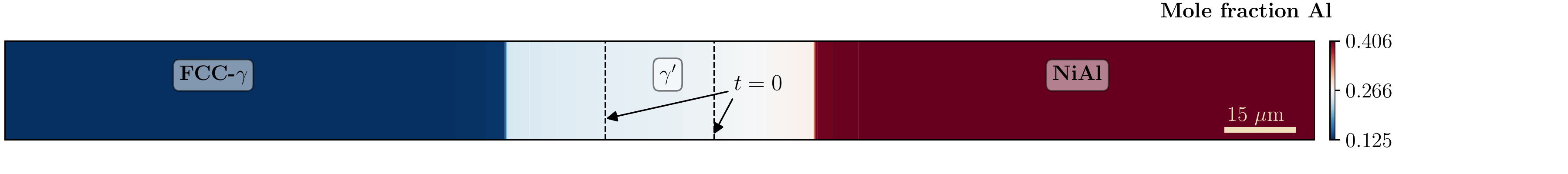}
\caption{}
\label{Fig.0a}
\end{subfigure}
\end{center}
\begin{subfigure}{0.50\textwidth}
\includegraphics[keepaspectratio,width=\linewidth]{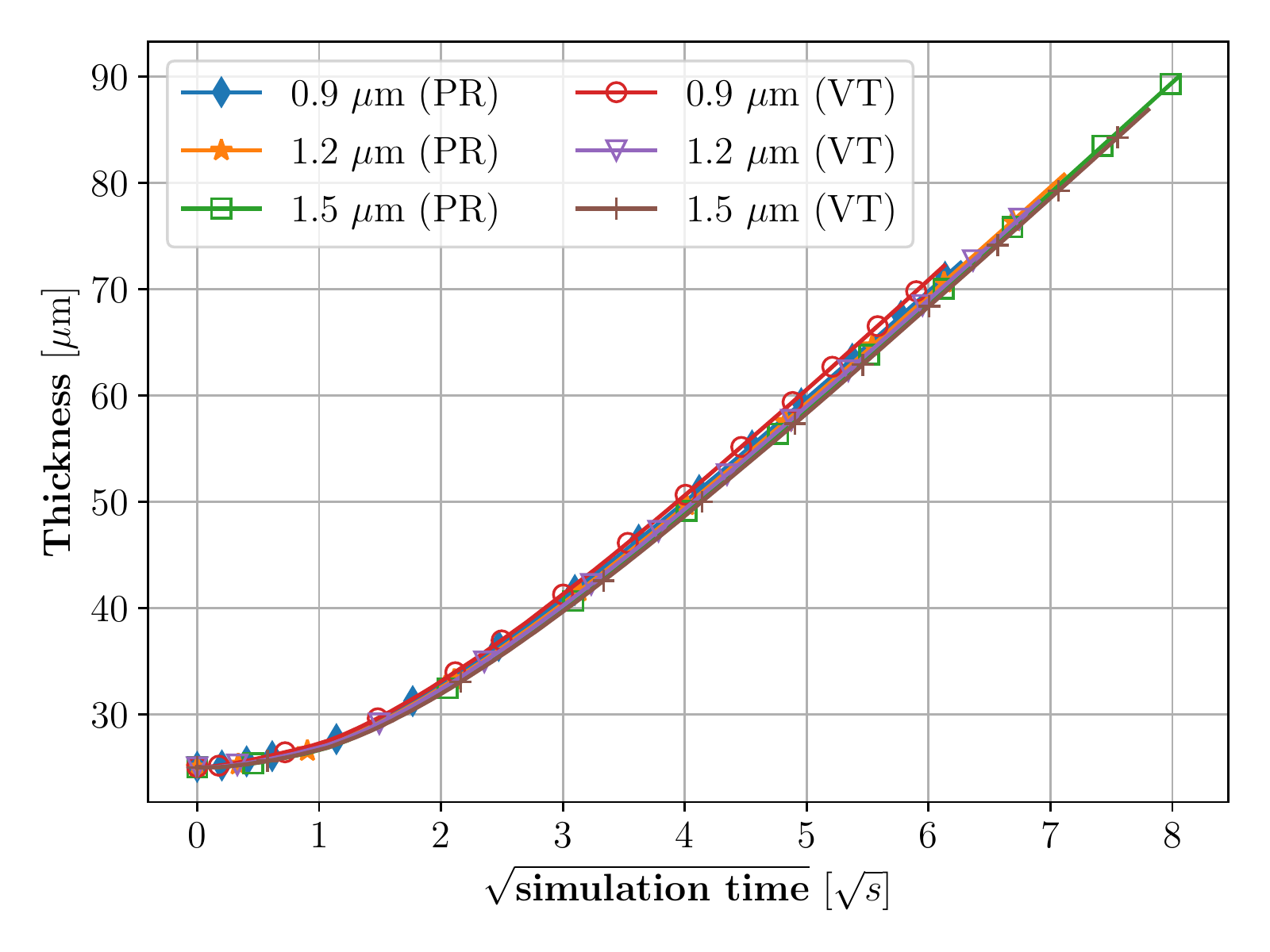}
\caption{}
\label{Fig.1a}
\end{subfigure}
\begin{subfigure}{0.50\textwidth}
\includegraphics[keepaspectratio,width=\linewidth]{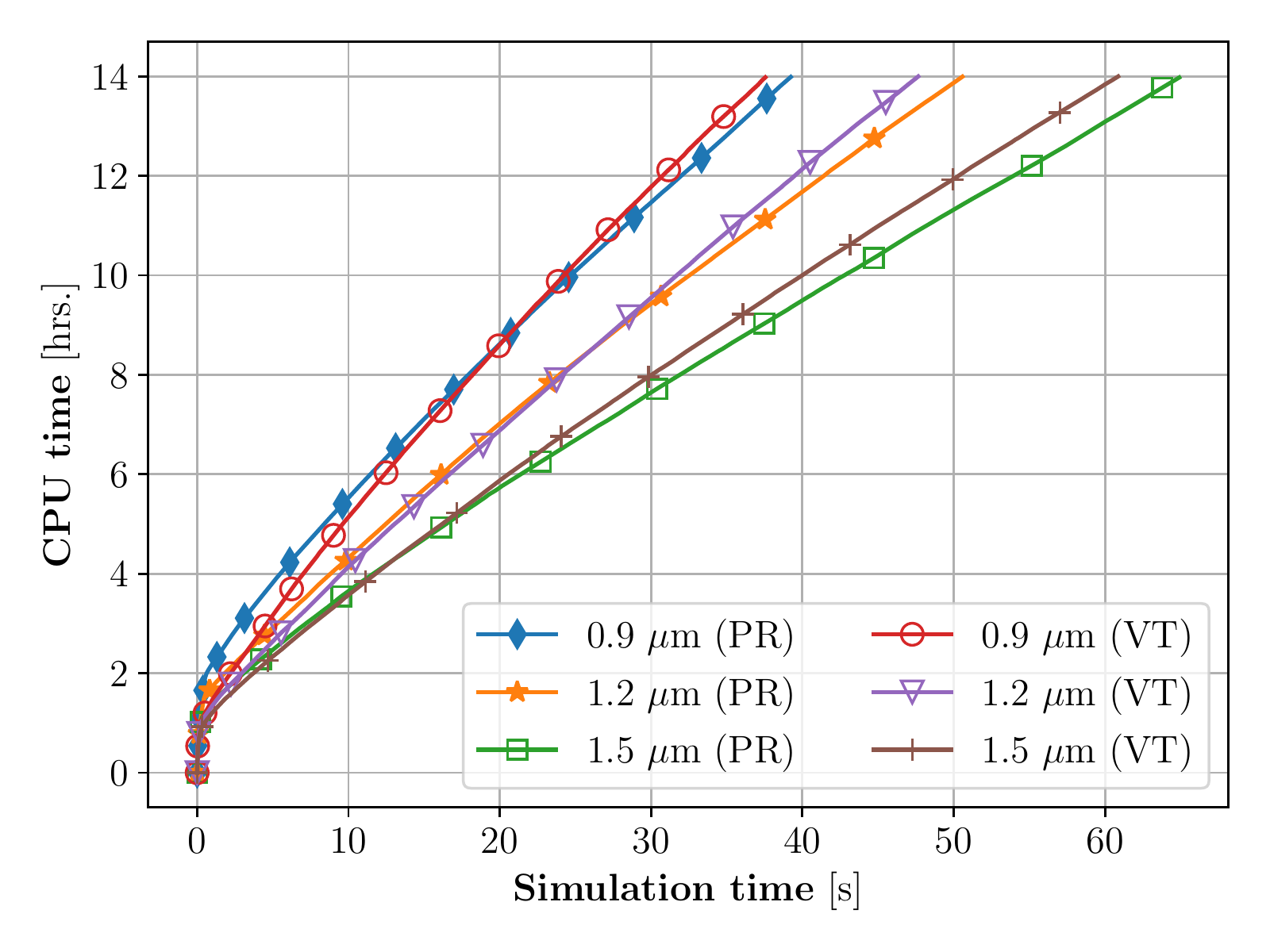}
\caption{}
\label{Fig.1b}
\end{subfigure}
\caption{For a Ni-Al fcc-$\gamma/\mathrm{Ni}_3\mathrm{Al-}\gamma^{\prime}/\mathrm{NiAl}$ coherently stressed planar diffusion couple: a) schematic of the simulation domain, eigenstrains and mechanical boundary conditions; b) simulated Al-mole fraction field at time $t=37$ s. For three different interface widths, the temporal variation in $\mathrm{Ni}_3\mathrm{Al}$ thickness as a function of the square root of simulation time using the partial rank-one (PR) and Voigt-Taylor (VT) homogenization schemes (c); and the CPU time as a function of simulation time for both these schemes (d).}
\end{figure}

First, we simulated a coherently stressed planar $\mathrm{fcc-}\gamma/\gamma^{\prime}\mathrm{-Ni}_3\mathrm{Al}/\mathrm{NiAl}$ alloy that is mechanically constrained at the left and right boundaries (Fig. \ref{Fig1New}). We have assumed periodic boundary conditions for the phase field, composition and displacement variables at the top and bottom boundaries. While homogeneous Neumann boundary conditions are applied at the left and right boundaries for the phase-field and composition variables, viz.
\begin{align}
\text{grad}\,\boldsymbol{\phi}\cdot \boldsymbol{n}^{\Gamma}(x=\pm L_x/2,y,t) &= 0,\label{ResEq1}\\
\text{grad}\,\tilde{\mu}_{Al}\cdot \boldsymbol{n}^{\Gamma}(x=\pm L_x/2,y,t) &= 0,\label{ResEq2}
\end{align}
where $\tilde{\mu}_{Al}$ is the Al-diffusion potential; $L_{x}$ is the length of the simulation domain, and $\boldsymbol{n}^{\Gamma}$ is the unit normal at the left and right boundaries. The displacement boundary conditions at these boundaries are (Table \ref{table:bc}):
 \begin{align}
u_{x}(x=\pm L_x/2,y,t) &= 0,\label{ResEq3}\\
u_{y}(x=\pm L_x/2,y,t) &= 0.\label{ResEq4}
\end{align}
Since the three phases cannot coexist, the intermediate $\gamma^{\prime}\mathrm{-}\mathrm{Ni}_3\mathrm{Al}$ phase grows at the expense of $\gamma$ and NiAl phases. Fig. \ref{Fig.0a} shows the Al mole fraction field at time $t=37$ s. Moreover, we find that the thickness of $\gamma^{\prime}$ phase increases linearly as a function of the square root of simulation time (Fig. \ref{Fig.1a}), thus indicating parabolic growth kinetics. This thickness is numerically determined by locating the $\gamma/\gamma^{\prime}$ and $\gamma^{\prime}/\mathrm{NiAl}$ interface positions as a function of time using the phase-field variables. To further test the influence of interface width on kinetics, we vary the interfacial parameters: $\kappa$, $m$ and $L_{\phi}$, using Eqs. (\ref{Eqn22}) \& (\ref{Eqn51_new}), for three different interface widths. We find that the thickness of the $\mathrm{Ni}_3\mathrm{Al}$ phase remains relatively unaltered with varying interface width using both schemes (Fig. \ref{Fig.1a}). Expectedly, for both PR and VT schemes, the CPU time decreases with increasing interface width; since the grid spacing, $\Delta x = l_{w}/6.0$, is directly proportional to interface width $l_{w}$ (Fig. \ref{Fig.1b}). However, we find that the PR scheme shows comparatively better convergence compared to the VT scheme (Fig. \ref{Fig.1b}). This shows that the proposed PR scheme is computationally efficient compared to the VT scheme for a longer simulation time.

\begin{figure}[!ht]
\begin{subfigure}{0.50\textwidth}
\includegraphics[keepaspectratio,width=\linewidth]{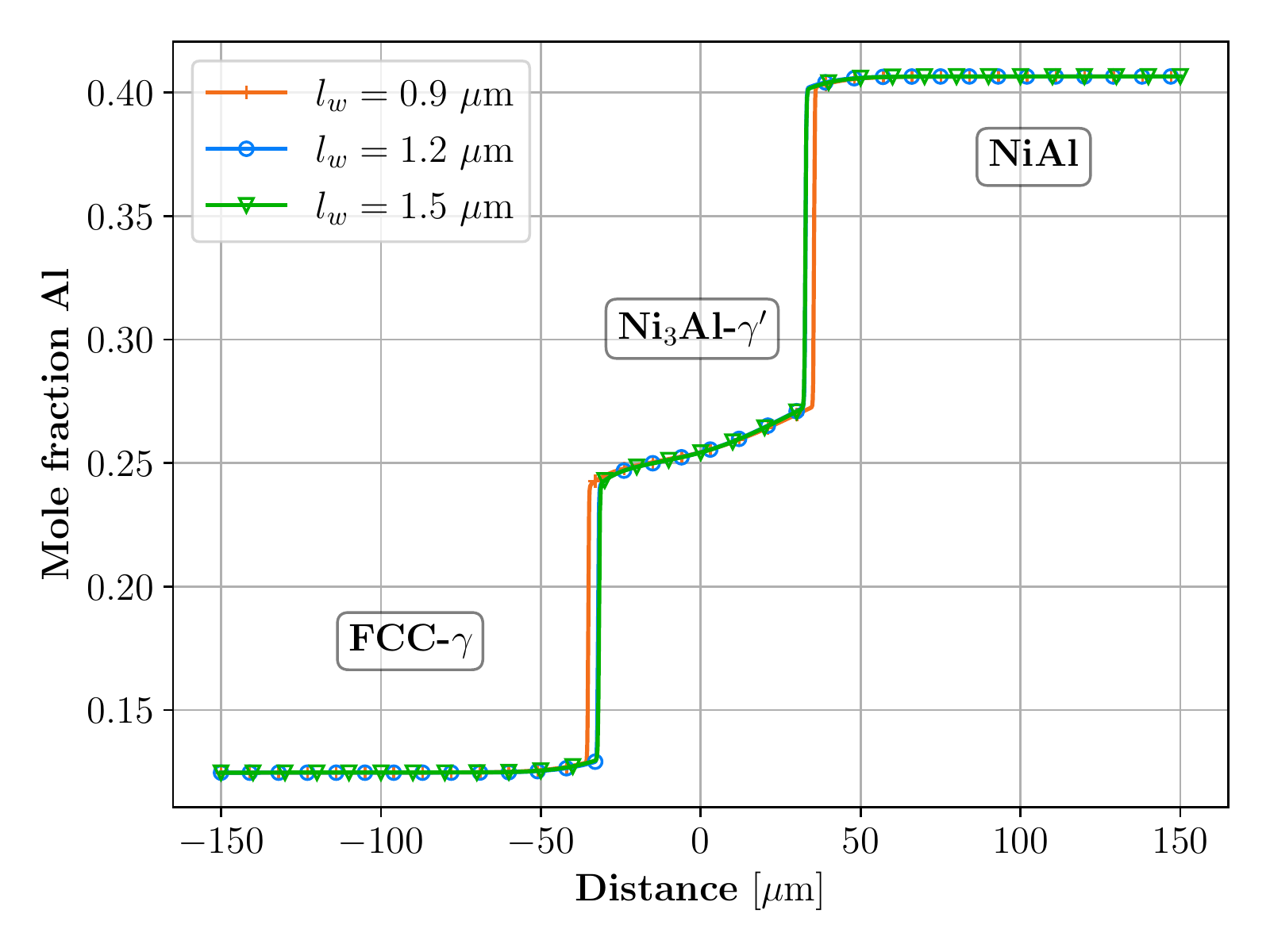}
\caption{}
\label{Fig.2a}
\end{subfigure}
\begin{subfigure}{0.50\textwidth}
\includegraphics[keepaspectratio,width=\linewidth]{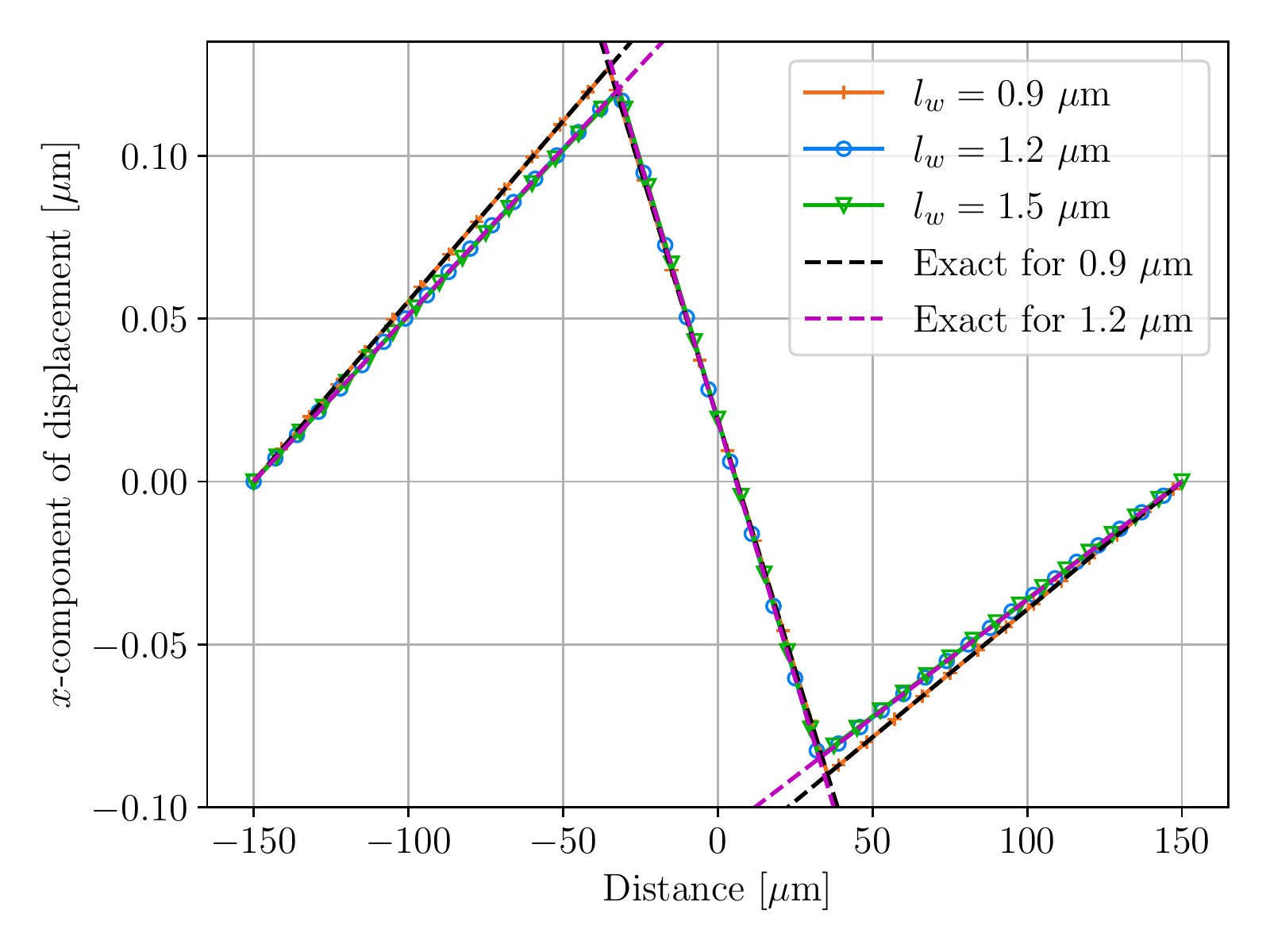}
\caption{}
\label{Fig.2b}
\end{subfigure}
\begin{subfigure}{0.50\textwidth}
\includegraphics[keepaspectratio,width=\linewidth]{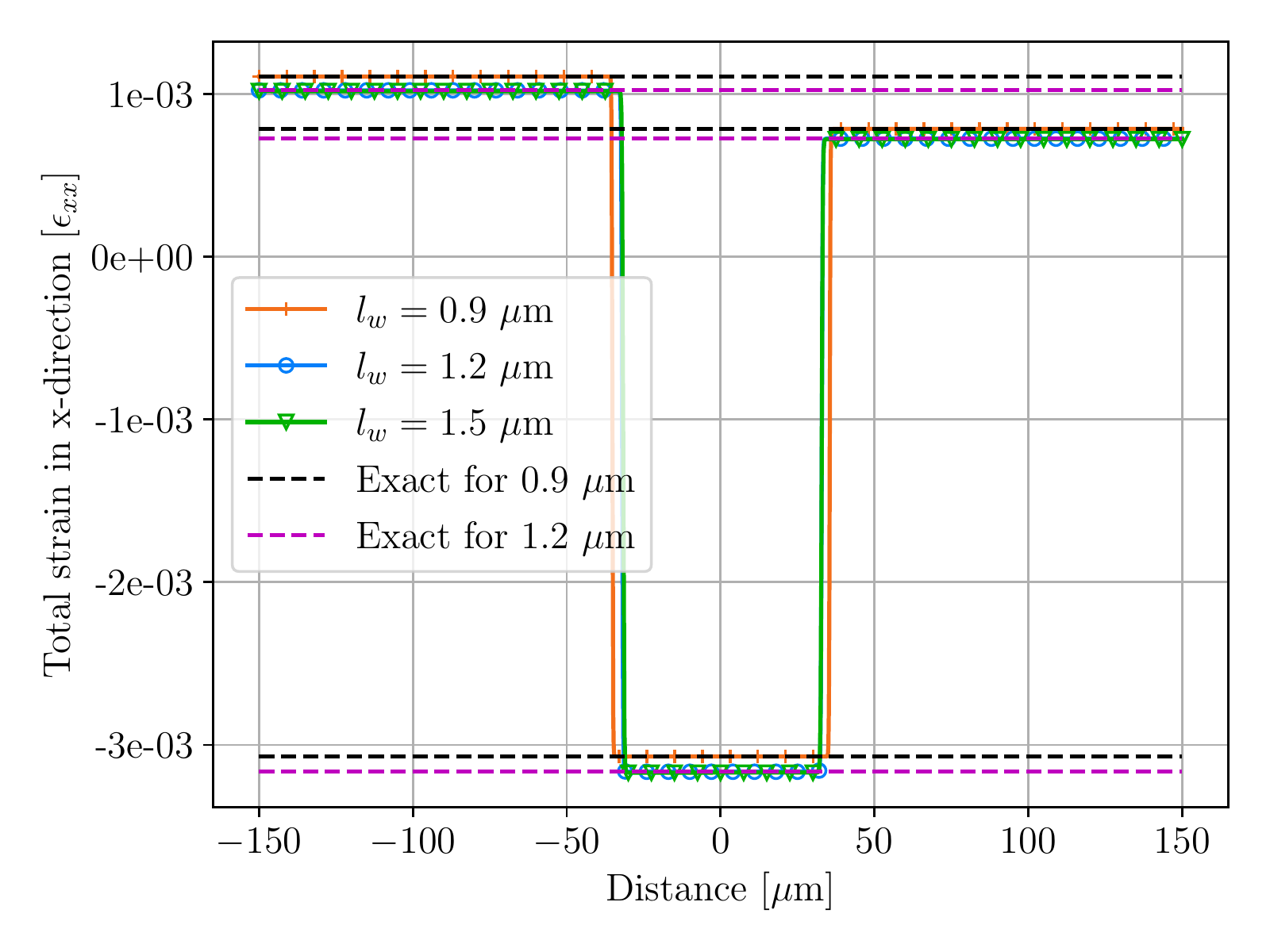}
\caption{}
\label{Fig.2c}
\end{subfigure}
\begin{subfigure}{0.50\textwidth}
\includegraphics[keepaspectratio,width=\linewidth]{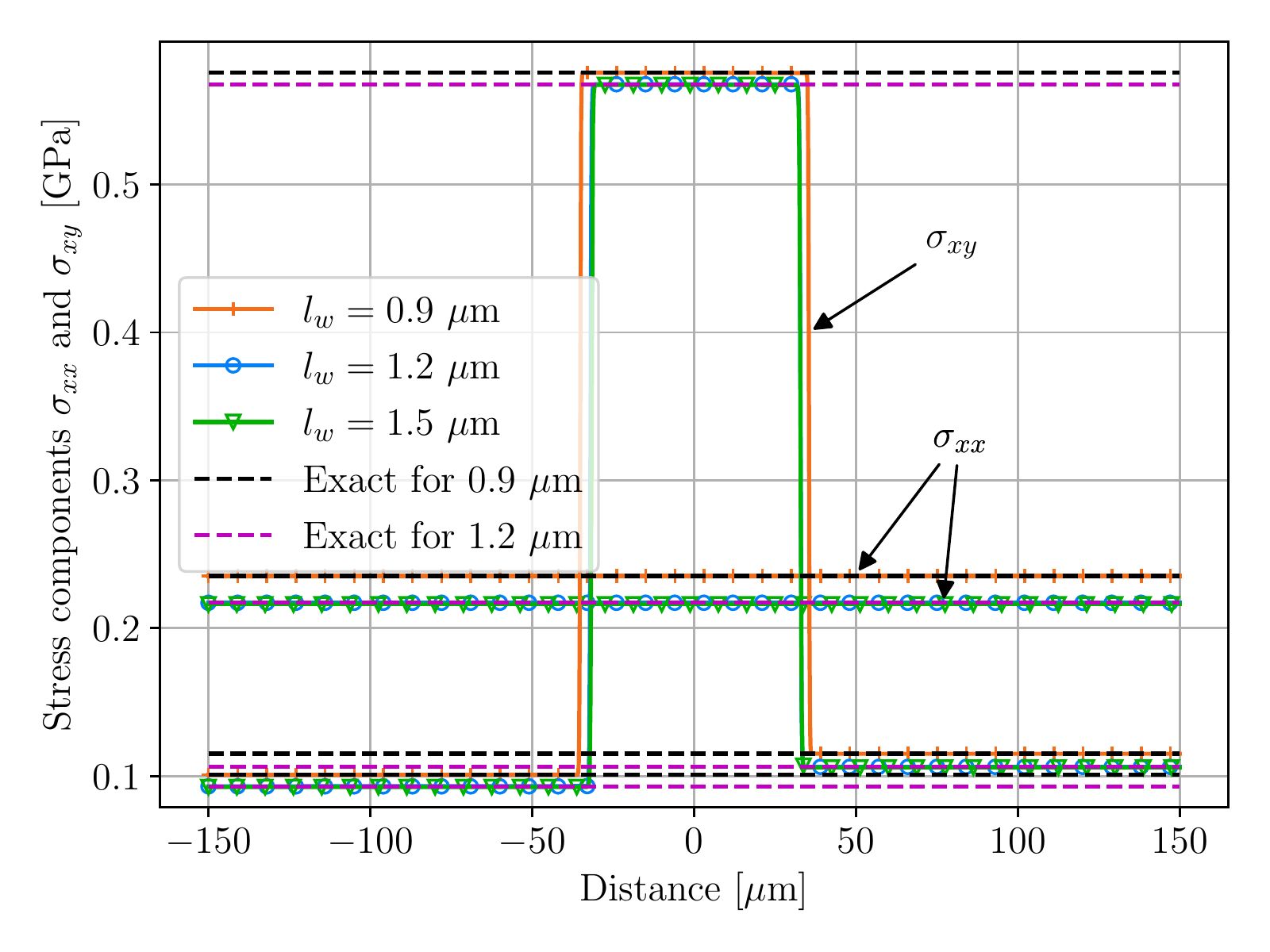}
\caption{}
\label{Fig.2d}
\end{subfigure}
\caption{For the Ni-Al fcc-$\gamma/\mathrm{Ni}_3\mathrm{Al-}\gamma^{\prime}/\mathrm{NiAl}$ planar diffusion couple case using the partial rank-one scheme and three different interface widths: a) Al-mole fraction profiles; b)  x-component of displacement field; c) total normal strain; and d) normal and shear stresses as functions of distance perpendicular to the interface. The superimposed black and magenta dotted lines are the analytically calculated elastic fields using interface width values of $1.2$ $\mu$m and $0.9$ $\mu$m, respectively. }
\label{Fig.2}
\end{figure}

To further verify the spatial accuracy, we sample the spatial variation of the composition field and the elastic quantities across a line normal to the interface at time $t=37$ s. Fig. \ref{Fig.2a} compares the Al mole fraction profile for three different interface widths using the PR scheme. We find that the simulated Al mole fraction profile remains independent of interface width in the bulk regions. Nevertheless, we find marginal deviations in the interfacial regions since the composition is interpolated in this region. We also find this deviation in the  $x$-component of the displacement field near the interfaces. Specifically, we find that the displacement fields using interface widths of $1.2$ $\mu$m and $1.5$ $\mu$m are in agreement with the analytically obtained solutions (Fig. \ref{Fig.2b}). It should be noted that the interface positions required in this analytical solution are obtained assuming an interface width of $1.2$ $\mu$m. Consequently, the simulated solution using an interface width of $0.9$ $\mu$m shows deviation from this analytical solution near the interfaces (Fig. \ref{Fig.2b}). This is expected because the analytical solution depends on the accuracy of the numerically determined interface positions (see Appendix F), which slightly depends on interface width (see the thickness variation in Fig. \ref{Fig.1a}). To verify this, we re-compare the simulated displacement field having an interface width of $0.9$ $\mu$m against an analytical solution that uses the interface positions determined from the same simulation. We then find good quantitative agreement between the two results (Fig.\ref{Fig.2b}). It should be noted that we will obtain similar quantitative agreement between the analytical and simulated elastic fields using the VT scheme. This is because the interface positions as a function of time are relatively independent of the scheme of homogenization. 

Moreover, from Fig.  \ref{Fig.2b}, we see that the maximum displacement is at the $\gamma/\gamma^{\prime}$ and $\gamma^{\prime}/$NiAl interfaces. This is because of the assumed eigenstrain in the $\gamma^{\prime}$ phase.  As shown in Appendix F, since the displacement field varies linearly as a function of distance, the total strain and stress normal to the interface are spatially constant within the three phases (Figs. \ref{Fig.2c} and Fig. \ref{Fig.2d}). Moreover, due to the deviation in the simulated displacement field having an interface width of $0.9$ $\mu$m from the analytical solution, we find similar disagreement in the total strain and stress normal to the interface from this analytical solution. However, by comparing this case against the analytical solution having an interface width of $0.9$ $\mu$m (shown as a dotted magenta coloured line in Figs.  \ref{Fig.2c} and  \ref{Fig.2d}), we find good quantitative agreement.

\subsection{Planar three-phase Al-Cr-Ni simulation}
\begin{figure}[!ht]
\begin{center}
\begin{subfigure}{0.8\textwidth}
\includegraphics[trim=0 0 0 0, clip, keepaspectratio,width=\linewidth]{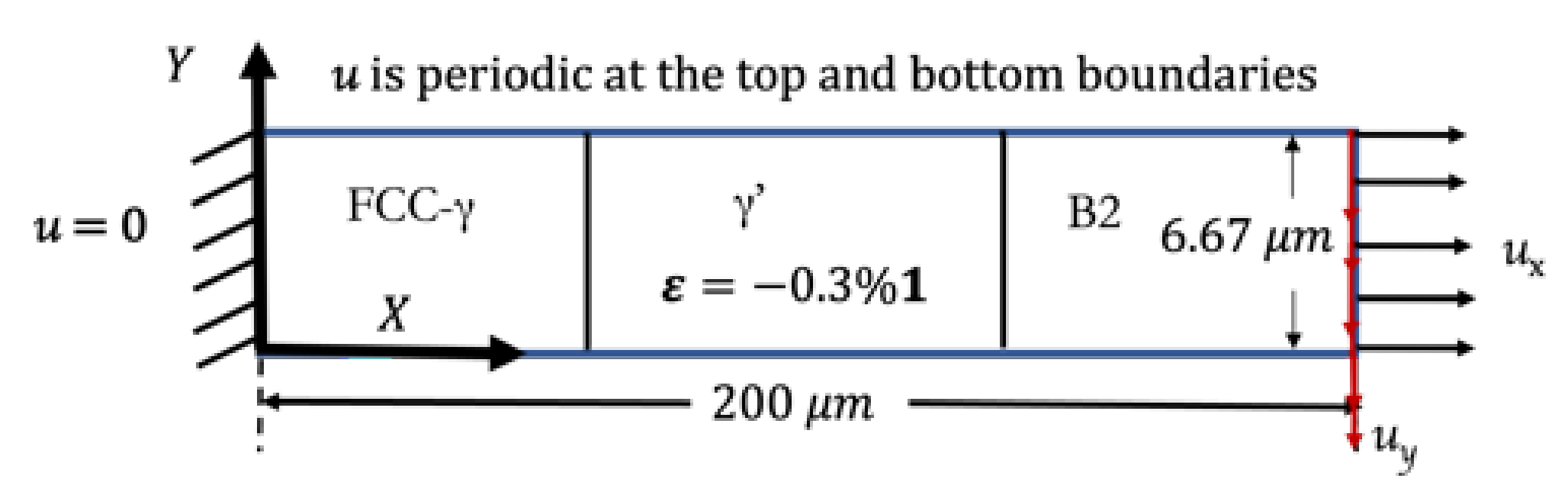}
\caption{}
\label{Fig3New}
\end{subfigure}
\end{center}
\begin{center}
\begin{subfigure}{1.0\textwidth}
\includegraphics[trim=0 10 60 0, clip, keepaspectratio,width=\linewidth]{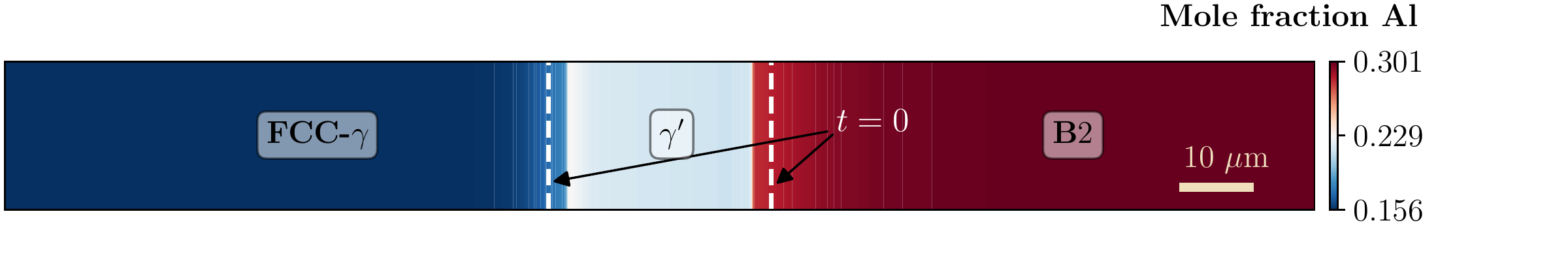}
\caption{}
\label{Fig.3a1}
\end{subfigure}\\
\begin{subfigure}{1.0\textwidth}
\includegraphics[trim=0 10 60 0, clip, keepaspectratio,width=\linewidth]{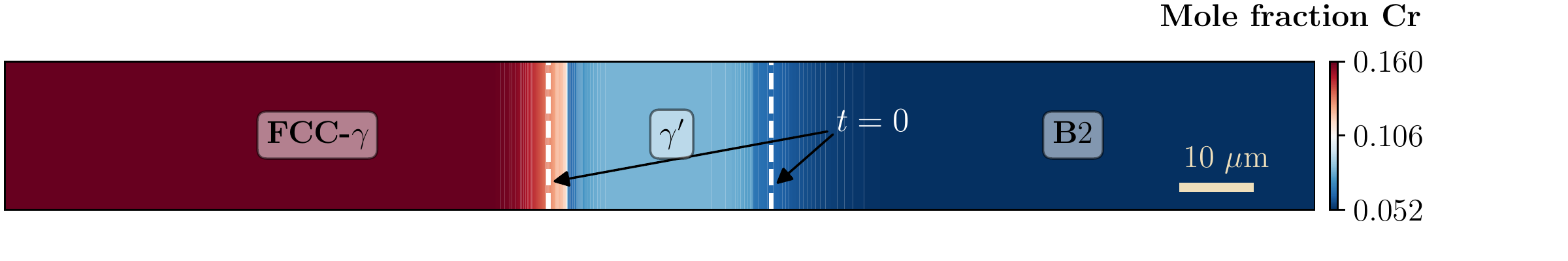}
\caption{}
\label{Fig.3a2}
\end{subfigure}
\end{center}
\begin{subfigure}{0.50\textwidth}
\includegraphics[keepaspectratio,width=\linewidth]{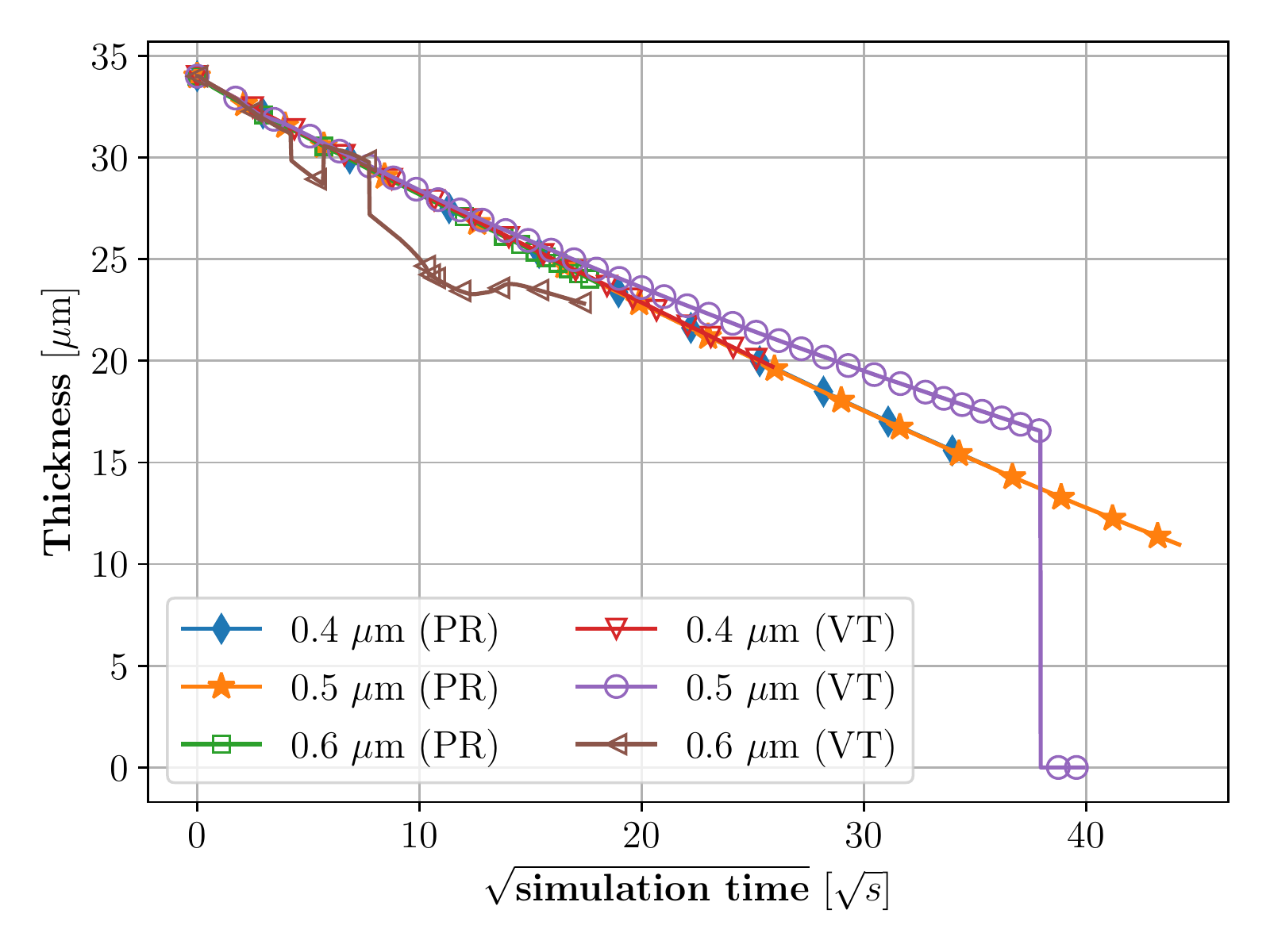}
\caption{}
\label{Fig.3b}
\end{subfigure}
\begin{subfigure}{0.50\textwidth}
\includegraphics[keepaspectratio,width=\linewidth]{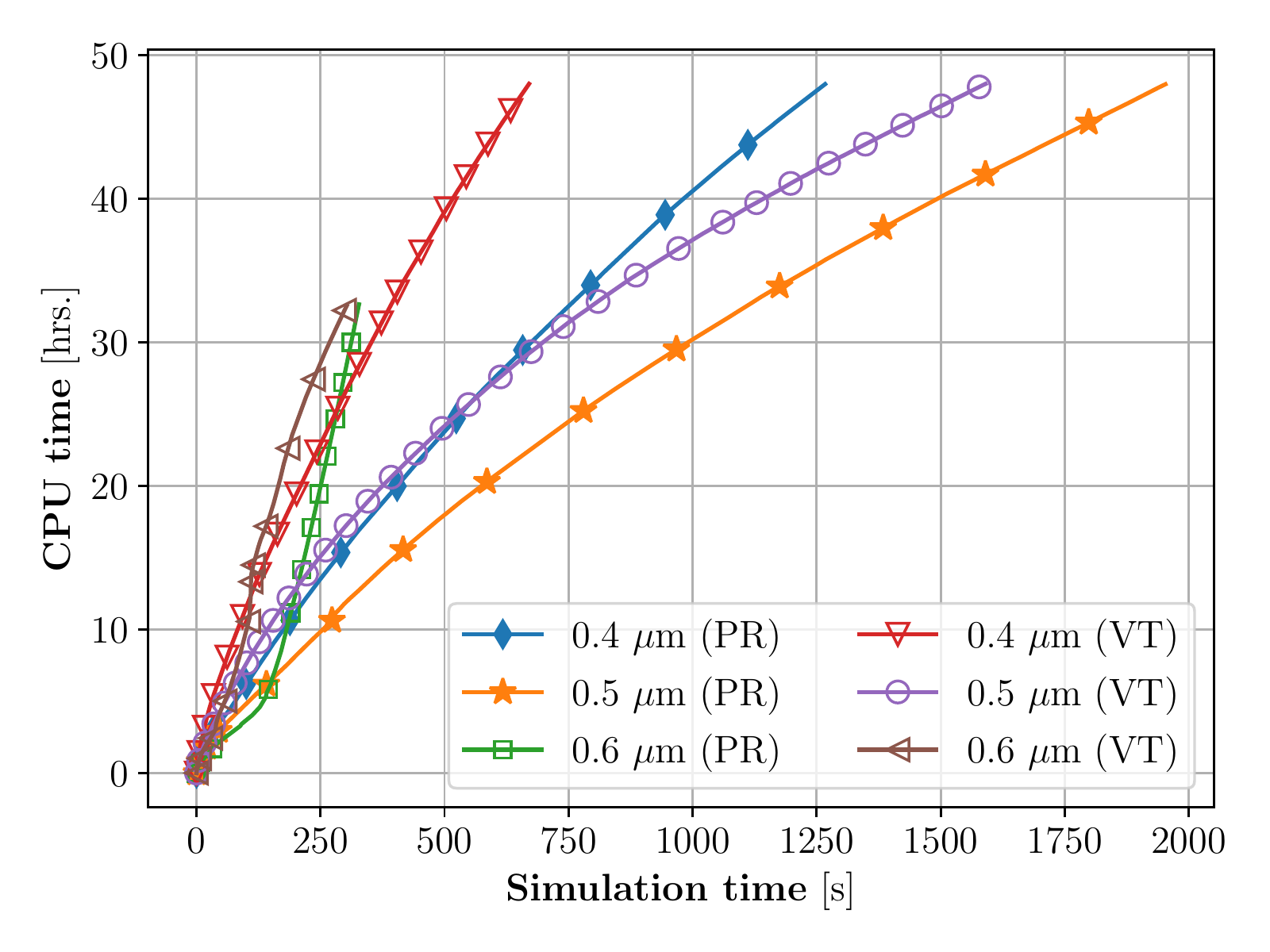}
\caption{}
\label{Fig.3c}
\end{subfigure}
\caption{For an Al-Cr-Ni fcc-$\gamma/\gamma^{\prime}/\mathrm{B2}$ coherently stressed planar diffusion couple: schematic of the simulation domain, eigenstrains and mechanical boundary conditions (a); the simulated Al-mole fraction field (b) and Cr-mole fraction field (c) at time $t=100$ s. For three different interface widths, the temporal variation in the ternary $\gamma^{\prime}$ thickness as a function of the square root of simulation time using the partial rank-one (PR) and Voigt-Taylor (VT) homogenization schemes (d); and change in CPU time with simulation time for both these schemes (e). }
\end{figure}

Secondly, we considered a planar ternary Al-Cr-Ni $\mathrm{fcc-}\gamma/\gamma^{\prime}/\mathrm{B2}$  alloy having a similar geometry compared to the previous case (Fig. \ref{Fig3New}). Moreover, the boundary conditions at the top, left, and bottom boundaries are identical to the previous case. However, the mechanical displacements at the right boundary are
\begin{align}
u_{x}(x=L_{x}/2,y,t) &= u_{x}^{R},\label{ResEq5}\\
u_{y}(x=L_{x/2},y,t) & = u_{y}^{R},\label{ResEq6}
\end{align}
where $u_{x}^{R}$ and $u_{y}^{R}$ are the imposed mechanical displacements (Table \ref{table:bc}). 

Unlike the previous case, the three phases, in this case, may coexist in equilibrium because the system is ternary. However, the initial conditions are set such that the system is out of equilibrium. Consequently, we find that $\gamma^{\prime}$ phase shrinks while the $\gamma$ and B$2$ phases grow. The simulated  Al and Cr mole fraction fields using the PR scheme at time $t= 100$ s are shown in Figs. \ref{Fig.3a1} and \ref{Fig.3a2}.   Moreover, we find that the thickness of the ternary $\gamma^{\prime}$ phase decreases linearly as a function of the square root of simulation time using the PR scheme (Fig. \ref{Fig.3b}). Moreover, we find that this variation remains independent of interface width using the partial rank-one (PR) scheme (Fig. \ref{Fig.3b}). This is, however, not true for simulations using the VT scheme.  Specifically, we find that as the interface width is increased from $0.4$ $\mu$m to $0.6$ $\mu$m,  the VT scheme shows deviation from the expected parabolic growth kinetics (Fig. \ref{Fig.3b}). Because this behaviour is a consequence of the increase in the interface width, this deviation from the parabolic kinetics may be attributed to the excess interfacial energy contribution arising in the case of the VT scheme. Surprisingly, we find that the CPU time is higher using both PR and VT schemes for the simulations with interface widths of $0.6$ $\mu$m compared to cases having interface widths of $0.4$ $\mu$m  and $0.5$ $\mu$m (Fig. \ref{Fig.3c}). Nevertheless, we find that the convergence of the PR scheme is significantly faster compared to the VT scheme for interface width values of $0.4$ $\mu$m and $0.5$ $\mu$m (Fig. \ref{Fig.3c}). 

\begin{figure}[!ht]
\begin{subfigure}{0.50\textwidth}
\includegraphics[keepaspectratio,width=\linewidth]{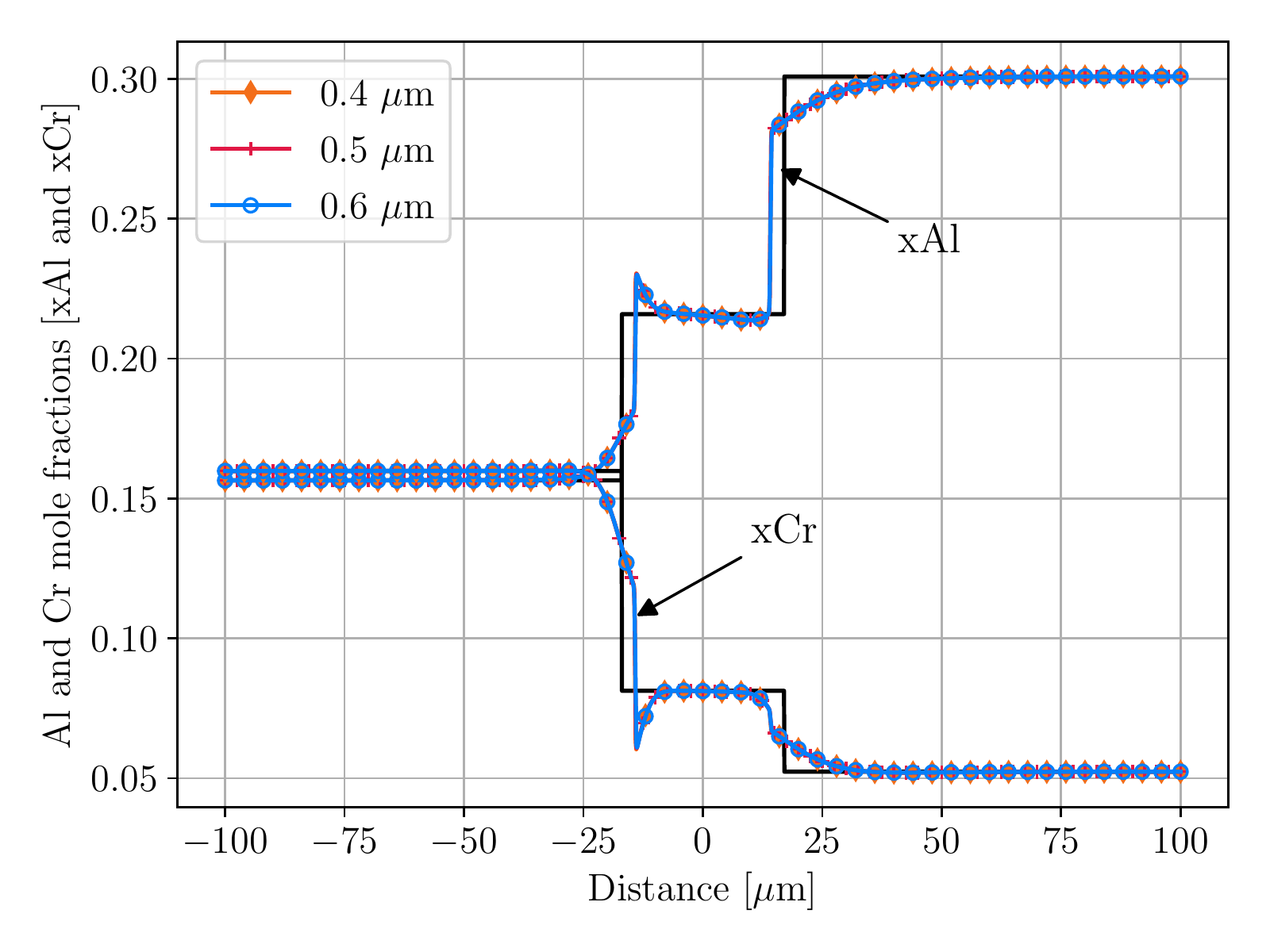}
\caption{}
\label{Fig.4a}
\end{subfigure}
\begin{subfigure}{0.50\textwidth}
\includegraphics[keepaspectratio,width=\linewidth]{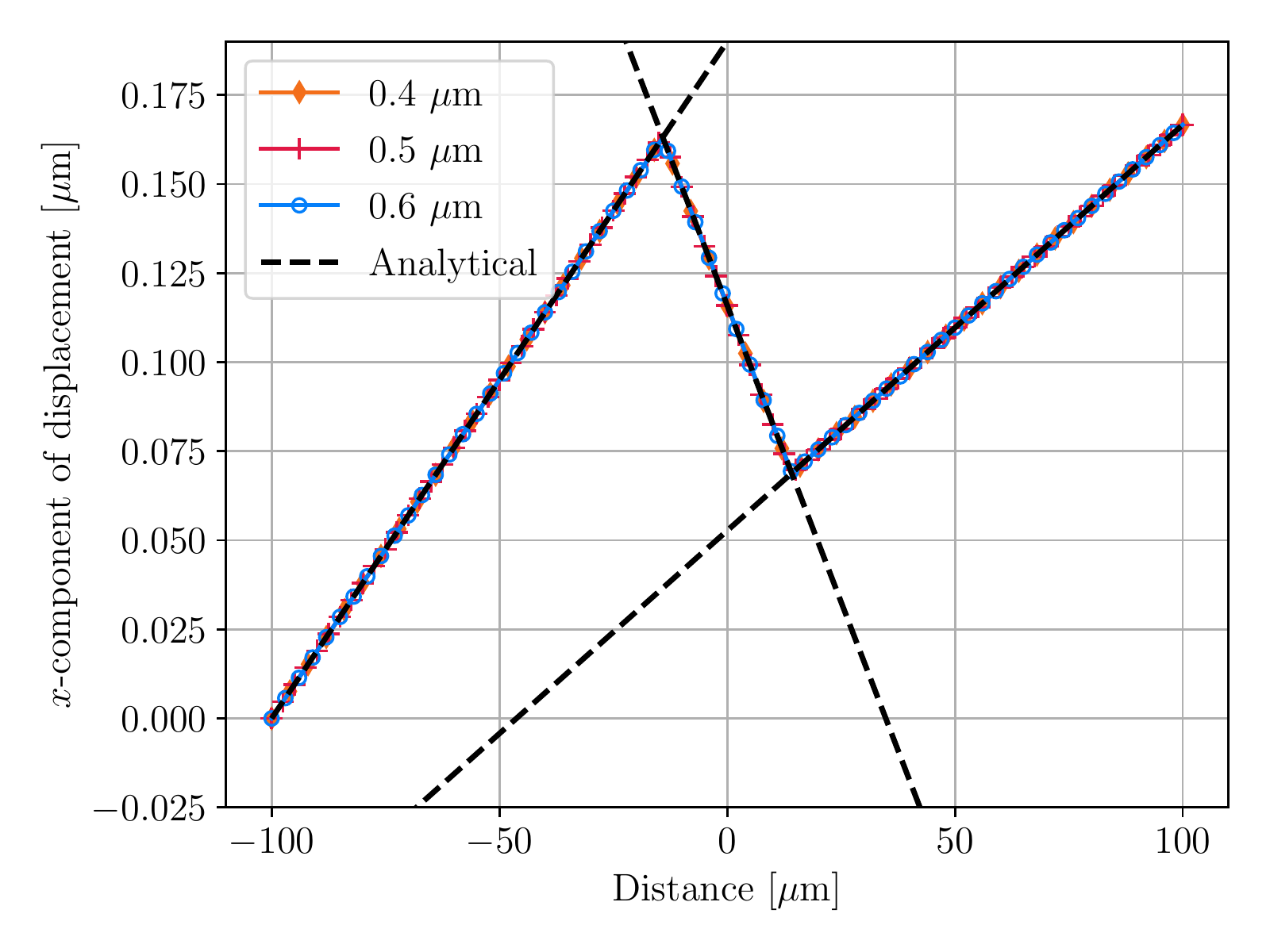}
\caption{}
\label{Fig.4b}
\end{subfigure}
\begin{subfigure}{0.50\textwidth}
\includegraphics[keepaspectratio,width=\linewidth]{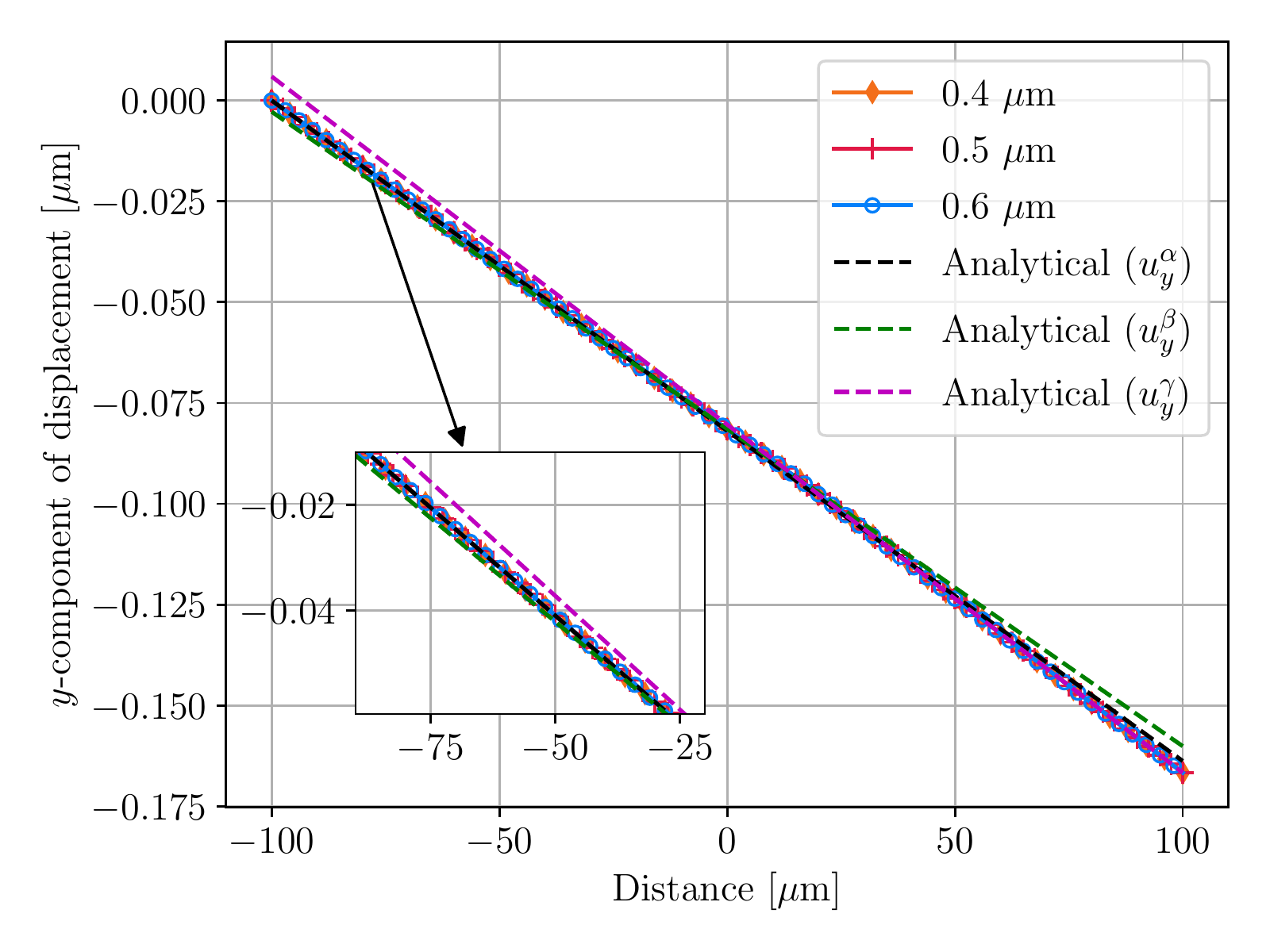}
\caption{}
\label{Fig.4c}
\end{subfigure}
\begin{subfigure}{0.50\textwidth}
\includegraphics[keepaspectratio,width=\linewidth]{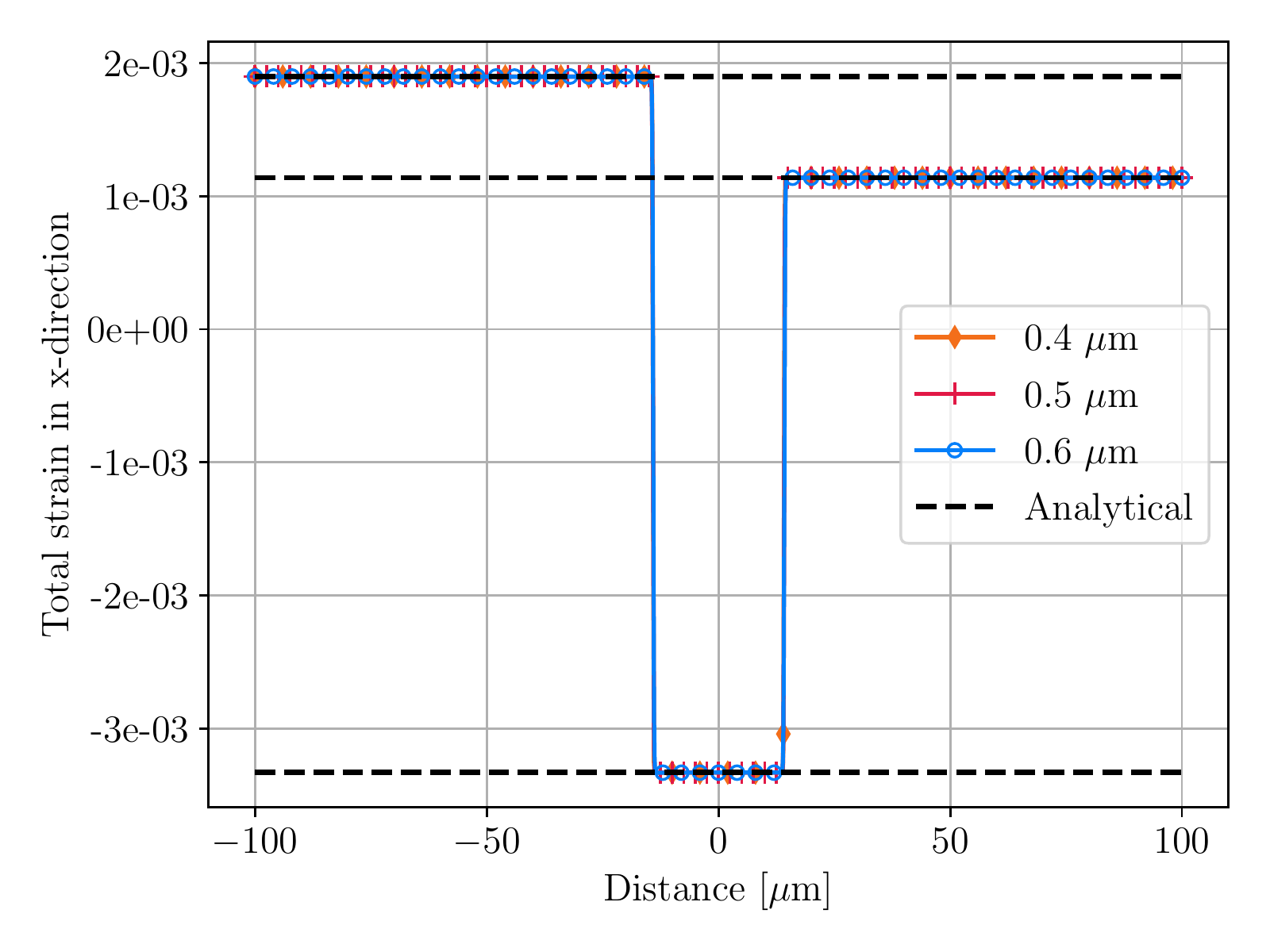}
\caption{}
\label{Fig.4d}
\end{subfigure}
\begin{subfigure}{0.50\textwidth}
\includegraphics[keepaspectratio,width=\linewidth]{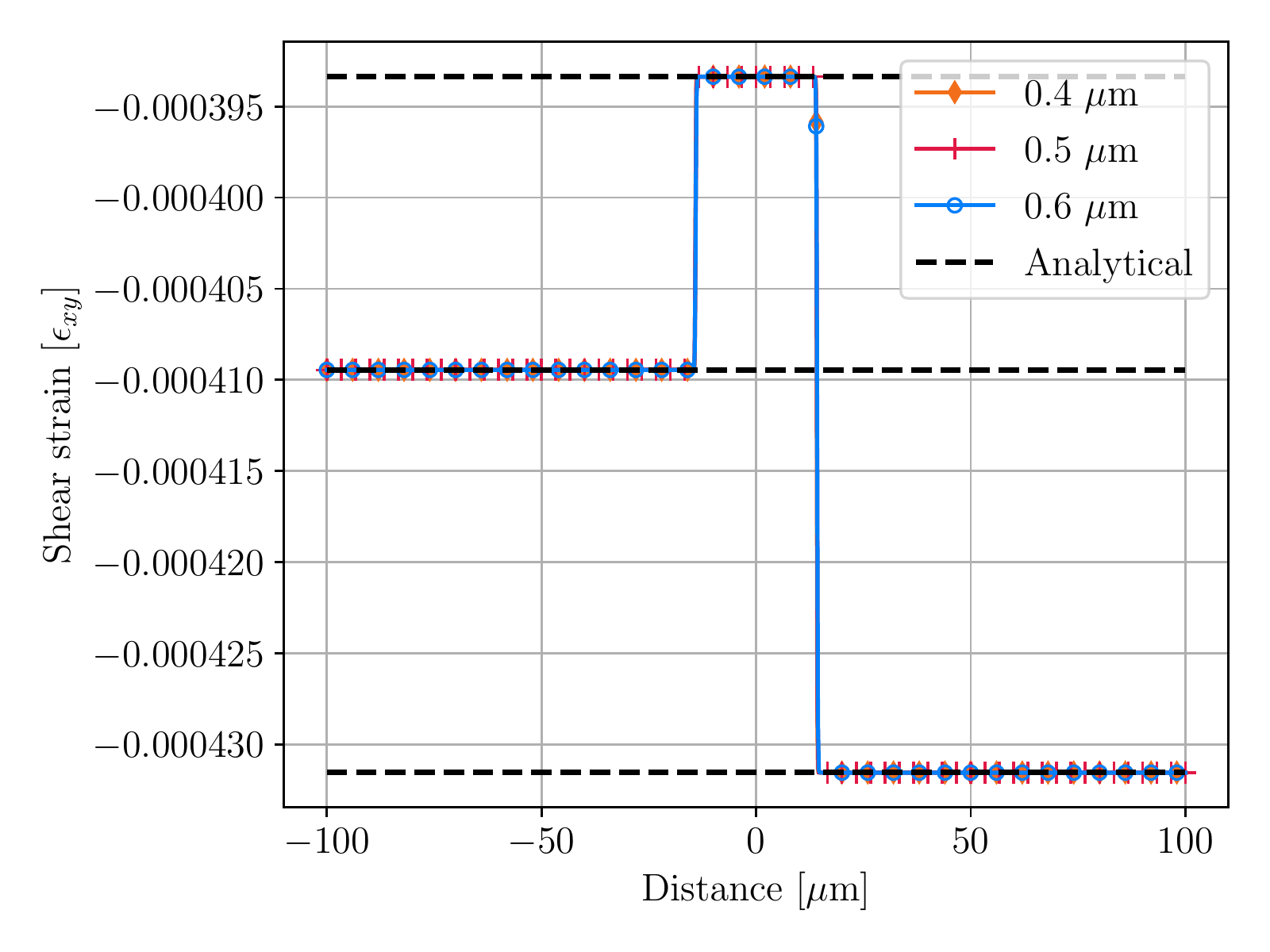}
\caption{}
\label{Fig.4e}
\end{subfigure}
\begin{subfigure}{0.50\textwidth}
\includegraphics[keepaspectratio,width=\linewidth]{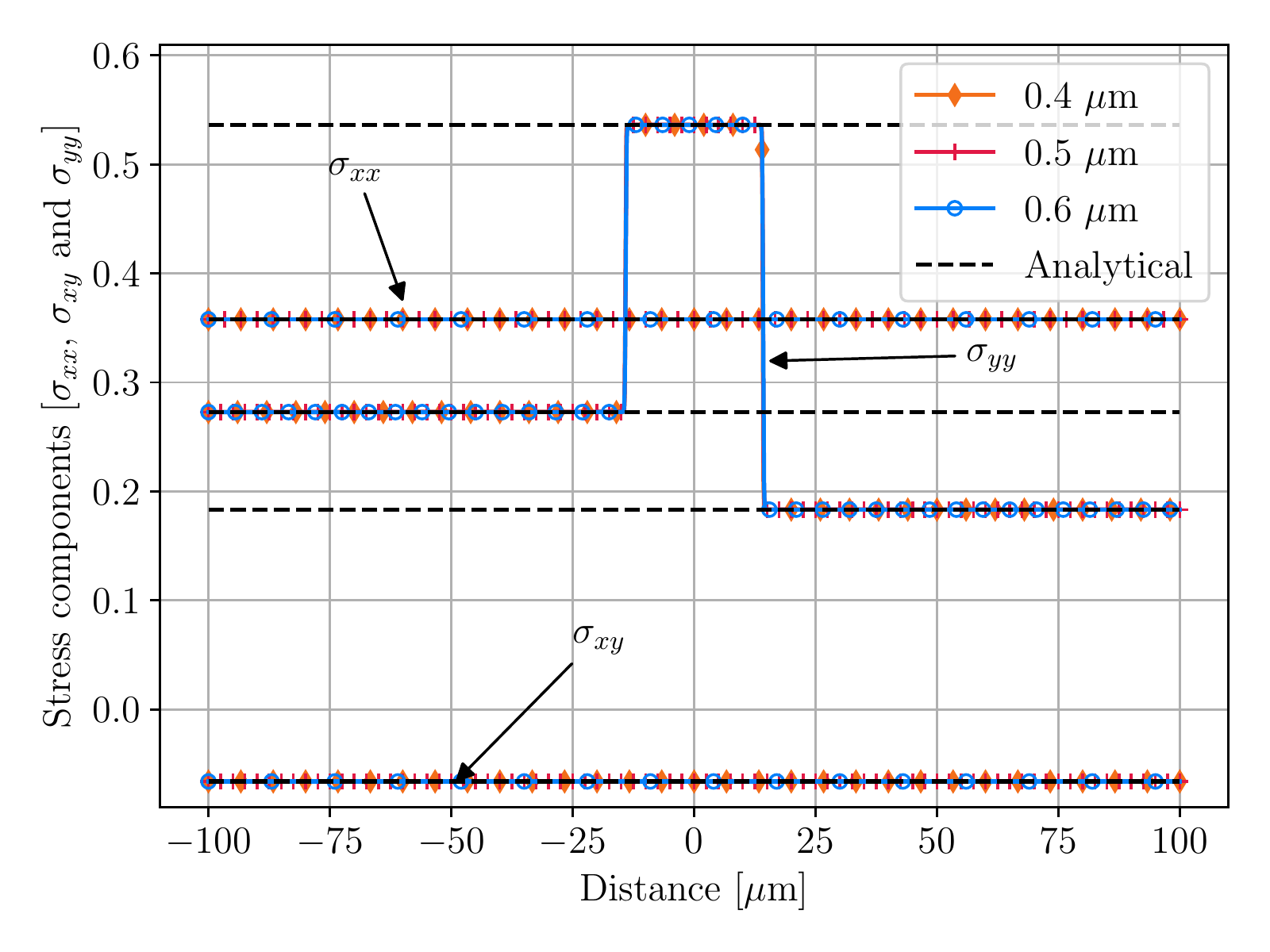}
\caption{}
\label{Fig.4f}
\end{subfigure}
\caption{For an Al-Cr-Ni fcc-$\gamma/\gamma^{\prime}/\mathrm{B2}$ coherently stressed planar diffusion couple, the spatial distribution of Al and Cr-mole fraction profiles (a); x and y-components of displacement field  (b) and (c);  total normal and shear strain (d) and (e); normal and shear stresses (f), as functions of distance perpendicular to the interface at time $t= 100$ s. Simulations using different interface widths are also superimposed on these figures. The superimposed dotted black lines indicate the analytical solution. For sake of clarity, the analytical solution for the $y$-component of displacement field within the bulk regions are denoted by different colours in Fig. \ref{Fig.4b}.}
\label{Fig.4}
\end{figure}

To verify the spatial accuracy, we calculate the composition and elastic fields along a line parallel to the interface normal at time $t=100$ s. We find that the spatial distribution of the simulated Al and Cr mole fraction fields normal to the interface is independent of the interface width (Fig. \ref{Fig.4a}). Moreover, the simulated x-component of the displacement field normal to the interface shows good quantitative agreement with the analytical solution, independent of the choice of interface width (Fig.\ref{Fig.4b}).  Due to the applied mechanical displacement at the right boundary, the $y$-component of the displacement field is also non-zero in this case. Fig. \ref{Fig.4c} shows that the simulated and analytically obtained solutions for the y-component displacement field are also in quantitative agreement in the bulk domains. Likewise, this agreement is not a function of the interface width. Expectedly, we find that the total strain normal to the interface is constant within the bulk phases and is in agreement with the analytical solution (Fig. \ref{Fig.4d}). Since the system is elastically anisotropic, the shear strains are non-zero and constant within the bulk phases  (Fig. \ref{Fig.4e}). Finally, the non-zero elastic stresses as a function of distance normal to the interface are shown in Fig. \ref{Fig.4f}.

\subsection{Non-planar three-phase Ni-Al alloy}
\begin{figure}[!ht]
\begin{subfigure}{0.55\textwidth}
\includegraphics[trim=24 15 10 10, clip, keepaspectratio,width=\linewidth]{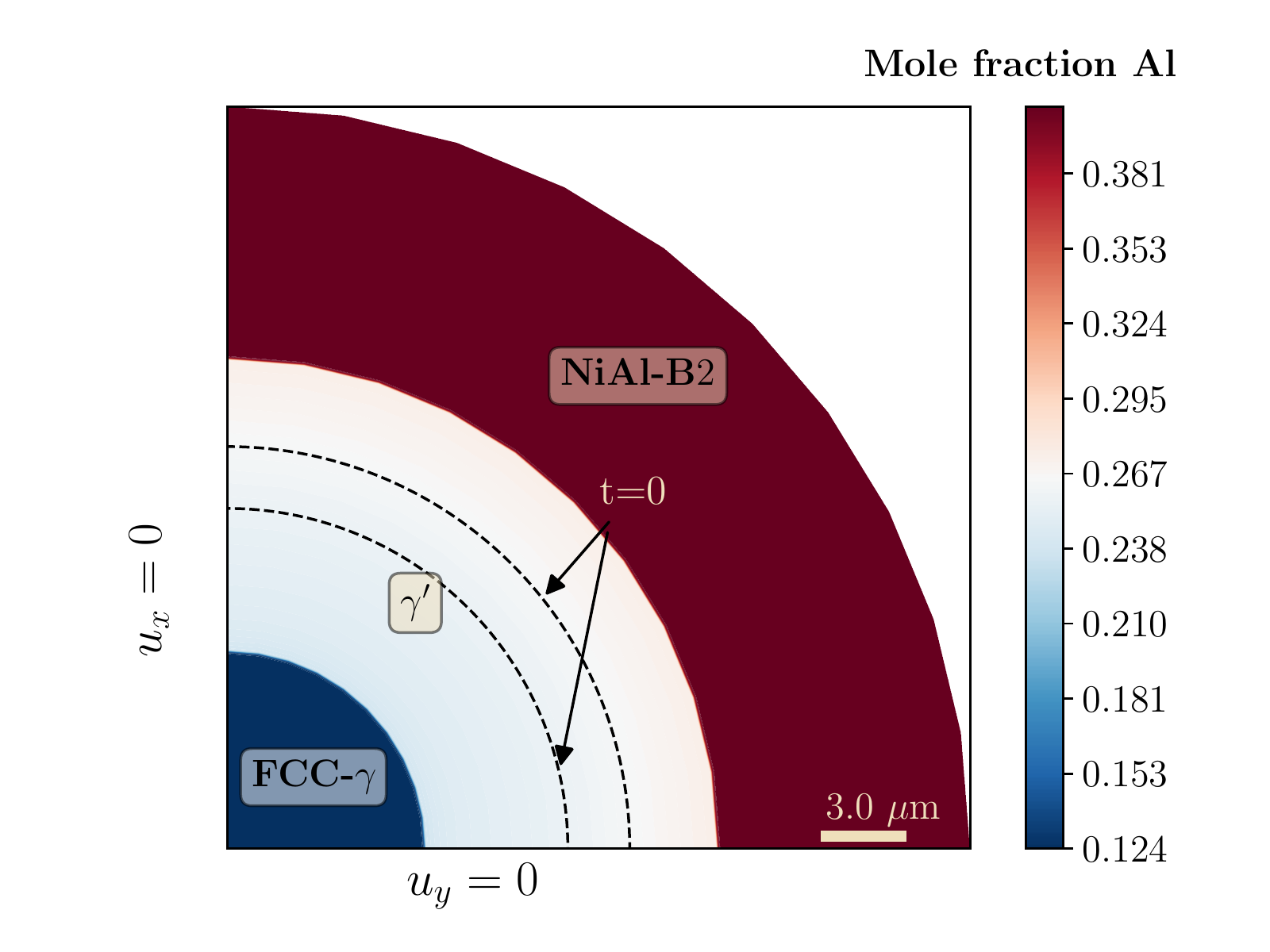}
\caption{}
\label{Fig.5a}
\end{subfigure}
\begin{subfigure}{0.5\textwidth}
\includegraphics[trim=60 15 12 10, clip, keepaspectratio,width=\linewidth]{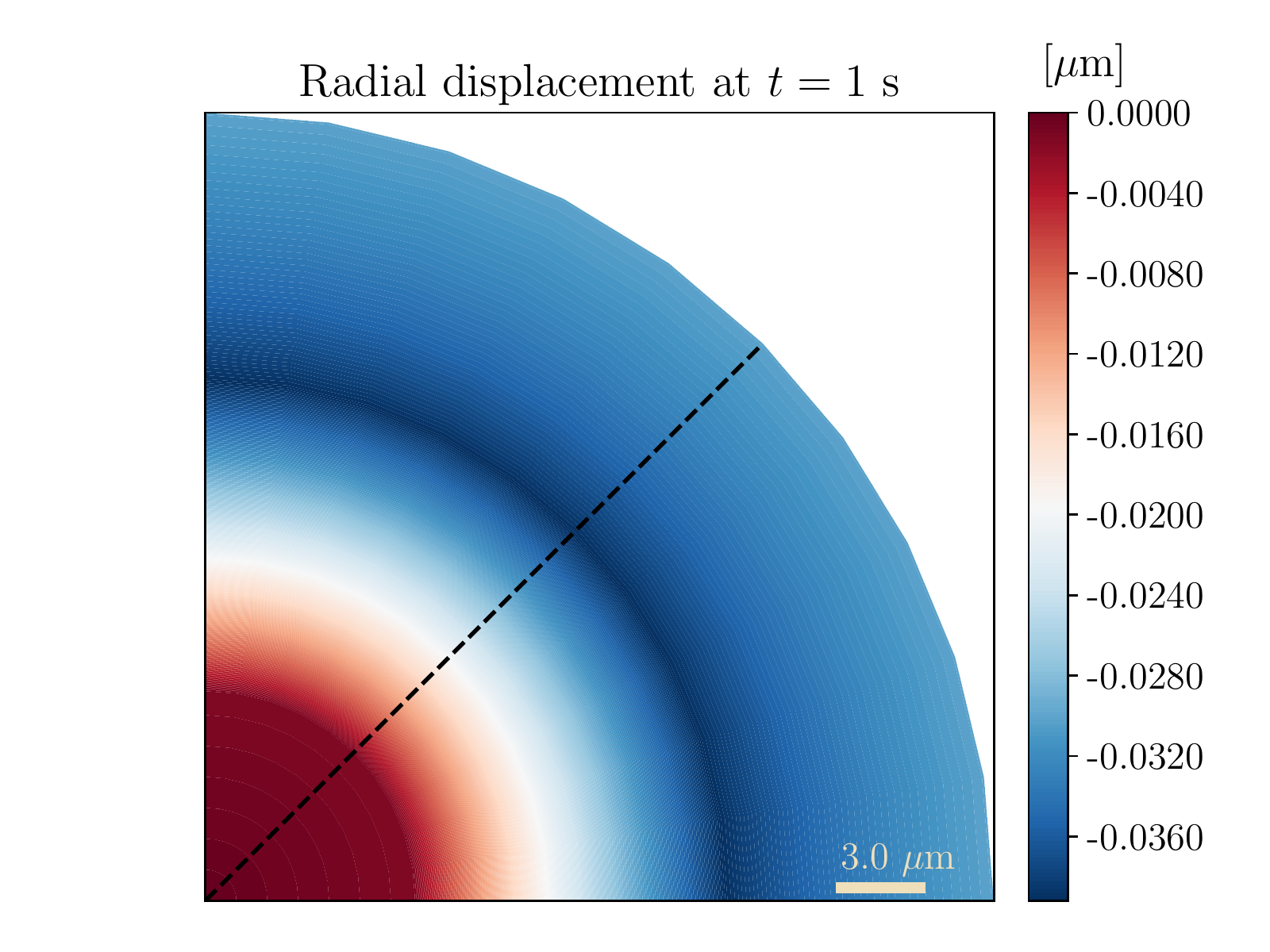}
\caption{}
\label{Fig.5b}
\end{subfigure}
\begin{subfigure}{0.50\textwidth}
\includegraphics[keepaspectratio,width=\linewidth]{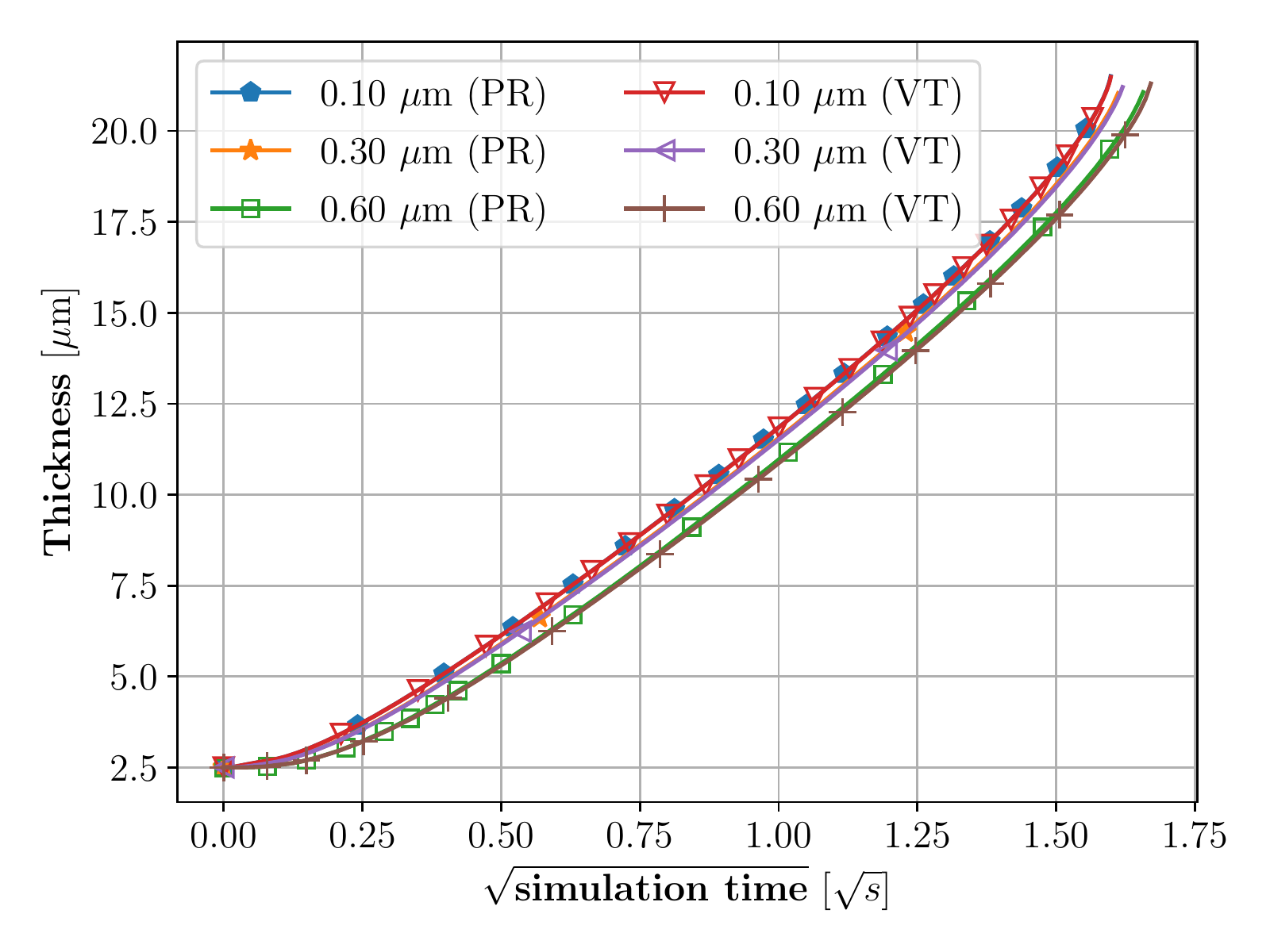}
\caption{}
\label{Fig.5c}
\end{subfigure}
\begin{subfigure}{0.50\textwidth}
\includegraphics[keepaspectratio,width=\linewidth]{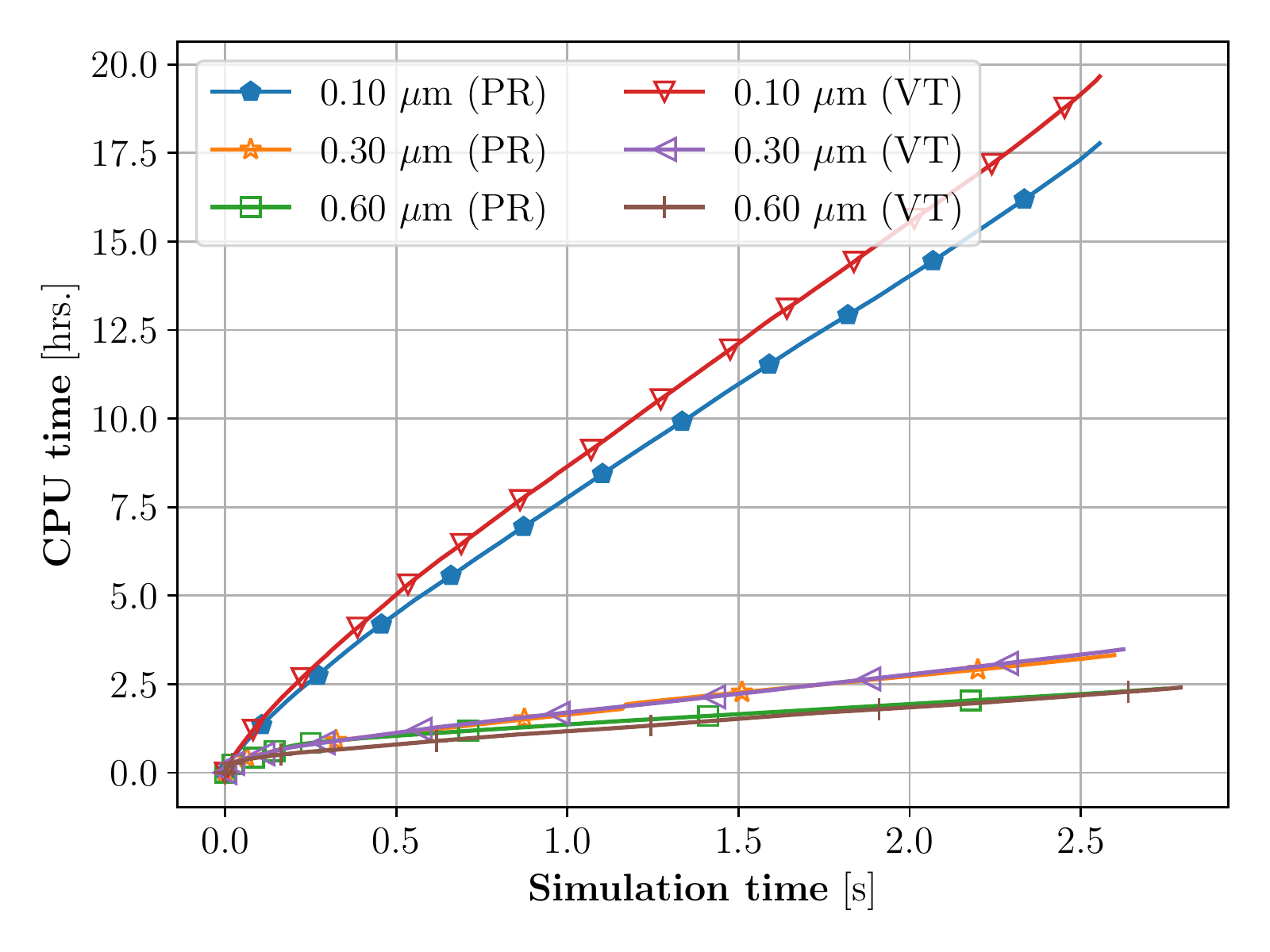}
\caption{}
\label{Fig.5d}
\end{subfigure}
\caption{For a non-planar Ni-Al fcc-$\gamma/\mathrm{Ni}_3\mathrm{Al}/\mathrm{NiAl}$ coherently stressed diffusion couple: simulation domain, eigenstrains, mechanical boundary conditions and the simulated Al-mole fraction field at time $t=1$ s (a); the simulated radial displacement field at the same time (b). For three different interface widths, variation in $\mathrm{Ni}_3\mathrm{Al}$ thickness as a function of the square root of simulation time using the partial rank-one (PR) and Voigt-Taylor (VT) homogenization schemes (c); the CPU time as a function of simulation time for both these schemes (d).}
\label{Fig.5}
\end{figure}

Thirdly, we simulated a $\mathrm{fcc-}\gamma/\gamma^{\prime}\mathrm{-Ni}_3\mathrm{Al}/\mathrm{NiAl}$ alloy having concentric ring interfaces (Fig. \ref{Fig.5a}). We have assumed homogeneous Neumann boundary conditions along the left, bottom and outer boundaries for the composition and phase-field variables. Further, we have imposed symmetry boundary conditions on displacements along the bottom and left boundaries, and zero traction boundary conditions at the outer surface, viz.
\begin{align}
u_{x} (x=0, y, t)&= 0,\\
u_{y} (x, y=0,t)&= 0,\\
\boldsymbol{\sigma}\boldsymbol{n}^{\Gamma}(x,y,t) &= \boldsymbol{0}\quad \mathrm{on} \quad x^{2} + y^2 = R_{0}^{2},
\end{align}
where $R_{0}$ is the radius of the domain.

Similar to our first case, the three phases cannot coexist in equilibrium. Thus, we find that the intermediate  $\gamma^{\prime}$ phase grows while the innermost $\gamma$ and outermost NiAl phases shrink. We run this simulation until the $\gamma/\gamma^{\prime}$ interface vanishes. Figs. \ref{Fig.5a} and \ref{Fig.5b} show the simulated contour map of the Al mole fraction and the radial displacement fields at time $t=1$ s for an interface thickness of  $0.15$ $\mu$m using the PR scheme. The variation in the $\gamma^{\prime}$ thickness as a function of the square root of simulation time using the PR and VT schemes are shown in Fig.\ref{Fig.5c}. As expected, we find parabolic growth kinetics using both these schemes. To check the influence of this result on interface width, we vary the interface width from $0.10$ $\mu$m to $0.30$ $\mu$m.

We find that the interface kinetics remains unaffected for interface width values of $0.10$ $\mu$m and $0.30$ $\mu$m using both PR and VT schemes (Fig. \ref{Fig.5c}). However, for the case with interface width value of $0.60$ $\mu$m, we find that the calculated thickness of $\mathrm{Ni}_3\mathrm{Al}$ is slightly lower in both schemes (Fig. \ref{Fig.5c}). Nevertheless, we find that the kinetics is still parabolic. Moreover, for a given simulation time, CPU time in the case of the PR scheme is always lower compared to the VT scheme for interface width values of $0.10$ $\mu$m and $0.30$ $\mu$m (Fig. \ref{Fig.5d}). The difference in CPU time is however negligible for an interface width of $0.60$ $\mu$m. This suggests that the PR scheme converges at a faster or nearly equal rate compared to the VT scheme for this system.

\begin{figure}[!ht]
\begin{subfigure}{0.5\textwidth}
\includegraphics[keepaspectratio,width=\linewidth]{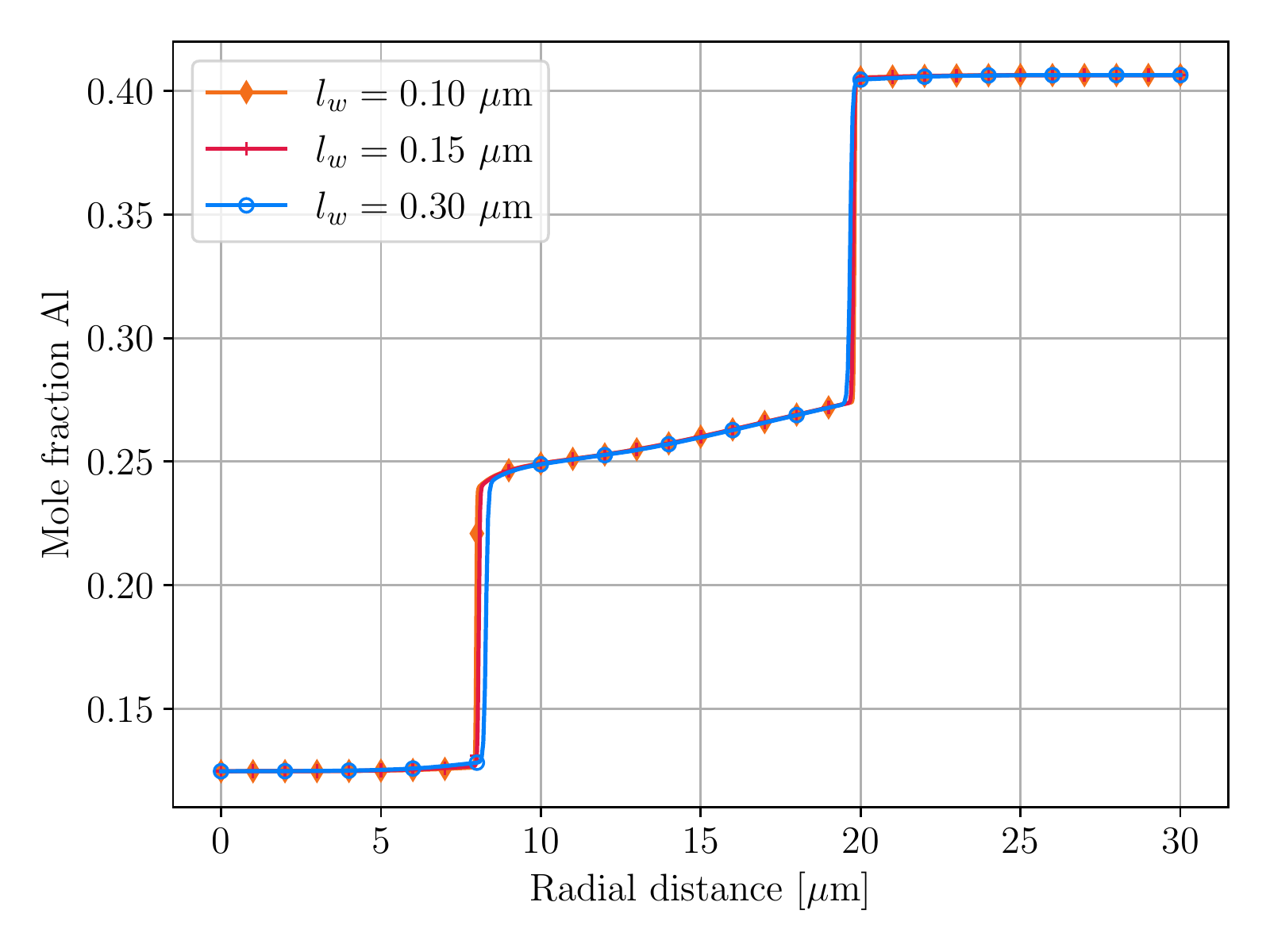}
\caption{}
\label{Fig.6a}
\end{subfigure}
\begin{subfigure}{0.5\textwidth}
\includegraphics[keepaspectratio,width=\linewidth]{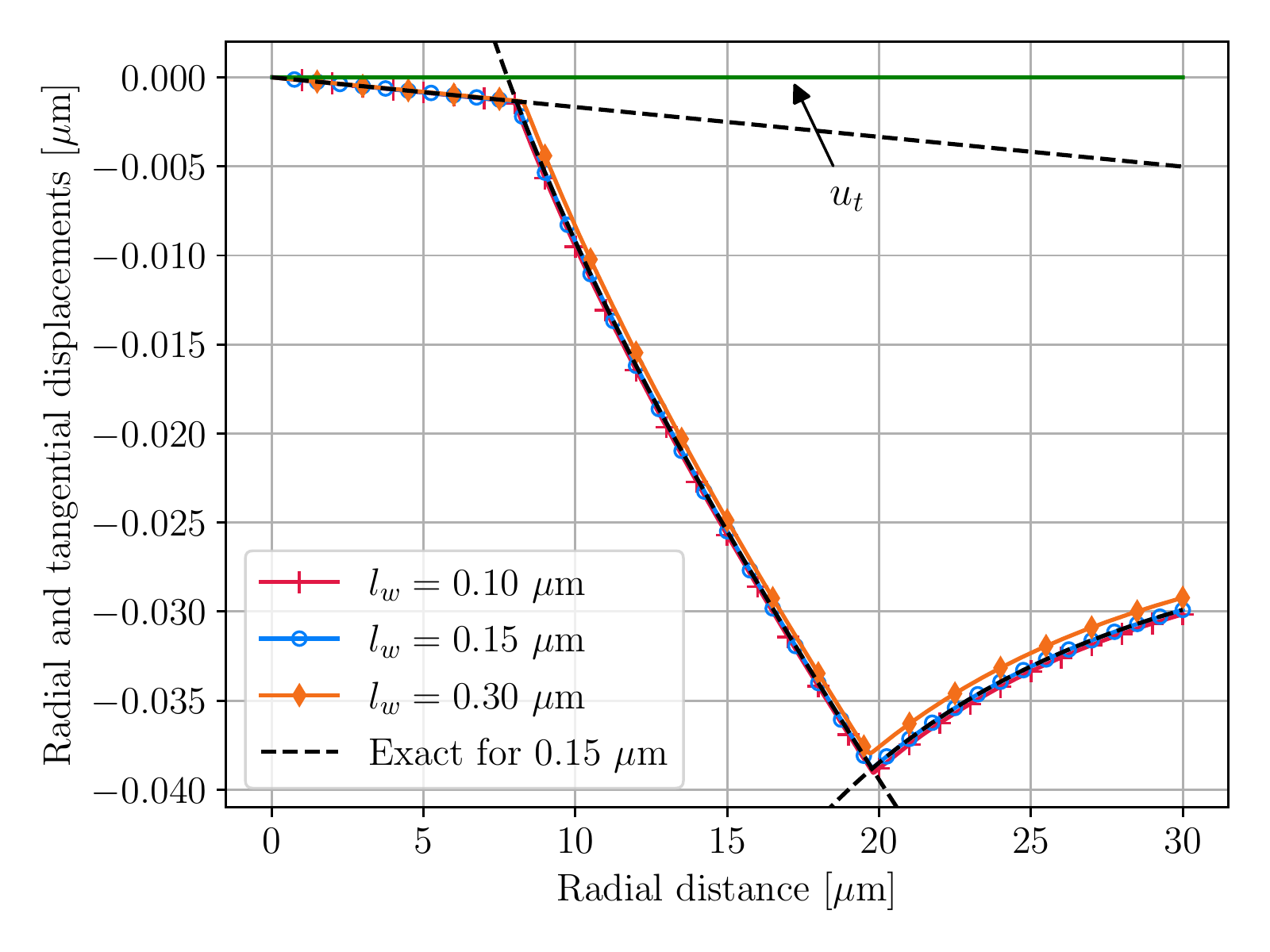}
\caption{}
\label{Fig.6b}
\end{subfigure}
\begin{subfigure}{0.5\textwidth}
\includegraphics[keepaspectratio,width=\linewidth]{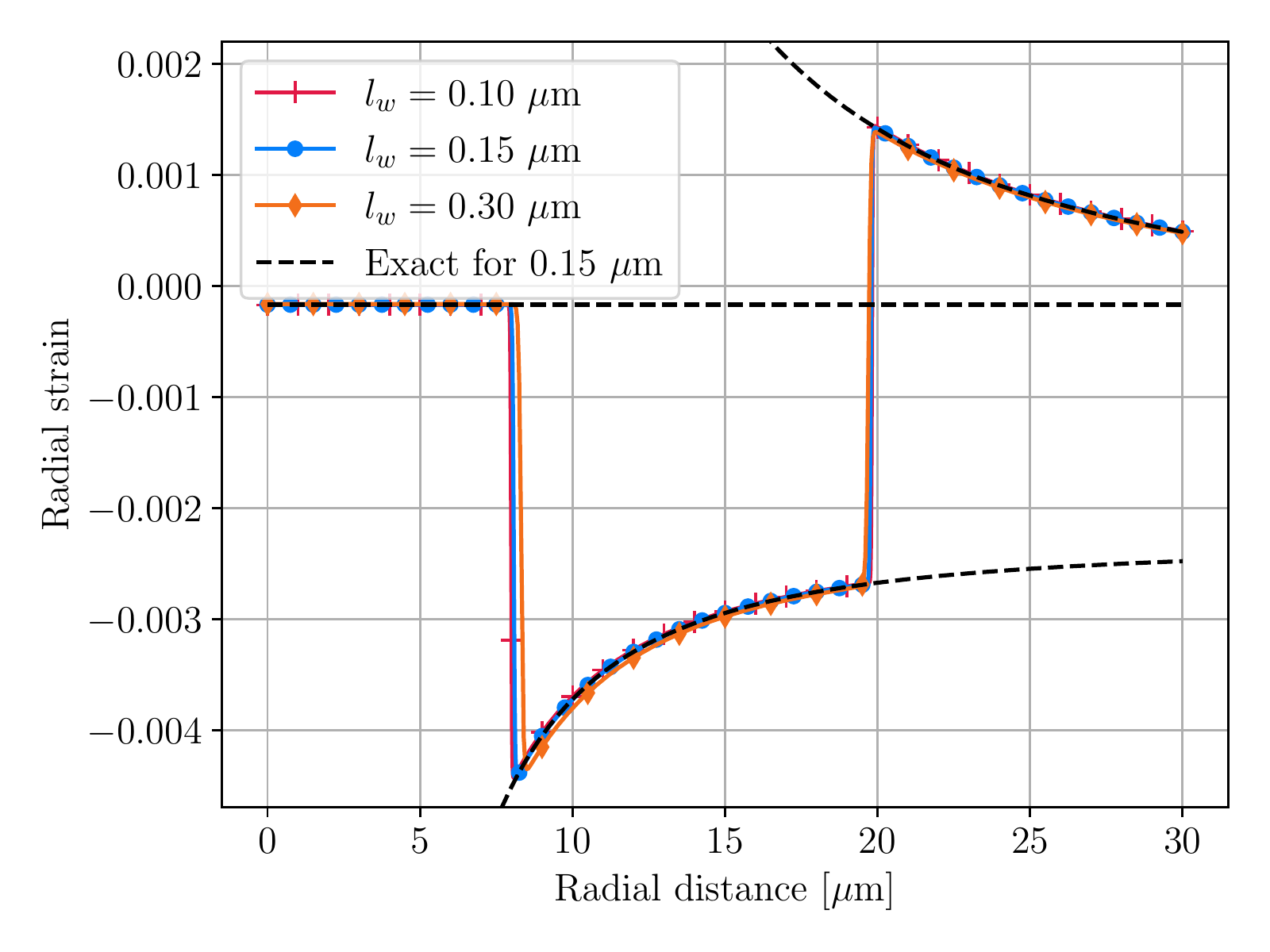}
\caption{}
\label{Fig.6c}
\end{subfigure}
\begin{subfigure}{0.5\textwidth}
\includegraphics[keepaspectratio,width=\linewidth]{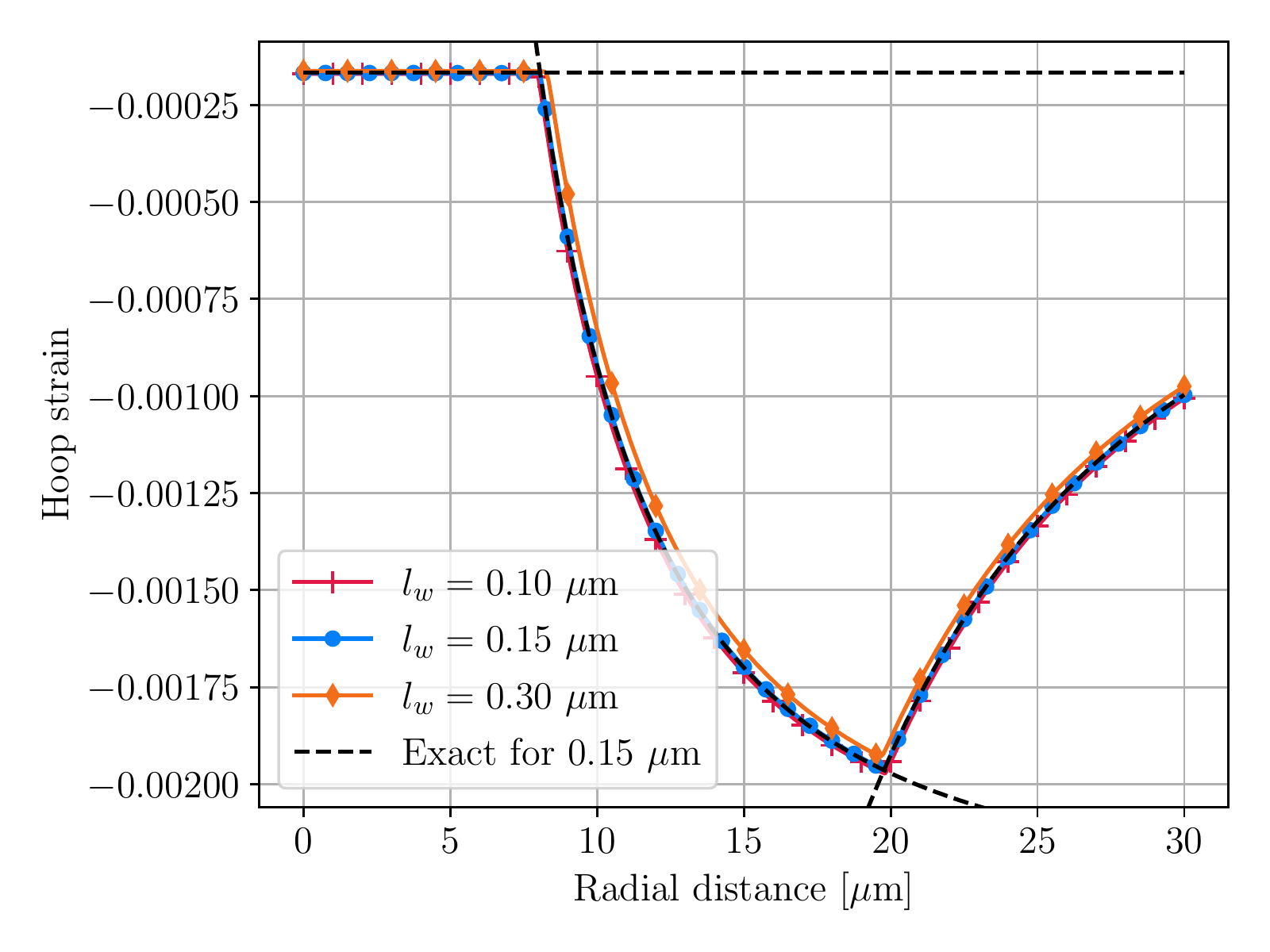}
\caption{}
\label{Fig.6d}
\end{subfigure}
\begin{subfigure}{0.5\textwidth}
\includegraphics[keepaspectratio,width=\linewidth]{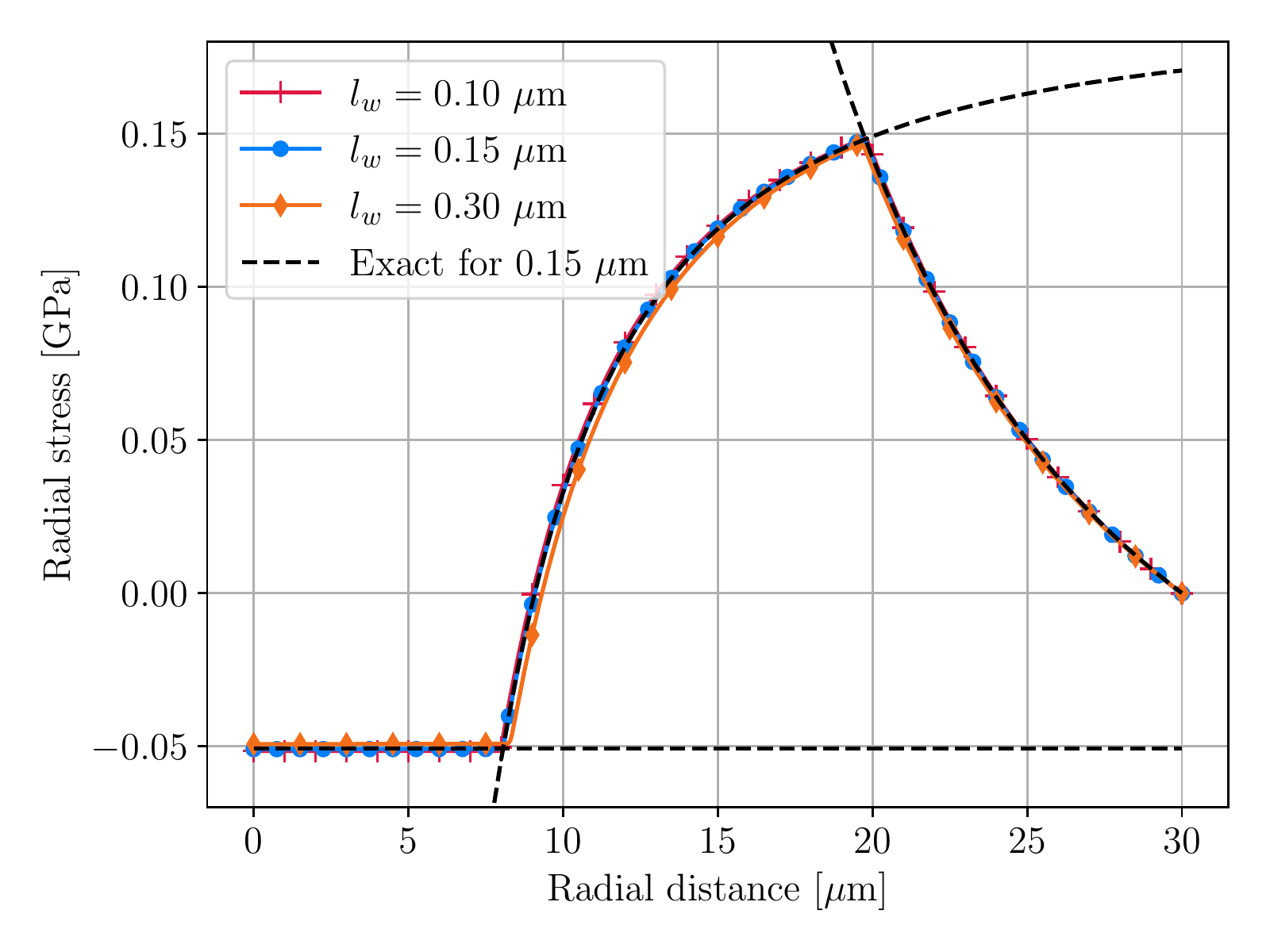}
\caption{}
\label{Fig.6e}
\end{subfigure}
\begin{subfigure}{0.5\textwidth}
\includegraphics[keepaspectratio,width=\linewidth]{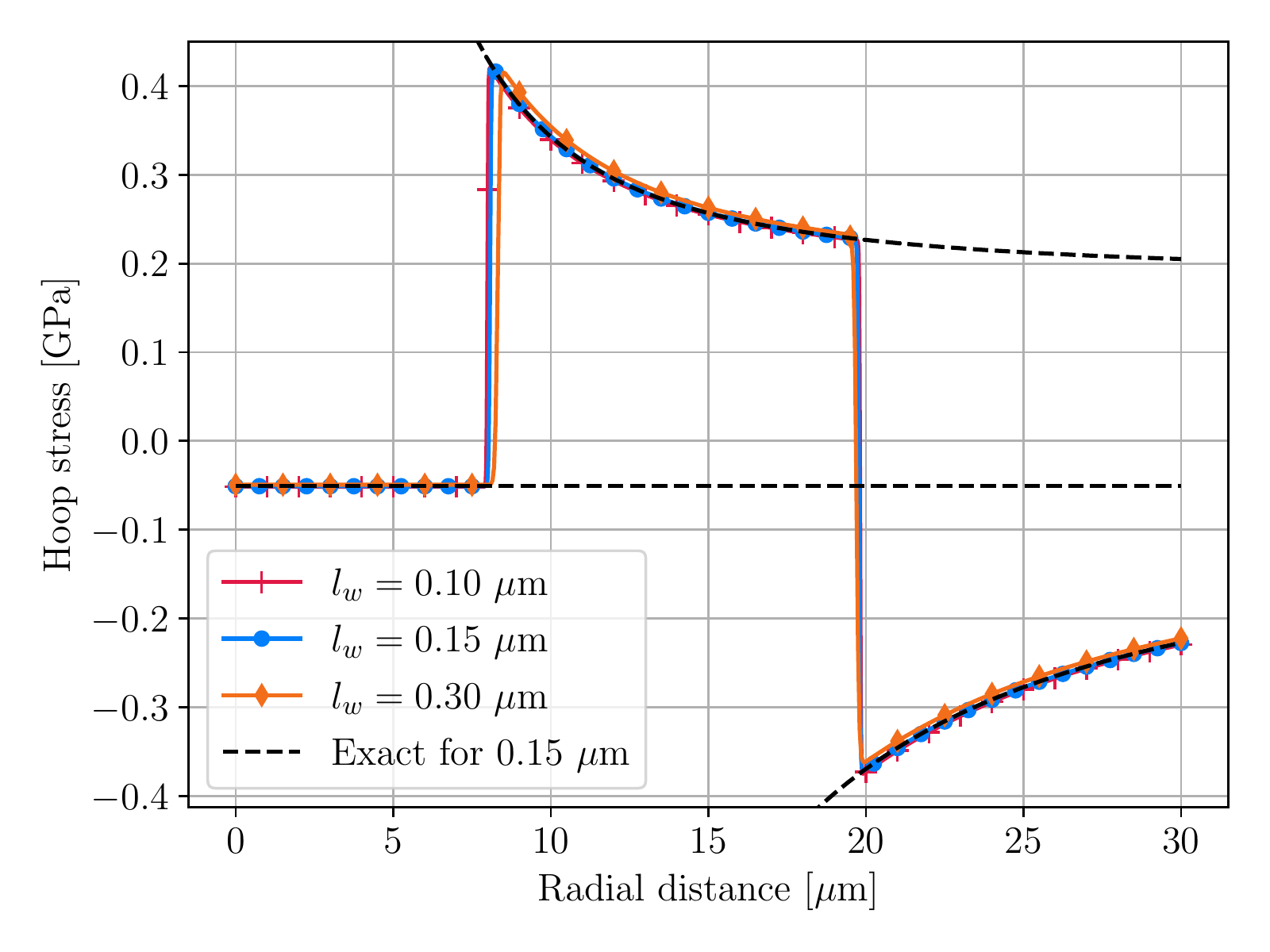}
\caption{}
\label{Fig.6f}
\end{subfigure}
\caption{For a non-planar Ni-Al fcc-$\gamma/\mathrm{Ni}_3\mathrm{Al}/\mathrm{NiAl}$ diffusion couple, the spatial variation in a) Al-mole fraction; b) radial and tangential displacements; c) radial strain; d) hoop strain; e) radial stress; and f) hoop stress as functions of radial distance at time $t=1$ s. The plots show this variation for three different interface widths $l_{w}$ using the partial rank-one scheme. The dotted and discontinuous black lines are the analytically obtained solutions.}
\label{Fig.6}
\end{figure}

To verify the spatial accuracy of the simulated solution, we calculated the composition and elastic fields along the radius at time $t=100$ s (Fig.{\ref{Fig.5b}}). We find that the radial distribution of the Al mole fraction field within the bulk domains remains independent of interface width for values between $0.10$ $\mu$m and $0.30$ $\mu$m (Fig. \ref{Fig.6a}). Likewise, the radial displacement within the bulk phases remains unaltered with varying interface width (Fig. \ref{Fig.6b}). Also, note that the tangential displacement is negligible within the bulk phases (Fig. \ref{Fig.6b}). We also find excellent quantitative agreement between the simulated and analytically obtained radial displacement in the bulk $\gamma$ and $\mathrm{Ni}_3\mathrm{Al}$ phases. It should be emphasized that the analytical solution uses the interface positions calculated from the simulation with an interface width of $0.15$ $\mu$m (Fig. \ref{Fig.6b}).

However, for the simulation with an interface width of $0.30$ $\mu$m, we see that the radial displacement near the Ni$_3$Al/NiAl interface deviates marginally from this analytical solution. It should be noted that a similar observation was made for the planar Ni-Al case. Furthermore, we explained this deviation by accounting for the inaccuracy caused by numerically determining the interface positions. As discussed before, the analytical solution is sensitive to the calculated interface positions, which are, in turn, dependent on the interface width. Further, we have verified this assertion by matching the simulated field for this case with an analytical solution where the interface positions are calculated using the same interface thickness.

We also find quantitative agreement between the simulated and the analytically obtained radial and hoop strains (Figs. \ref{Fig.6c} and \ref{Fig.6d}). As expected, the radial and hoop strains are equal and constant in the bulk $\gamma$ phase. However, the radial and hoop strains are dependent on the radius in the $\gamma^{\prime}$-Ni$_3$Al and NiAl phases. Likewise, for the radial and hoop stress fields, we also obtained a good match between the analytical and simulated fields (Figs. \ref{Fig.6e} and \ref{Fig.6f}).

\subsection{Non-planar three-phase Al-Cr-Ni alloy}
\begin{figure}[!ht]
\begin{subfigure}{0.55\textwidth}
\includegraphics[trim=24 15 25 10, clip, keepaspectratio,width=\linewidth]{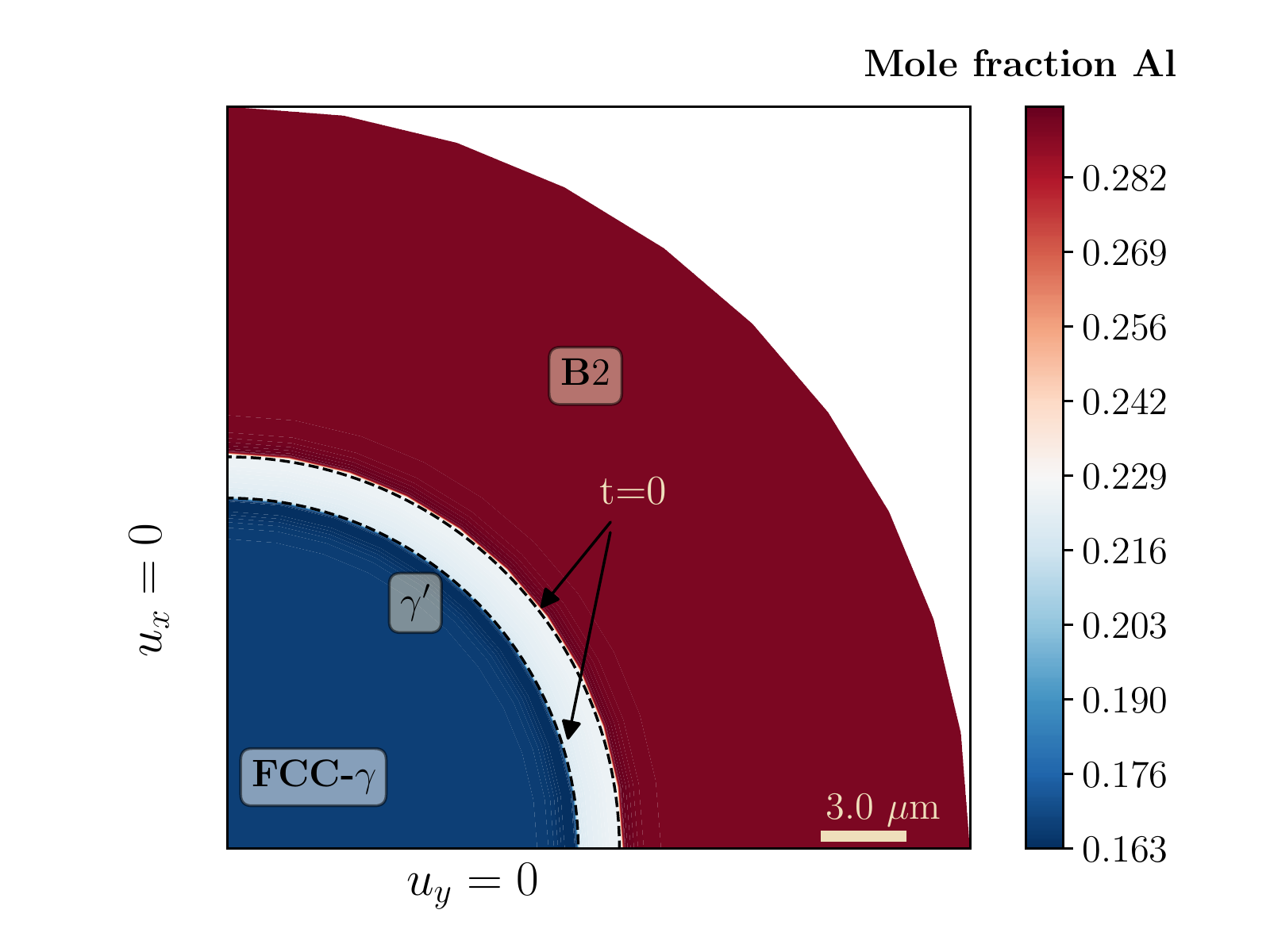}
\caption{}
\label{Fig.7a}
\end{subfigure}
\begin{subfigure}{0.5\textwidth}
\includegraphics[trim=60 15 10 12, clip, keepaspectratio,width=\linewidth]{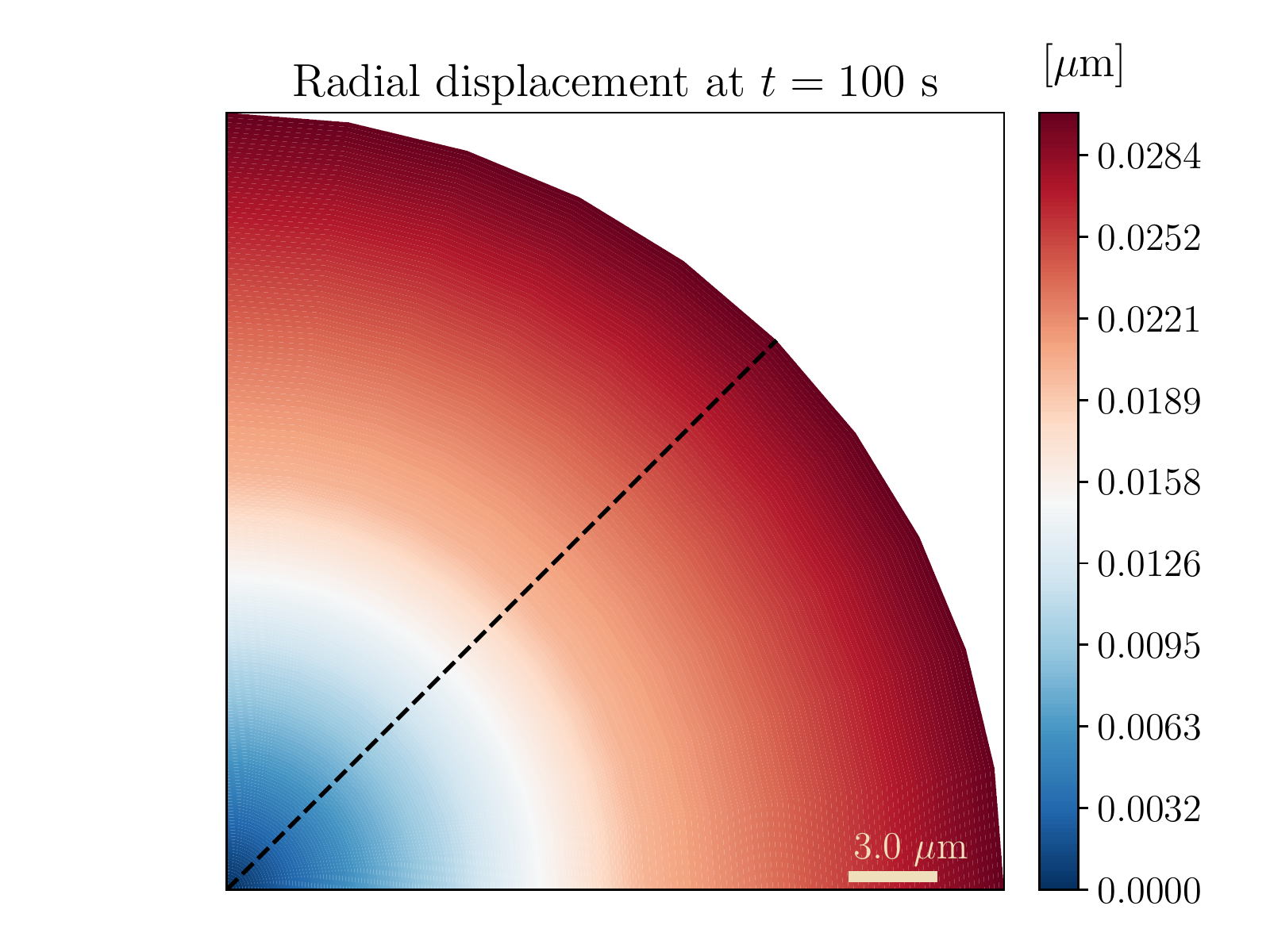}
\caption{}
\label{Fig.7b}
\end{subfigure}
\begin{subfigure}{0.50\textwidth}
\includegraphics[keepaspectratio,width=\linewidth]{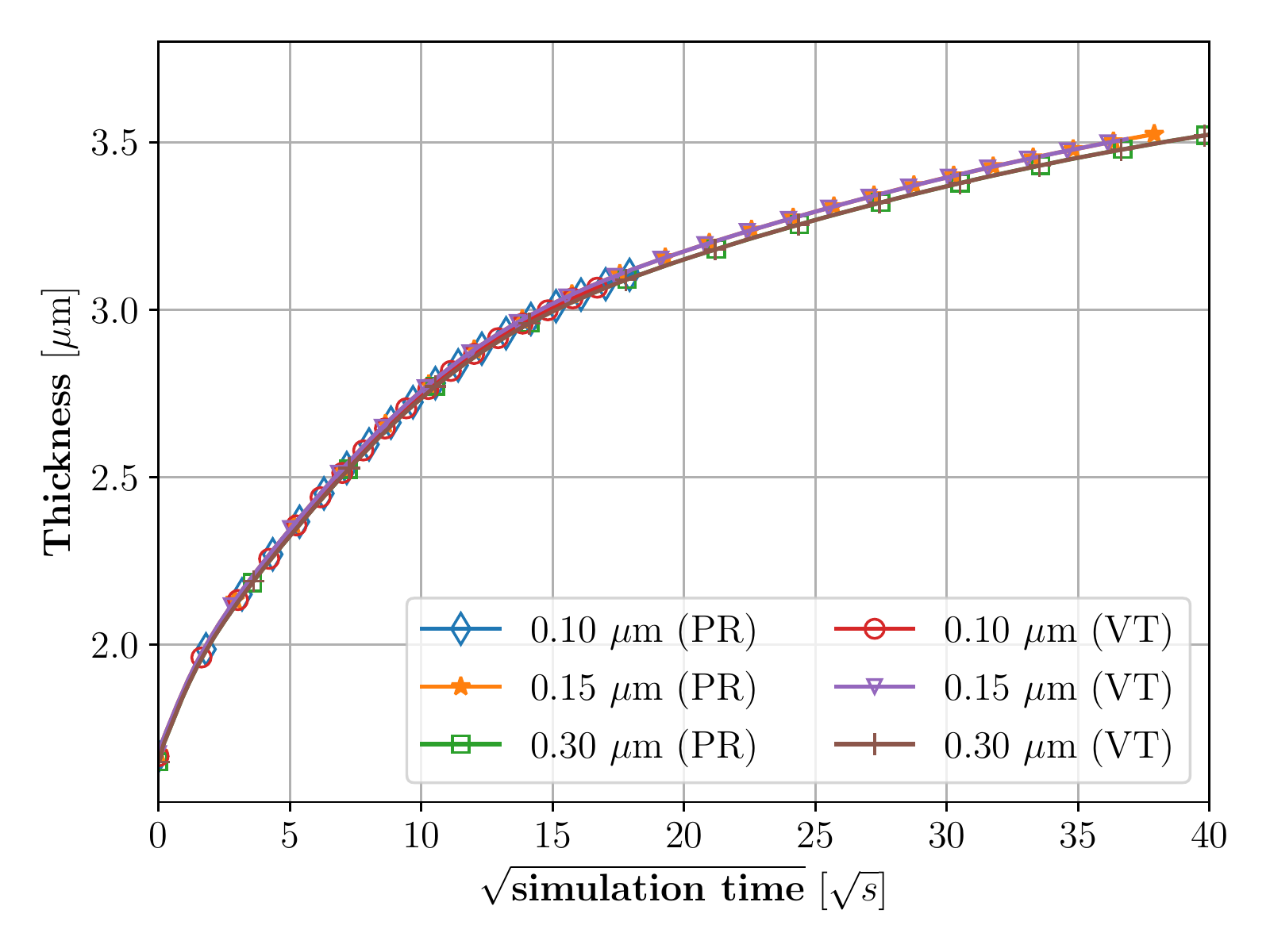}
\caption{}
\label{Fig.7c}
\end{subfigure}
\begin{subfigure}{0.50\textwidth}
\includegraphics[keepaspectratio,width=\linewidth]{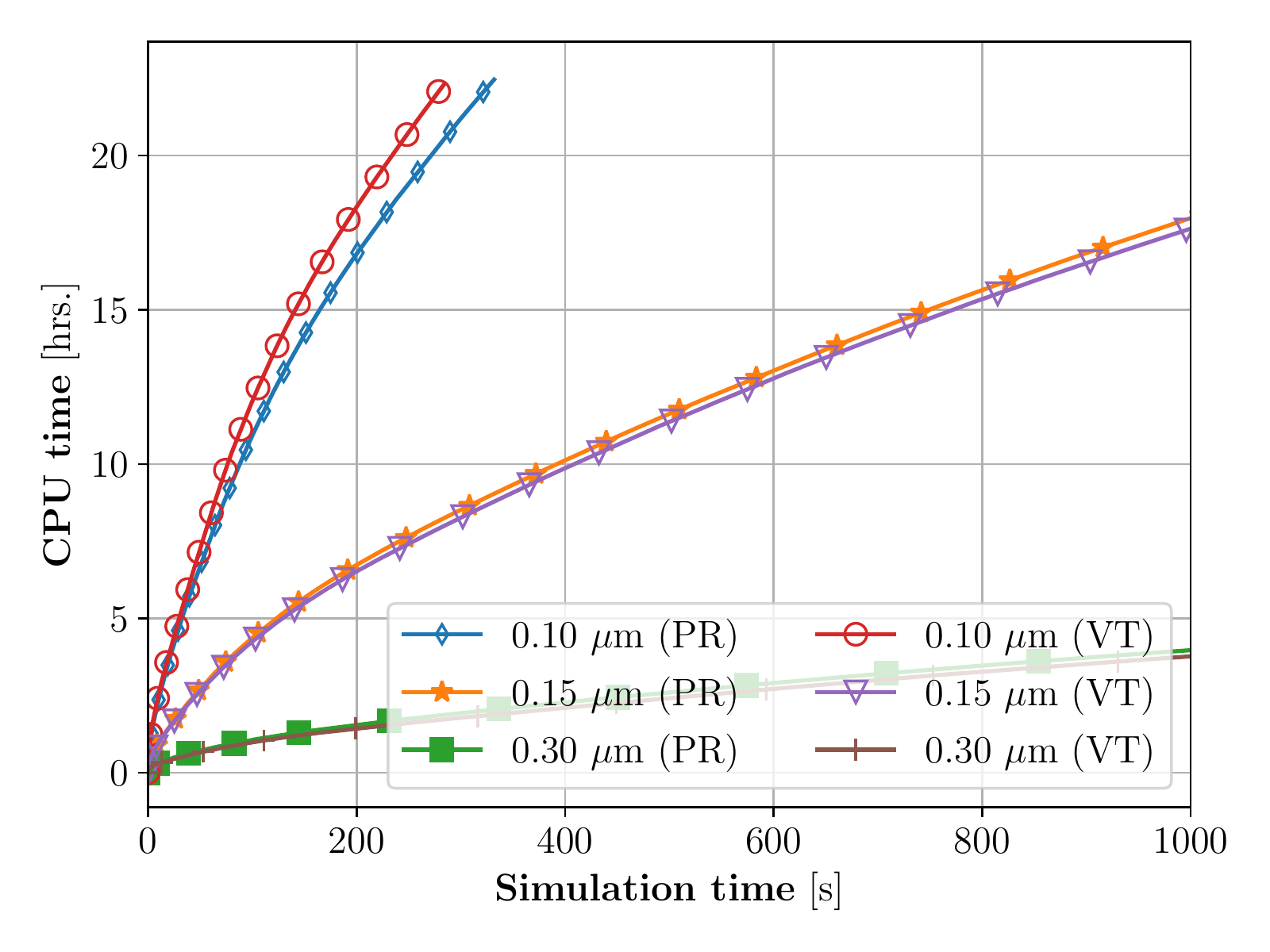}
\caption{}
\label{Fig.7d}
\end{subfigure}
\caption{For a non-planar Al-Cr-Ni fcc-$\gamma/\gamma^{\prime}/\mathrm{B2}$ coherently stressed diffusion couple: simulation domain, mechanical boundary conditions and the simulated Al-mole fraction field at time $t=100$ s (a); the simulated radial displacement field at the same time (b). For three different interface widths, variation in $\gamma^{\prime}$ thickness as a function of the square root of simulation time using the partial rank-one (PR) and Voigt-Taylor (VT) homogenization schemes (c); the CPU time as a function of simulation time for both these schemes (d).}
\label{Fig.7}
\end{figure}

Lastly, we simulated a ternary Al-Cr-Ni $\mathrm{fcc-}\gamma/\gamma^{\prime}/\mathrm{B2}$ alloy having concentric  interfaces (Fig. \ref{Fig.7a}). As listed in Table \ref{table:bc}, compared to the previous case, the mechanical displacements at the outer boundary are different in this case. Specifically,
\begin{align}
u_{x}(x,y,t) = \epsilon_{R}^{g}x \quad \text{on}\quad x^{2} + y^2 = R_{0}^{2} \label{ResEq7},\\
u_{y}(x,y,t) = \epsilon_{R}^{g}y\quad \text{on} \quad x^{2} + y^2 = R_{0}^{2} \label{ResEq8},
\end{align}
where $\epsilon_{R}^{g}=0.1\%$ is the imposed hoop strain. For sake of completeness, we note that the boundary conditions at the remaining boundaries are identical to the previous case. Moreover, we have assumed that the eigenstrains in the bulk phases are zero. Because of this, VT and Khachaturayan schemes become identical for this special case \cite{SC2}. Since there are no eigenstrains, the mechanical stresses are simply due to the imposed boundary conditions. 

As noted previously for the planar case, the three phases may coexist since the overall alloy composition lies in the three-phase region. However, the initial conditions are set such that the $\gamma^{\prime}$ grows at the expense of the ternary $\gamma$ and B$2$ phases. Fig. \ref{Fig.7a} shows the spatial variation in the Al-mole fraction field at time $t=100$ s. As shown in Fig. \ref{Fig.7b}, the radial displacement field is symmetric due to the imposed boundary conditions and isotropic elastic properties.

Fig. \ref{Fig.7c} shows the variation in the thickness of the $\gamma^{\prime}$ phase as a function of the square root of simulation time. Unlike the previous case, we find that the $\gamma^{\prime}$ phase first grows parabolically as a function of time. Eventually, its growth slows down as the system reaches towards the equilibrium state. This parabolic growth behaviour of the $\gamma^{\prime}$ phase is due to the fact that the process is diffusion-controlled. Moreover, we find that the accuracy of the temporal variation in the $\gamma^{\prime}$ phase thickness is independent of the interface width choice and the homogenization scheme (Fig. \ref{Fig.7c}). However, in contrast to the previous three simulations, we find that the convergence of the VT scheme is marginally faster in this case compared to the PR scheme for two interface width values of $0.15$ $\mu$m and $0.30$ $\mu$m  (Fig. \ref{Fig.7d}). We think this is possibly due to the absence of eigenstrains in this simulation compared to all other previous cases. Nevertheless, we find that the PR scheme converges faster compared to the VT scheme only for the case with an interface width of $0.10$ $\mu$m. 

\begin{figure}[!ht]
\begin{subfigure}{0.5\textwidth}
\includegraphics[keepaspectratio,width=\linewidth]{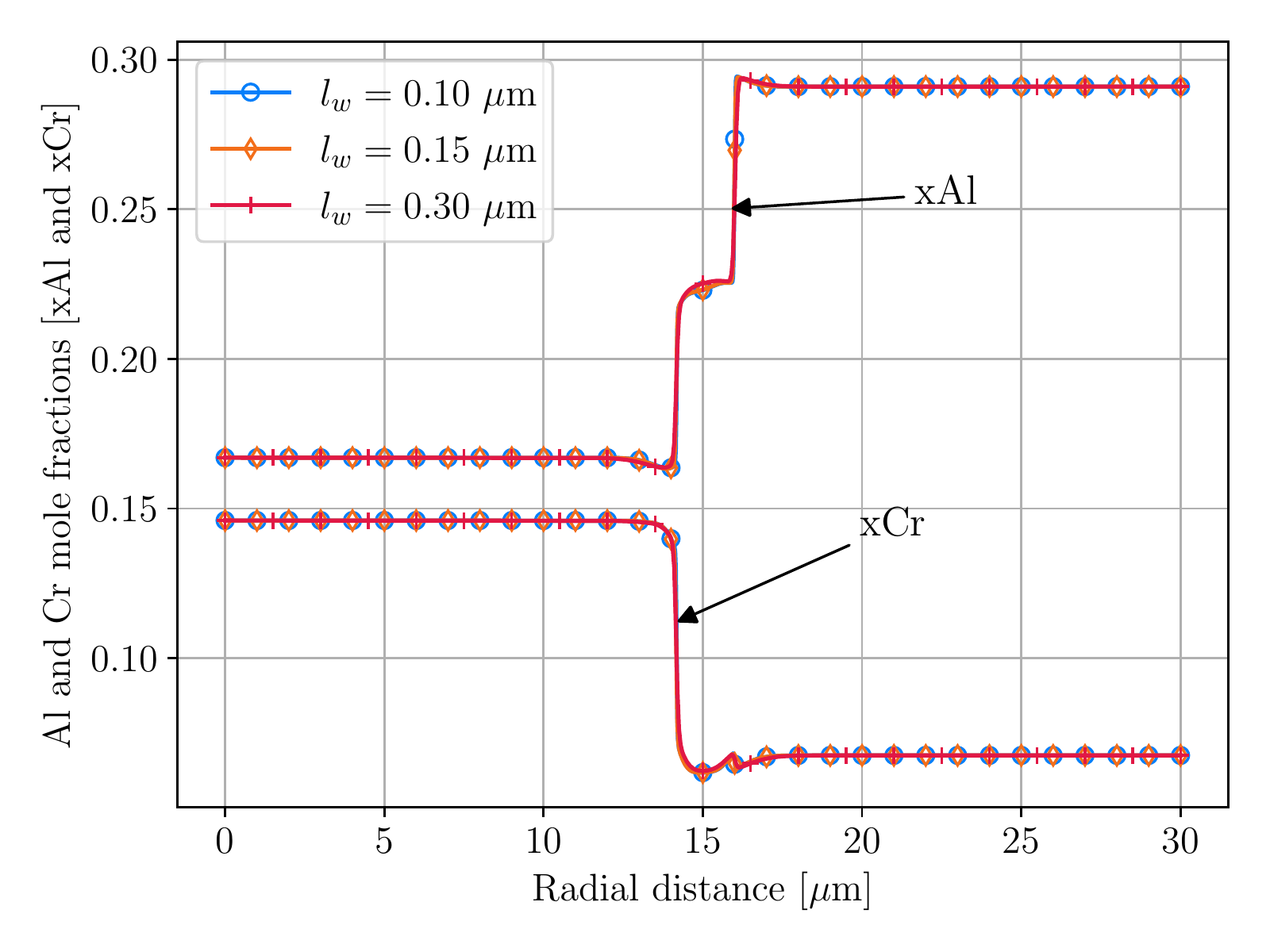}
\caption{}
\label{Fig.8a}
\end{subfigure}
\begin{subfigure}{0.5\textwidth}
\includegraphics[keepaspectratio,width=\linewidth]{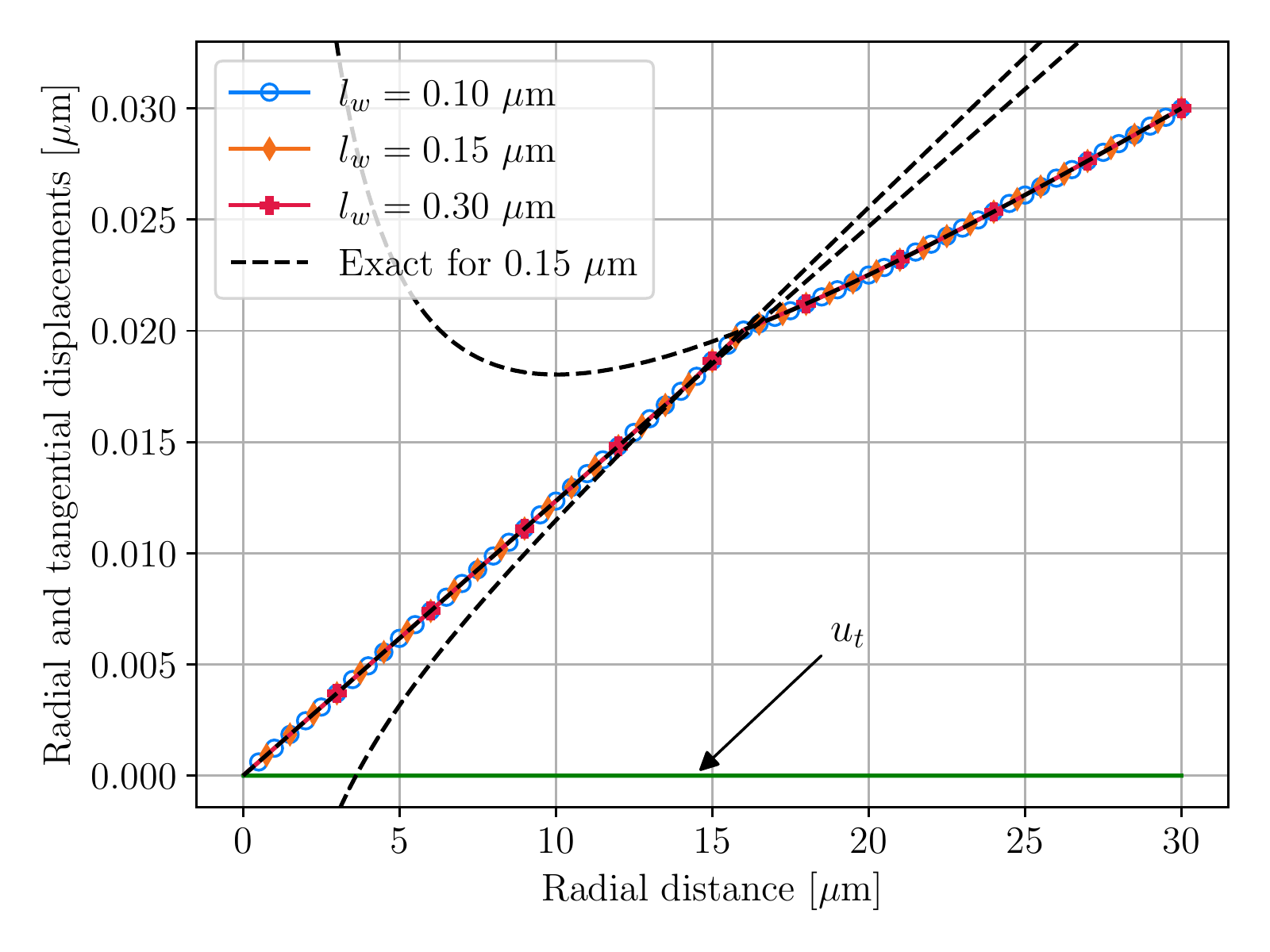}
\caption{}
\label{Fig.8b}
\end{subfigure}
\begin{subfigure}{0.5\textwidth}
\includegraphics[keepaspectratio,width=\linewidth]{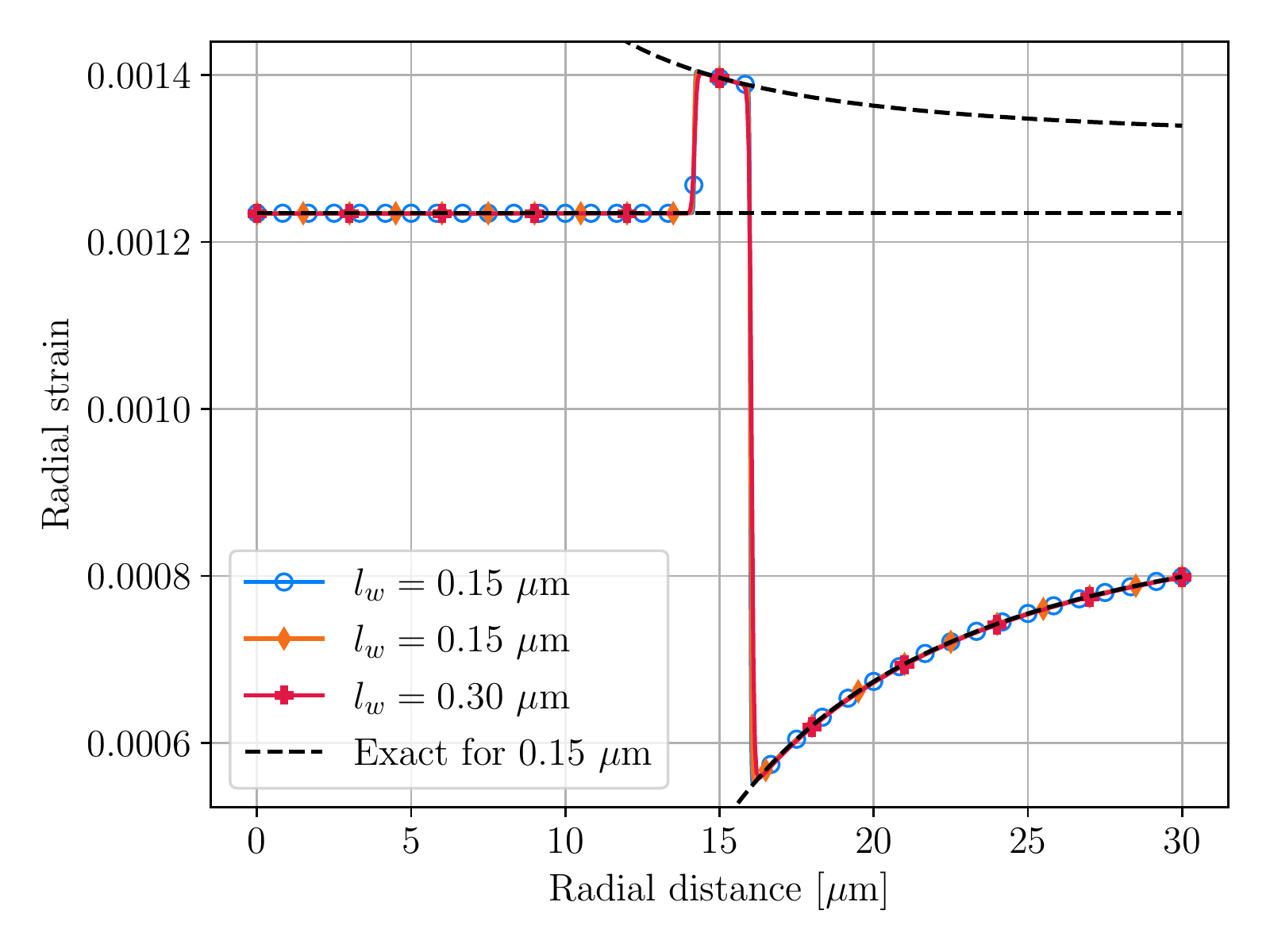}
\caption{}
\label{Fig.8c}
\end{subfigure}
\begin{subfigure}{0.5\textwidth}
\includegraphics[keepaspectratio,width=\linewidth]{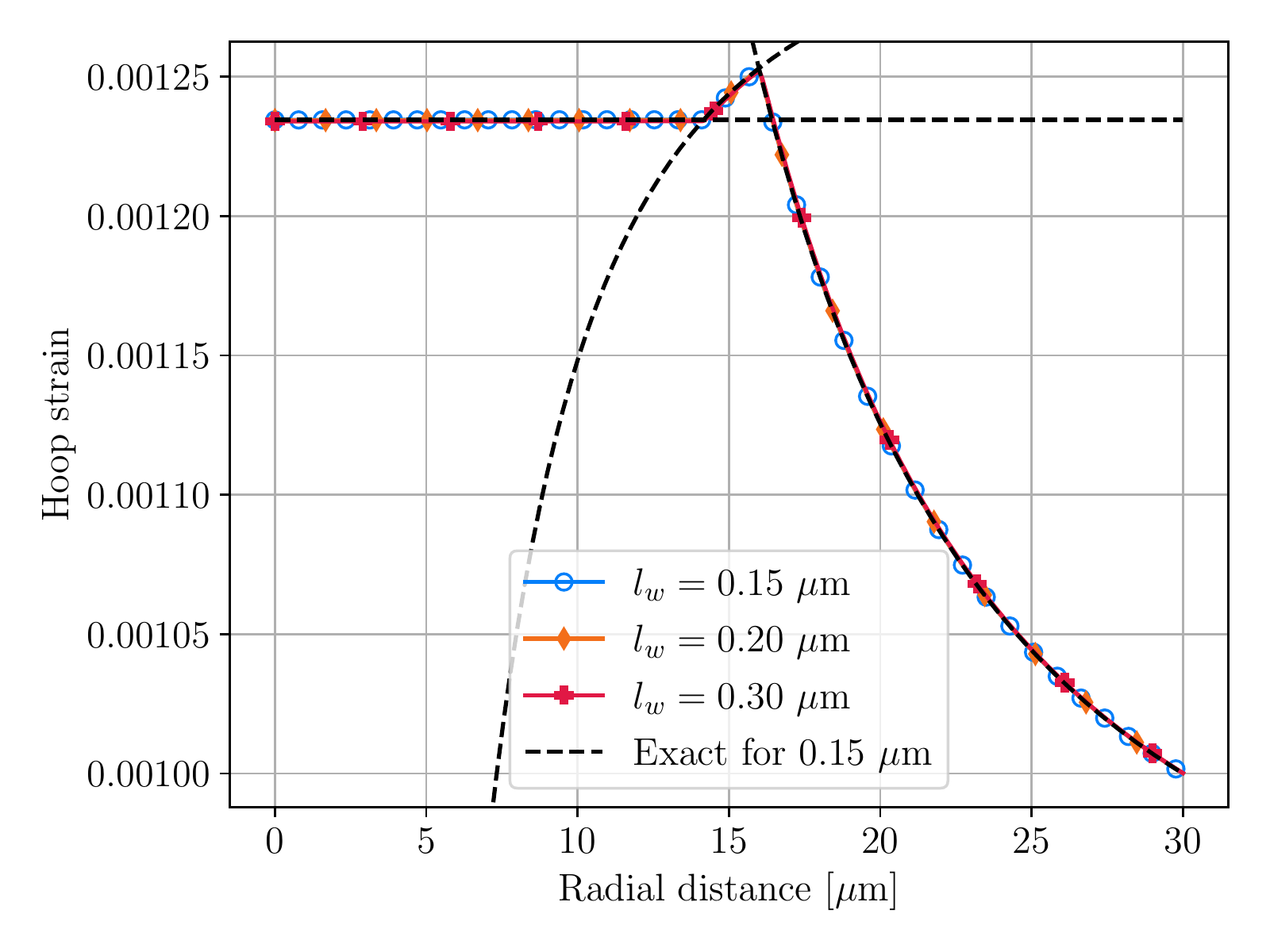}
\caption{}
\label{Fig.8d}
\end{subfigure}
\begin{subfigure}{0.5\textwidth}
\includegraphics[keepaspectratio,width=\linewidth]{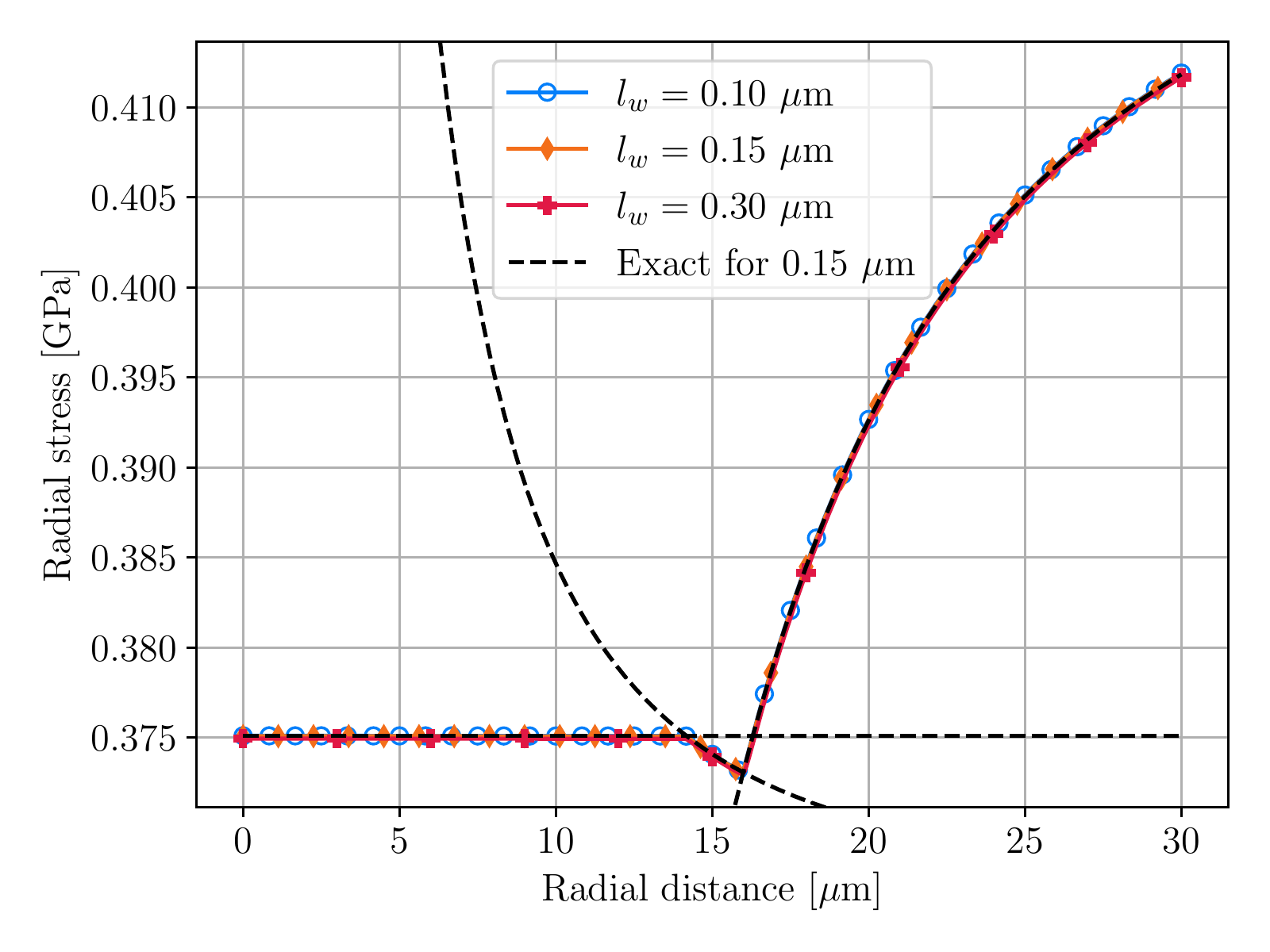}
\caption{}
\label{Fig.8e}
\end{subfigure}
\begin{subfigure}{0.5\textwidth}
\includegraphics[keepaspectratio,width=\linewidth]{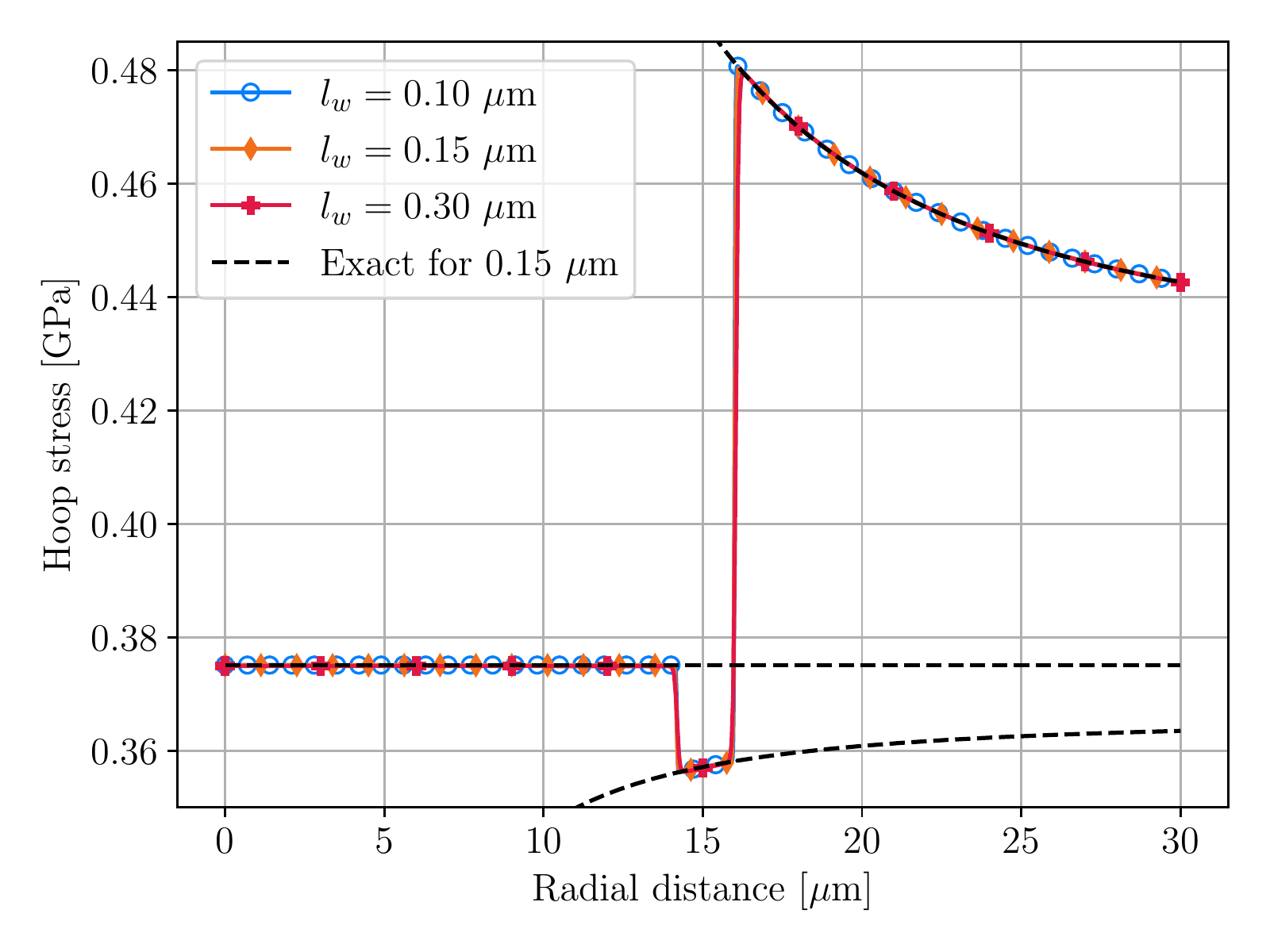}
\caption{}
\label{Fig.8f}
\end{subfigure}
\caption{For a non-planar Al-Cr-Ni fcc-$\gamma/\gamma^{\prime}/\mathrm{B2}$ diffusion couple, the spatial variation in a) Al and Cr mole fraction fields; b) radial and tangential displacements; c) radial strain; d) hoop strain; e) radial stress;  and f) hoop stress as functions of radial distance at time $t=100$ s. The plots show this variation for three different interface widths $l_{w}$ using the partial rank-one scheme. The dotted and discontinuous black lines are the analytically obtained solutions.}
\label{Fig.6}
\end{figure}

To test the accuracy of our simulations, we calculate the spatial variation of elastic and composition fields along the radial direction based on the PR scheme (Fig. \ref{Fig.7b}). In the bulk phases, we find that the spatial variation in the Al and Cr mole fraction fields along the radius is independent of the choice of interface width (Fig. \ref{Fig.8a}). Moreover, our simulated radial displacement field is consistent with the analytically obtained solution (Fig. \ref{Fig.8b}).  We also find that the accuracy remains unaltered for three different interface widths (Fig. \ref{Fig.8b}). As shown in Fig. \ref{Fig.8b}, the tangential displacement is negligible for this case. Figs. \ref{Fig.8c} and \ref{Fig.8d} show the variation in the total radial and hoop strains as functions of radial distance. Similar to the previous case, notice that the radial and hoop strains in the $\gamma$ phase are constant and equal. However, the radial and hoop strains in the $\gamma^{\prime}$ and B$2$ phases depend on the radius. Moreover, our simulated radial and hoop stresses in the bulk phases are also consistent with the analytical solution (Figs. \ref{Fig.8e} and \ref{Fig.8f}). Finally, we emphasize that the simulated stress and strain fields are independent of the choice of interface width.

\section{Conclusions}
This paper first generalizes the partial rank-one homogenization scheme to multi-phase systems. Subsequently, it implements this scheme for a three-phase system by analytically solving the static compatibility equations, thereby ensuring both static and kinematic compatibilities in the interfacial regions. Following this, a multi-phase-field grand-potential-based model is formulated using the rank-one scheme for solids undergoing small-strain deformations. To demonstrate its application for real alloys, a coupling technique is utilized to extract the prerequisite properties directly from CALPHAD databases.

Specifically, we test the model for two three-phase Ni-based alloys having either planar or concentric ring interfaces. We verify the accuracy of the simulated elastic fields against analytical solutions for all simulated cases. Our results show that the simulation accuracy using the rank-one scheme remains independent of the choice of interface width. Except for one case, we find that the rank-one scheme shows improved or nearly equal convergence compared to the Voigt-Taylor homogenization scheme, which ensures only kinematic compatibility. Nevertheless, the current implementation is still limited to linear elastic deformation, and in the future numerical approaches to solving the static compatibility equations, as demonstrated in \cite{Schneider2018}, will be explored.

\section{CRediT authorship contribution statement}
\textbf{Sourav Chatterjee:} Conceptualization, Methodology, Software, Validation, Writing - Original Draft, Writing - Review \& Editing, Visualization.
\textbf{Daniel Schwen:} Software, Writing - Review \& Editing.
\textbf{Nele Moelans:} Conceptualization, Methodology, Resources, Writing - Review \& Editing, Supervision, Funding acquisition.

\section{Acknowledgement}
This work was supported by the European Research Council (ERC) under the European Union’s Horizon 2020 research and innovation program (INTERDIFFUSION, grant agreement no. $714754$). The computational resources and services used in this work were provided by the VSC (Flemish Supercomputer Center), funded by the Research Foundation - Flanders (FWO) and the Flemish Government - department EWI. 
\clearpage

\renewcommand{\theequation}{A.\arabic{equation}}
\renewcommand{\thesubsection}{A.\arabic{subsection}}
\setcounter{subsection}{0}
\setcounter{equation}{0}
\section*{Appendix A}
\label{Appendix_A}
\section*{Calculation of phase strains}
In this section, we provide an analytical approach to calculate the \textit{phase} strains for a $p$-phase system as functions of the total strain $\boldsymbol{\epsilon}(\boldsymbol{u})$, interpolations functions $h(\boldsymbol{\phi})$ and strain jumps. To this end, we first begin by deriving the \textit{phase} strains for two-phase and three-phase systems and then extend it to multi-phase systems. 
\subsection{Phase strains for a two-phase system}
For a two-phase $\alpha/\beta$ system, Eqs. (\ref{Eqn1}) and (\ref{Eqn3}) reduces to
\begin{align}
\boldsymbol{\epsilon}(\boldsymbol{u}) =  \boldsymbol{\epsilon}^{\alpha}h_{\alpha} + \boldsymbol{\epsilon}^{\beta}h_{\beta}\label{EqnE1}\\
\boldsymbol{\epsilon}^{\alpha} - \boldsymbol{\epsilon}^{\beta} = \llbracket \epsilon\rrbracket^{\alpha\beta}
\label{EqnE2}
\end{align}
Multiplying Eq. (\ref{EqnE2}) with $h_{\beta}$ and then adding Eq. (\ref{EqnE1}) yields
\begin{align}
\boldsymbol{\epsilon}^{\alpha}\left(h_{\alpha} + h_{\beta}\right) = \boldsymbol{\epsilon} +  h_{\beta}\llbracket\boldsymbol{\epsilon}\rrbracket^{\alpha\beta}
\label{EqnE3}
\end{align}
Since for a two-phase system $h_{\alpha} + h_{\beta}=1$, Eq. (\ref{EqnE3}) simplifies to
\begin{align}
\boldsymbol{\epsilon}^{\alpha} = \boldsymbol{\epsilon} +  h_{\beta}\llbracket\boldsymbol{\epsilon}\rrbracket^{\alpha\beta}
\label{EqnE4}
\end{align}
Since $h_{\alpha} = 1-h_{\beta}$, it follows from Eqs. (\ref{EqnE2}) and (\ref{EqnE4}) that
\begin{align}
\boldsymbol{\epsilon}^{\beta} = \boldsymbol{\epsilon} -  h_{\alpha}\llbracket\boldsymbol{\epsilon}\rrbracket^{\alpha\beta}
\label{EqnE5}
\end{align}
We simply note that Eqs. (\ref{EqnE4}) and (\ref{EqnE5}) are completely equivalent to Eqs. (\ref{EqnE1}) and (\ref{EqnE2}). Next, we attempt to use the two-phase relations to extend the model to three-phase systems.

\subsection{Phase strains for a three-phase system}
For a system consisting of three phases, say $\alpha$, $\beta$ \& $\gamma$, Eqs. (\ref{Eqn1}) and (\ref{Eqn3}) reduces to
\begin{align}
\boldsymbol{\epsilon} =  \boldsymbol{\epsilon}^{\alpha}h_{\alpha} + \boldsymbol{\epsilon}^{\beta}h_{\beta} + \boldsymbol{\epsilon}^{\gamma}h_{\gamma}, \label{EqnE6}\\
\boldsymbol{\epsilon}^{\alpha} - \boldsymbol{\epsilon}^{\beta} = \llbracket \epsilon\rrbracket^{\alpha\beta},\label{EqnE7}\\
\boldsymbol{\epsilon}^{\beta} - \boldsymbol{\epsilon}^{\gamma} = \llbracket \epsilon\rrbracket^{\beta\gamma}.
\label{EqnE8}
\end{align}
By defining $\boldsymbol{\epsilon}^{\prime} = \boldsymbol{\epsilon} - \boldsymbol{\epsilon}^{\gamma}h_{\gamma}$, Eq. (\ref{EqnE6}) may be written as
\begin{align}
\boldsymbol{\epsilon}^{\prime} = \boldsymbol{\epsilon}^{\alpha}h_{\alpha} + \boldsymbol{\epsilon}^{\beta}h_{\beta}.
\label{EqnE9}
\end{align}
Notice that Eqs. (\ref{EqnE9}) and (\ref{EqnE7}) are similar to Eqs. (\ref{EqnE1}) and (\ref{EqnE2}). Because of this similarity, we can directly use Eq. (\ref{EqnE3}) to write 
\begin{align}
\boldsymbol{\epsilon}^{\alpha}\left(h_{\alpha} + h_{\beta}\right) = \boldsymbol{\epsilon}^{\prime} +  h_{\beta}\llbracket\boldsymbol{\epsilon}\rrbracket^{\alpha\beta} = \boldsymbol{\epsilon} - \boldsymbol{\epsilon}^{\gamma}h_{\gamma} + h_{\beta}\llbracket\boldsymbol{\epsilon}\rrbracket^{\alpha\beta}.
\label{EqnE10}
\end{align}
It should be noticed from the right-hand side of Eq. (\ref{EqnE10})  that $\left(h_{\alpha} + h_{\beta}\right) \neq 1$ since this is a three-phase system. Consequently, adding Eqs. (\ref{EqnE7}) \& (\ref{EqnE8}) and substituting: $\boldsymbol{\epsilon}^{\gamma} = \boldsymbol{\epsilon}^{\alpha} - \llbracket \epsilon\rrbracket^{\alpha\beta} - \llbracket \epsilon\rrbracket^{\beta\gamma}$, in Eq. (\ref{EqnE10}) gives
\begin{align}
\boldsymbol{\epsilon}^{\alpha}(h_{\alpha} + h_\beta + h_{\gamma}) = \boldsymbol{\epsilon} + h_{\beta} \llbracket\boldsymbol{\epsilon}\rrbracket^{\alpha\beta} + h_{\gamma} \left(\llbracket\boldsymbol{\epsilon}\rrbracket^{\alpha\beta} + \llbracket \boldsymbol{\epsilon}\rrbracket^{\beta\gamma}\right).
\label{EqnE11}
\end{align}
Now, since $(h_{\alpha} + h_\beta + h_{\gamma}) = 1$ for a three-phase system, Eq. (\ref{EqnE11}) reduces to
\begin{align}
\boldsymbol{\epsilon}^{\alpha}= \boldsymbol{\epsilon} + h_{\beta} \llbracket\boldsymbol{\epsilon}\rrbracket^{\alpha\beta} + h_{\gamma} \left(\llbracket\boldsymbol{\epsilon}\rrbracket^{\alpha\beta} + \llbracket \boldsymbol{\epsilon}\rrbracket^{\beta\gamma}\right).
\label{EqnE12}
\end{align}
Next, substituting Eq. (\ref{EqnE12}) in Eq. (\ref{EqnE7}) and then using $(1 - h_{\beta}- h_{\gamma}) = h_{\alpha}$ gives 
\begin{align}
\boldsymbol{\epsilon}^{\beta} = \boldsymbol{\epsilon} - h_{\alpha} \llbracket\boldsymbol{\epsilon}\rrbracket^{\alpha\beta} + h_{\gamma}\llbracket \boldsymbol{\epsilon}\rrbracket^{\beta\gamma}.
\label{EqnE13}
\end{align}
Finally, substituting Eq. (\ref{EqnE13}) in Eq. (\ref{EqnE8}) yields
\begin{align}
\boldsymbol{\epsilon}^{\gamma} = \boldsymbol{\epsilon} - h_{\alpha} \llbracket\boldsymbol{\epsilon}\rrbracket^{\alpha\beta} -  \left(h_{\alpha} + h_{\beta}\right)\llbracket\boldsymbol{\epsilon}\rrbracket^{\beta\gamma}.
\label{EqnE14}
\end{align}
Thus, we have obtained the \textit{phase} strains as functions of the total strain, interpolation functions and strain jumps for a three-phase system. We again note that Eqs. (\ref{EqnE12}), (\ref{EqnE13}) \& (\ref{EqnE14}) are completely equivalent to Eqs. (\ref{EqnE6}), (\ref{EqnE7}) \& (\ref{EqnE8}).

\subsection{Phase strains for a multi-phase system}
Based on the previous two derivations, it is worth noting that once the phase strain pertaining to a particular phase, say $\alpha$, is determined, the phase strains of the remaining $(p-1)$ phases may be obtained  using the $(p-1)$ compatibility equations, i.e., Eqs. (\ref{Eqn3}). For sake of concretness, if phase strain pertaining to $\alpha$-phase is known, then the phase strains in $\beta, \gamma, \hdots, (p-1), p$ phases are
\begin{align}
\begin{split}
\boldsymbol{\epsilon}^{\beta} &= \boldsymbol{\epsilon}^{\alpha} - \llbracket \boldsymbol{\epsilon}\rrbracket^{\alpha\beta},\\
\boldsymbol{\epsilon}^{\gamma} &= \boldsymbol{\epsilon}^{\beta} - \llbracket \boldsymbol{\epsilon}\rrbracket^{\beta\gamma},\\
\boldsymbol{\epsilon}^{\delta} &= \boldsymbol{\epsilon}^{\gamma} - \llbracket \boldsymbol{\epsilon}\rrbracket^{\gamma\delta},\\
&\vdots\\
\boldsymbol{\epsilon}^{p} &= \boldsymbol{\epsilon}^{p-1} - \llbracket \boldsymbol{\epsilon}\rrbracket^{(p-1),p}.
\end{split}
\label{EqnE15}
\end{align}
Therefore, if an analytical expression for the $\alpha$-phase strain in a system consisting of $p$ phases is known, \textit{all} remaining phase strains can be calculated.

It can be observed from Eqs. (\ref{EqnE4}) \& (\ref{EqnE12}) that the $\alpha$-phase strain for a three-phase system differs from a two-phase system by just one term. Specifically, this term is equal to the product of the interpolation function associated with the new phase and the sum of all jump vectors in that system, i.e., $h_{\gamma} \left(\llbracket\boldsymbol{\epsilon}\rrbracket^{\alpha\beta} + \llbracket \boldsymbol{\epsilon}\rrbracket^{\beta\gamma}\right)$. Consequently, $\boldsymbol{\epsilon}^{\alpha}$ for a multi-phase system may be written as
\begin{align}
\boldsymbol{\epsilon}^{\alpha} = \boldsymbol{\epsilon}(\boldsymbol{u}) + h_{\beta} \llbracket\boldsymbol{\epsilon}\rrbracket^{\alpha\beta} + h_{\gamma} \left(\llbracket\boldsymbol{\epsilon}\rrbracket^{\alpha\beta} + \llbracket \boldsymbol{\epsilon}\rrbracket^{\beta\gamma}\right) + \hdots + h_{p}\left(\sum_{i=\alpha\beta}^{(p-1),p}\llbracket\boldsymbol{\epsilon}\rrbracket^{i}\right),
\label{EqnE16}
\end{align}
where
\begin{align}
\sum_{i=\alpha\beta}^{(p-1),p}\llbracket\boldsymbol{\epsilon}\rrbracket^{i} = \llbracket\boldsymbol{\epsilon}\rrbracket^{\alpha\beta} + \llbracket \boldsymbol{\epsilon}\rrbracket^{\beta\gamma} +  \llbracket\boldsymbol{\epsilon}\rrbracket^{\gamma\delta} +  \hdots + \llbracket\boldsymbol{\epsilon}\rrbracket^{(p-1),p}.
\end{align}
By substituting Eq. (\ref{EqnE16}) in the first of the set of Eqs. (\ref{EqnE15}) and using the relation $ 1-\left(h_{\beta} + h_{\gamma} + \hdots +h_{p}\right)= h_{\alpha}$, it follows that 
\begin{align}
\boldsymbol{\epsilon}^{\beta} = \boldsymbol{\epsilon}(\boldsymbol{u}) - h_{\alpha} \llbracket\boldsymbol{\epsilon}\rrbracket^{\alpha\beta} + h_{\gamma}\llbracket \boldsymbol{\epsilon}\rrbracket^{\beta\gamma} + \hdots + h_{p}\left(\llbracket\boldsymbol{\epsilon}\rrbracket^{\beta\gamma} + \llbracket\boldsymbol{\epsilon}\rrbracket^{\gamma\delta} + \hdots + \llbracket\boldsymbol{\epsilon}\rrbracket^{(p-1),p}\right).
\label{EqnE18}
\end{align}
Similarly, by substituting Eq. (\ref{EqnE18}) in the second of the set of Eqs. (\ref{EqnE15}) and using $1- (h_{\gamma} + h_{\delta} + \hdots + h_{p}) = \left(h_{\alpha} + h_{\beta}\right)$, it follows that
\begin{align}
\boldsymbol{\epsilon}^{\gamma} = \boldsymbol{\epsilon}(\boldsymbol{u}) - h_{\alpha} \llbracket\boldsymbol{\epsilon}\rrbracket^{\alpha\beta} - \left(h_{\alpha} + h_{\beta}\right)\llbracket \boldsymbol{\epsilon}\rrbracket^{\beta\gamma} + \hdots + h_{p}\left(\llbracket\boldsymbol{\epsilon}\rrbracket^{\gamma\delta} + \hdots + \llbracket\boldsymbol{\epsilon}\rrbracket^{(p-1),p}\right).
\label{EqnE19}
\end{align}
Thus, by using Eqs. (\ref{EqnE15}) and (\ref{EqnE16}) we can calculate the phase strains for an arbitrary multi-phase system.

\renewcommand{\theequation}{B.\arabic{equation}}
\renewcommand{\thesubsection}{B.\arabic{subsection}}
\setcounter{equation}{0}
\setcounter{subsection}{0}
\section*{Appendix B}
\label{Appendix_B}
\section*{Some useful relations}
\subsection{Derivatives with respect to phase-field variables}
Since $h_{\alpha} + h_{\beta} + h_{\gamma} = 1$, it follows that 
\begin{align}
\frac{\partial h_{\alpha}}{\partial \phi_{\alpha}} = -\left(\frac{\partial h_{\beta}}{\partial \phi_{\alpha}} + \frac{\partial h_{\gamma}}{\partial \phi_{\alpha}}\right)\label{EqnA01}\\
\frac{\partial h_{\gamma}}{\partial \phi_{\alpha}} = -\left(\frac{\partial h_{\beta}}{\partial \phi_{\alpha}} + \frac{\partial h_{\alpha}}{\partial \phi_{\alpha}}\right)\label{EqnA02}
\end{align}
Differentiating Eqs. (\ref{Eqn4}), (\ref{Eqn5}) and (\ref{Eqn6}) with respect to $\phi_{\alpha}$ yields
\begin{align}
\frac{\partial \epsilon_{ij}^{\alpha}}{\partial \phi_{\alpha}} &= \left(\frac{\partial h_{\beta}}{\partial \phi_{\alpha}} + \frac{\partial h_{\gamma}}{\partial \partial \phi_{\alpha}}\right)\llbracket\epsilon_{ij}\rrbracket^{\alpha\beta} + (h_{\beta} + h_{\gamma})\frac{\partial \llbracket \epsilon_{ij}\rrbracket^{\alpha\beta}}{\partial \phi_{\alpha}} + \frac{\partial h_{\gamma}}{\partial \phi_{\alpha}}\llbracket \epsilon_{ij}\rrbracket^{\beta\gamma} + h_{\gamma}\frac{\partial \llbracket \epsilon_{ij} \rrbracket^{\beta\gamma}}{\partial \phi_{\alpha}}\label{EqnA1}\\
\frac{\partial \epsilon_{ij}^{\beta}}{\partial \phi_{\alpha}} &= - \frac{\partial h_{\alpha}}{\partial \phi_{\alpha}}\llbracket \epsilon_{ij}\rrbracket^{\alpha\beta} - h_{\alpha}\frac{\partial \llbracket \epsilon_{ij}\rrbracket^{\alpha\beta}}{\partial \phi_{\alpha}} + \frac{\partial h_{\gamma}}{\partial \phi_{\alpha}}\llbracket \epsilon_{ij}\rrbracket^{\beta}_{\gamma} + h_{\gamma}\frac{\partial \llbracket \epsilon_{ij}\rrbracket^{\beta\gamma}}{\partial \phi_{\alpha}}\label{EqnA2}\\
\frac{\partial \epsilon_{ij}^{\gamma}}{\partial \phi_{\alpha}} &= - \frac{\partial h_{\alpha}}{\partial \phi_{\alpha}}\llbracket \epsilon_{ij} \rrbracket^{\alpha\beta} - h_{\alpha}\frac{\partial \llbracket \epsilon_{ij} \rrbracket^{\alpha\beta}}{\partial \phi_{\alpha}} - \left(\frac{\partial h_{\beta}}{\partial \phi_{\alpha}} + \frac{\partial h_{\alpha}}{\partial \phi_{\alpha}}\right)\llbracket\epsilon_{ij} \rrbracket^{\beta\gamma} - (h_{\beta} + h_{\alpha})\frac{\partial \llbracket \epsilon_{ij}\rrbracket^{\beta\gamma}}{\partial \phi_{\alpha}}
\label{EqnA3}
\end{align}
Multiplying Eqs. (\ref{EqnA1}), (\ref{EqnA2}) and (\ref{EqnA3}) with $h_{\alpha}{\sigma}_{ij}^{\alpha}$, $h_{\beta}\sigma_{ij}^{\beta}$ and $h_{\gamma}\sigma_{ij}^{\gamma}$, respectively, and then setting the terms premultiplied by $h_{\gamma}h_{\alpha}$ to zero, and  using  Eqs. (\ref{EqnA01}) and (\ref{EqnA02}) yields
\begin{align}
h_{\alpha}\sigma_{ij}^{\alpha}\frac{\partial \epsilon_{ij}^{\alpha}}{\partial \phi_{\alpha}} &= - h_{\alpha}\frac{\partial h_{\alpha}}{\partial \phi_{\alpha}}\sigma_{ij}^{\alpha}\llbracket \epsilon_{ij}\rrbracket^{\alpha}_{\beta} + h_{\alpha}h_{\beta}\sigma_{ij}^{\alpha}\frac{\partial \llbracket\epsilon_{ij}\rrbracket^{\alpha}_{\beta}}{\partial \phi_{\alpha}} + h_{\alpha}\frac{\partial h_{\gamma}}{\partial \phi_{\alpha}}\sigma_{ij}^{\alpha}\llbracket \epsilon_{ij}\rrbracket^{\beta}_{\gamma}\label{EqnA4}\\
h_{\beta}\sigma_{ij}^{\beta}\frac{\partial \epsilon_{ij}^{\beta}}{\partial \partial \phi_{\alpha}} &= -h_{\beta}\frac{\partial h_{\alpha}}{\partial \phi_{\alpha}}\sigma_{ij}^{\beta}\llbracket\epsilon_{ij}\rrbracket^{\alpha}_{\beta} - h_{\alpha}h_{\beta}\sigma_{ij}^{\beta}\frac{\partial\llbracket\epsilon_{ij}\rrbracket^{\alpha}_{\beta}}{\partial \phi_{\alpha}} + h_{\beta}\frac{\partial h_{\gamma}}{\partial \phi_{\alpha}}\sigma_{ij}^{\beta}\llbracket \epsilon_{ij}\rrbracket^{\beta}_{\gamma} + h_{\beta}h_{\gamma}\sigma_{ij}^{\beta}\frac{\partial \llbracket \epsilon_{ij}\rrbracket^{\beta}_{\gamma}}{\partial \phi_{\alpha}}\label{EqnA5}\\
h_{\gamma}\sigma_{ij}^{\gamma}\frac{\partial \epsilon_{ij}^{\gamma}}{\partial \phi_{\alpha}} &= - h_{\gamma}\frac{\partial h_{\alpha}}{\partial \phi_{\alpha}}\sigma_{ij}^{\gamma}\llbracket\epsilon_{ij}\rrbracket^{\alpha}_{\beta} + h_{\gamma}\frac{\partial h_{\gamma}}{\partial \phi_{\alpha}}\sigma_{ij}^{\gamma}\llbracket\epsilon_{ij} \rrbracket^{\beta}_{\gamma} - h_{\gamma}h_{\beta}\sigma_{ij}^{\gamma}\frac{\partial \llbracket\epsilon_{ij}\rrbracket^{\beta}_{\gamma}}{\partial \phi_{\alpha}}\label{EqnA6}
\end{align}
Adding Eqs. (\ref{EqnA4}), (\ref{EqnA5}) and (\ref{EqnA6}) and using Eqs. (\ref{Eqn6a1}) and (\ref{Eqn6a2})  yields
\begin{align}
\begin{split}
\sum_{\theta=\alpha}^{\gamma}h_{\theta}(\boldsymbol{\phi})\sigma_{ij}^{\theta}\frac{\partial \epsilon_{ij}^{\theta}}{\partial \phi_{\alpha}} &= h_{\alpha}h_{\beta}\left\{\sigma_{ij}^{\alpha} - \sigma_{ij}^{\beta}\right\}n^{\alpha\beta}_{j}\frac{\partial a_{i}^{\alpha\beta}}{\partial \phi_{\alpha}} + h_{\beta}h_{\gamma}\left\{ \sigma_{ij}^{\beta} - \sigma_{ij}^{\gamma}\right\}n_{j}^{\beta\gamma}\frac{\partial a_{i}^{\beta\gamma}}{\partial \phi_{\alpha}}
\\
&-\frac{\partial h_{\alpha}}{\partial \phi_{\alpha}}\left\{\sum_{\theta=\alpha}h_{\theta}\sigma_{ij}^{\theta}\right\}\llbracket \epsilon_{ij}\rrbracket^{\alpha}_{\beta} + \frac{\partial h_{\gamma}}{\partial \phi_{\alpha}}\left\{\sum_{\theta=\alpha}^{\gamma}h_{\theta}\sigma_{ij}^{\theta}\right\}\llbracket \epsilon_{ij} \rrbracket^{\beta}_{\gamma}
\end{split}
\label{EqnA7}
\end{align}
It follows from Eqs. (\ref{Eqn10}) and (\ref{Eqn11}), that the first two terms on the right hand side of Eq. (\ref{EqnA7}) are zero. Thus, Eq. (\ref{EqnA7}) reduces to
\begin{align}
\sum_{\theta=\alpha}^{\gamma}h_{\theta}(\boldsymbol{\phi})\sigma_{ij}^{\theta}\frac{\partial \epsilon_{ij}^{\theta}}{\partial \phi_{\alpha}} = -\frac{\partial h_{\alpha}}{\partial \phi_{\alpha}}\left\{\sum_{\theta=\alpha}h_{\theta}\sigma_{ij}^{\theta}\right\}\llbracket \epsilon_{ij}\rrbracket^{\alpha}_{\beta} + \frac{\partial h_{\gamma}}{\partial \phi_{\alpha}}\left\{\sum_{\theta=\alpha}^{\gamma}h_{\theta}\sigma_{ij}^{\theta}\right\}\llbracket \epsilon_{ij} \rrbracket^{\beta}_{\gamma}
\label{EqnA8a}
\end{align}
Following a similar procedure, it can be shown that
\begin{align}
\sum_{\theta=\alpha}^{\gamma}h_{\theta}(\boldsymbol{\phi})\sigma_{ij}^{\theta}\frac{\partial \epsilon_{ij}^{\theta}}{\partial \phi_{\beta}} = -\frac{\partial h_{\alpha}}{\partial \phi_{\beta}}\left\{\sum_{\theta=\alpha}h_{\theta}\sigma_{ij}^{\theta}\right\}\llbracket \epsilon_{ij}\rrbracket^{\alpha}_{\beta} + \frac{\partial h_{\gamma}}{\partial \phi_{\beta}}\left\{\sum_{\theta=\alpha}^{\gamma}h_{\theta}\sigma_{ij}^{\theta}\right\}\llbracket \epsilon_{ij} \rrbracket^{\beta}_{\gamma}\label{EqnA8b}\\
\sum_{\theta=\alpha}^{\gamma}h_{\theta}(\boldsymbol{\phi})\sigma_{ij}^{\theta}\frac{\partial \epsilon_{ij}^{\theta}}{\partial \phi_{\gamma}} = -\frac{\partial h_{\alpha}}{\partial \phi_{\gamma}}\left\{\sum_{\theta=\alpha}h_{\theta}\sigma_{ij}^{\theta}\right\}\llbracket \epsilon_{ij}\rrbracket^{\alpha}_{\beta} + \frac{\partial h_{\gamma}}{\partial \phi_{\gamma}}\left\{\sum_{\theta=\alpha}^{\gamma}h_{\theta}\sigma_{ij}^{\theta}\right\}\llbracket \epsilon_{ij} \rrbracket^{\beta}_{\gamma}\label{EqnA8c}
\end{align}

Now, we obtain another similar relation by multiplying Eqs. (\ref{EqnA1}), (\ref{EqnA2}) and (\ref{EqnA3}) with $h_{\alpha}\mathcal{C}_{klij}^{\alpha}$, $h_{\beta}\mathcal{C}_{klij}^{\beta}$ and $h_{\gamma}\mathcal{C}_{klij}^{\gamma}$, respectively, and setting the terms premultiplied by $h_{\gamma}h_{\alpha}$ to zero yields 
\begin{align}
h_{\alpha}\mathcal{C}_{klij}^{\alpha}\frac{\partial \epsilon_{ij}^{\alpha}}{\partial \phi_{\alpha}} &= h_{\alpha}\left(\frac{\partial h_{\beta}}{\partial \phi_{\alpha}} + \frac{\partial h_{\gamma}}{\partial \phi_{\alpha}}\right)\mathcal{C}_{klij}^{\alpha}\llbracket \epsilon_{ij}\rrbracket^{\alpha}_{\beta} + h_{\alpha}h_{\beta}\mathcal{C}_{klij}^{\alpha}\frac{\partial \llbracket \epsilon_{ij}\rrbracket^{\alpha}_{\beta}}{\partial \phi_{\alpha}} + h_{\alpha}\frac{\partial h_{\gamma}}{\partial \phi_{\alpha}}\mathcal{C}_{klij}^{\alpha}\llbracket \epsilon_{ij}\rrbracket^{\beta}_{\gamma}\label{EqnA8}\\
h_{\beta}\mathcal{C}_{klij}^{\beta}\frac{\partial \epsilon_{ij}^{\beta}}{\partial \phi_{\alpha}} &= -h_{\beta}\frac{\partial h_{\alpha}}{\partial \phi_{\alpha}}\mathcal{C}_{klij}^{\beta}\llbracket \epsilon_{ij}\rrbracket^{\alpha}_{\beta} - h_{\beta}h_{\alpha}\mathcal{C}_{klij}^{\beta}\frac{\partial \llbracket \epsilon_{ij}\rrbracket^{\alpha}_{\beta}}{\partial \phi_{\alpha}} + h_{\beta}\frac{\partial h_{\gamma}}{\partial \phi_{\alpha}}\mathcal{C}_{klij}^{\beta}\llbracket \epsilon_{ij}\rrbracket^{\beta}_{\gamma} + h_{\beta}h_{\gamma}\mathcal{C}_{klij}^{\beta}\frac{\partial \llbracket \epsilon_{ij}\rrbracket^{\beta}_{\gamma}}{\partial \phi_{\alpha}}\label{EqnA9}\\
h_{\gamma}\mathcal{C}_{klij}^{\gamma}\frac{\partial \epsilon_{ij}^{\gamma}}{\partial \phi_{\alpha}} &= -h_{\gamma}\frac{\partial h_{\alpha}}{\partial \phi_{\alpha}}\mathcal{C}_{klij}^{\gamma}\llbracket\epsilon_{ij}\rrbracket^{\alpha}_{\beta} - h_{\gamma}\left(\frac{\partial h_{\beta}}{\partial \phi_{\alpha}} + \frac{\partial h_{\alpha}}{\partial \phi_{\alpha}}\right)\mathcal{C}_{klij}^{\gamma}\llbracket\epsilon_{ij} \rrbracket^{\beta}_{\gamma} - h_{\beta}h_{\gamma}\mathcal{C}_{klij}^{\gamma}\frac{\partial \llbracket \epsilon_{ij}\rrbracket^{\beta}_{\gamma}}{\partial \phi_{\alpha}}\label{EqnA10}
\end{align}
Substituting Eqs. (\ref{EqnA01}) and (\ref{EqnA02}) in Eqs. (\ref{EqnA8}) and (\ref{EqnA10}) gives
\begin{align}
h_{\alpha}\mathcal{C}_{klij}^{\alpha}\frac{\partial \epsilon_{ij}^{\alpha}}{\partial \phi_{\alpha}} &= -h_{\alpha}\frac{\partial h_{\alpha}}{\partial \phi_{\alpha}}\mathcal{C}_{klij}^{\alpha}\llbracket \epsilon_{ij}\rrbracket^{\alpha}_{\beta} + h_{\alpha}h_{\beta}\mathcal{C}_{klij}^{\alpha}\frac{\partial \llbracket \epsilon_{ij}\rrbracket^{\alpha}_{\beta}}{\partial \phi_{\alpha}} + h_{\alpha}\frac{\partial h_{\gamma}}{\partial \phi_{\alpha}}\mathcal{C}_{klij}^{\alpha}\llbracket \epsilon_{ij}\rrbracket^{\beta}_{\gamma}\label{EqnA11}\\
h_{\beta}\mathcal{C}_{klij}^{\beta}\frac{\partial \epsilon_{ij}^{\beta}}{\partial \phi_{\alpha}} &= -h_{\beta}\frac{\partial h_{\alpha}}{\partial \phi_{\alpha}}\mathcal{C}_{klij}^{\beta}\llbracket \epsilon_{ij}\rrbracket^{\alpha}_{\beta} - h_{\beta}h_{\alpha}\mathcal{C}_{klij}^{\beta}\frac{\partial \llbracket \epsilon_{ij}\rrbracket^{\alpha}_{\beta}}{\partial \phi_{\alpha}} + h_{\beta}\frac{\partial h_{\gamma}}{\partial \phi_{\alpha}}\mathcal{C}_{klij}^{\beta}\llbracket \epsilon_{ij}\rrbracket^{\beta}_{\gamma} + h_{\beta}h_{\gamma}\mathcal{C}_{klij}^{\beta}\frac{\partial \llbracket \epsilon_{ij}\rrbracket^{\beta}_{\gamma}}{\partial \phi_{\alpha}}\label{EqnA12}\\
h_{\gamma}\mathcal{C}_{klij}^{\gamma}\frac{\partial \epsilon_{ij}^{\gamma}}{\partial \phi_{\alpha}} &= -h_{\gamma}\frac{\partial h_{\alpha}}{\partial \phi_{\alpha}}\mathcal{C}_{klij}^{\gamma}\llbracket\epsilon_{ij}\rrbracket^{\alpha}_{\beta} + h_{\gamma}\frac{\partial h_{\gamma}}{\partial \phi_{\alpha}}\mathcal{C}_{klij}^{\gamma}\llbracket\epsilon_{ij} \rrbracket^{\beta}_{\gamma} - h_{\beta}h_{\gamma}\mathcal{C}_{klij}^{\gamma}\frac{\partial \llbracket \epsilon_{ij}\rrbracket^{\beta}_{\gamma}}{\partial \phi_{\alpha}}\label{EqnA13}
\end{align}
Adding Eqs. (\ref{EqnA11}), (\ref{EqnA12}), and (\ref{EqnA13}) yields
\begin{align}
\begin{split}
\sum_{\theta=\alpha}^{\gamma}h_{\theta}\mathcal{C}_{klij}^{\theta}\frac{\partial \epsilon_{ij}^{\theta}}{\partial \phi_{\alpha}} &= h_{\alpha}h_{\beta}\left\{\mathcal{C}_{klij}^{\alpha} - \mathcal{C}_{klij}^{\beta}\right\}\frac{\partial \llbracket \epsilon_{ij}\rrbracket^{\alpha}_{\beta}}{\partial \phi_{\alpha}} + h_{\beta}h_{\gamma}\left\{\mathcal{C}_{klij}^{\beta} - \mathcal{C}_{klij}^{\gamma}\right\}\frac{\partial \llbracket \epsilon_{ij} \rrbracket^{\beta}_{\gamma}}{\partial \phi_{\alpha}}\\
-&\frac{\partial h_{\alpha}}{\partial \phi_{\alpha}}\left\{ \sum_{\theta=\alpha}^{\gamma}h_{\theta}\mathcal{C}_{klij}^{\theta}\right\}\llbracket\epsilon_{ij} \rrbracket^{\alpha}_{\beta} + \frac{\partial h_{\gamma}}{\partial \phi_{\alpha}}\left\{\sum_{\theta=\alpha}^{\gamma}h_{\theta}\mathcal{C}_{klij}^{\theta}\right\}\llbracket\epsilon_{ij} \rrbracket^{\beta}_{\gamma}
\label{EqnA17}
\end{split}
\end{align}
By replacing $\partial \phi_{\alpha}$ with $\partial \phi_{\beta}$ and $\partial \phi_{\gamma}$ with $\partial \phi_{\gamma}$ equivalent expressions for $\phi_{\beta}$ and $\phi_{\gamma}$ can be easily obtained.

\subsection{Derivatives with respect to total strain}

Differentiating Eqs. (\ref{Eqn4}), (\ref{Eqn5}) and (\ref{Eqn6}) with respect to total strain $\boldsymbol{\epsilon}$ and then multiplying with $h_{\alpha}\sigma_{ij}^{\alpha}$, $h_{\beta}\sigma_{ij}^{\beta}$ and $h_{\gamma}\sigma_{ij}^{\gamma}$, respectively, yields
\begin{align}
h_{\alpha}\sigma_{ij}^{\alpha}\frac{\partial \epsilon_{ij}^{\alpha}}{\partial \epsilon_{mn}} &= h_{\alpha}\sigma_{mn}^{\alpha} + h_{\alpha}\left(h_{\beta} + h_{\gamma}\right)\sigma_{ij}^{\alpha}\frac{\partial \llbracket \epsilon_{ij}\rrbracket^{\alpha}_{\beta}}{\partial \epsilon_{mn}} +  h_{\gamma}h_{\alpha}\sigma_{ij}^{\alpha}\frac{\partial \llbracket \epsilon_{ij}\rrbracket^{\beta}_{\gamma}}{\partial \epsilon_{mn}}\label{EqnA18}\\
h_{\beta}\sigma_{ij}^{\beta}\frac{\partial \epsilon_{ij}^{\beta}}{\partial \epsilon_{mn}} &= h_{\beta}\sigma_{mn}^{\beta} - h_{\alpha}h_{\beta}\sigma_{ij}^{\beta}\frac{\partial \llbracket \epsilon_{ij}\rrbracket^{\alpha}_{\beta}}{\partial \epsilon_{mn}} +  h_{\gamma}h_{\beta}\sigma_{ij}^{\beta}\frac{\partial \llbracket \epsilon_{ij}\rrbracket^{\beta}_{\gamma}}{\partial \epsilon_{mn}}\label{EqnA19}\\
h_{\gamma}\sigma_{ij}^{\gamma}\frac{\partial \epsilon_{ij}^{\gamma}}{\partial \epsilon_{mn}} &= h_{\gamma}\sigma_{mn}^{\gamma} - h_{\alpha}h_{\gamma}\sigma_{ij}^{\gamma}\frac{\partial \llbracket \epsilon_{ij}\rrbracket^{\alpha}_{\beta}}{\partial \epsilon_{mn}} -h_{\gamma}\left(h_{\beta} + h_{\alpha}\right)\sigma_{ij}^{\gamma}\frac{\partial \llbracket \epsilon_{ij}\rrbracket^{\beta}_{\gamma}}{\partial \epsilon_{mn}}\label{EqnA20}
\end{align}
Now, we note that $h_{\gamma}h_{\alpha}$ is non-zero only near the $\gamma/\alpha$ interface boundary and the terms $\llbracket\boldsymbol{\epsilon}\rrbracket^{\alpha}_{\beta}$ and $\llbracket\boldsymbol{\epsilon}\rrbracket^{\beta}_{\gamma}$ are also non-zero only within the interfacial regions of $\beta/\gamma$ and $\alpha/\beta$ boundaries. We therefore set all terms premultiplied by $h_{\gamma}h_{\alpha}$ to zero in Eqs. (\ref{EqnA18}), (\ref{EqnA19}) and (\ref{EqnA20}). Now adding these equations yields
\begin{align}
\begin{split}
\sum_{\theta=\alpha}^{\gamma}h_{\theta}\sigma_{ij}^{\theta}\frac{\partial \epsilon_{ij}^{\theta}}{\partial \epsilon_{mn}} &= \sum_{\theta=\alpha}^{\gamma}h_{\theta}\sigma_{mn}^{\theta} + h_{\alpha}h_{\beta}\left\{ \sigma_{ij}^{\alpha} - \sigma_{ij}^{\beta}\right\}n_{j}^{\alpha\beta}\frac{\partial a_{i}^{\alpha\beta}}{\partial \epsilon_{mn}} \\ 
&+h_{\gamma}h_{\beta}\left\{\sigma_{ij}^{\beta} - \sigma_{ij}^{\gamma}\right\}n_{j}^{\beta\gamma}\frac{\partial a_{i}^{\beta\gamma}}{\partial \epsilon_{mn}}
\label{EqnA21}
\end{split}
\end{align}

Due to Eqs. (\ref{Eqn10}) and (\ref{Eqn11}), the last two terms in Eq. (\ref{EqnA21}) must be zero. This gives
\begin{align}
\begin{split}
\sum_{\theta=\alpha}^{\gamma}h_{\theta}\sigma_{ij}^{\theta}\frac{\partial \epsilon_{ij}^{\theta}}{\partial \epsilon_{mn}} &= \sum_{\theta=\alpha}^{\gamma}h_{\theta}\sigma_{mn}^{\theta} 
\label{EqnA22}
\end{split}
\end{align}
Next, differentiating Eqs. (\ref{Eqn4}), (\ref{Eqn5}) and (\ref{Eqn6}) with respect to total strain $\boldsymbol{\epsilon}$ and multiplying with $h_{\alpha}\mathcal{C}_{ijkl}^{\alpha}$, $h_{\beta}\mathcal{C}_{mnkl}^{\beta}$ and $h_{\gamma}\mathcal{C}_{mnkl}^{\gamma}$, yields

\begin{align}
h_{\alpha}\mathcal{C}_{mnkl}^{\alpha}\frac{\partial \epsilon_{kl}^{\alpha}}{\partial \epsilon_{rs}} &= h_{\alpha}\mathcal{C}_{mnrs}^{\alpha} + h_{\alpha}\left(h_{\beta} + h_{\gamma}\right)\mathcal{C}_{mnkl}^{\alpha}\frac{\partial \llbracket\epsilon_{kl}\rrbracket^{\alpha}_{\beta}}{\partial \epsilon_{rs}} + h_{\gamma}h_{\alpha}\mathcal{C}_{mnkl}^{\alpha}\frac{\partial \llbracket \epsilon_{kl}\rrbracket^{\beta}_{\gamma}}{\partial \epsilon_{rs}}\label{EqnA23}\\
h_{\beta}\mathcal{C}_{mnkl}^{\beta}\frac{\partial \epsilon_{kl}^{\beta}}{\partial \epsilon_{rs}} &= h_{\beta}\mathcal{C}_{mnrs}^{\beta} - h_{\alpha}h_{\beta}\mathcal{C}_{mnkl}^{\beta}\frac{\partial \llbracket\epsilon_{kl}\rrbracket^{\alpha}_{\beta}}{\partial \epsilon_{rs}} + h_{\gamma}h_{\beta}\mathcal{C}_{mnkl}^{\beta}\frac{\partial \llbracket \epsilon_{kl}\rrbracket^{\beta}_{\gamma}}{\partial \epsilon_{rs}}\label{EqnA24}\\
h_{\gamma}\mathcal{C}_{mnkl}^{\gamma}\frac{\partial \epsilon_{kl}^{\gamma}}{\partial \epsilon_{rs}} &= h_{\gamma}\mathcal{C}_{mnrs}^{\gamma} - h_{\alpha}h_{\gamma}\mathcal{C}_{mnkl}^{\gamma}\frac{\partial \llbracket\epsilon_{kl}\rrbracket^{\alpha}_{\beta}}{\partial \epsilon_{rs}} - h_{\gamma}\left(h_{\beta} +  h_{\alpha}\right)\mathcal{C}_{mnkl}^{\gamma}\frac{\partial \llbracket \epsilon_{kl}\rrbracket^{\beta}_{\gamma}}{\partial \epsilon_{rs}}\label{EqnA25}
\end{align}
Again, we set the four terms in Eqs. (\ref{EqnA23}), (\ref{EqnA24}) and (\ref{EqnA25}) which are premultiplied by $h_{\gamma}h_{\alpha}$ to zero. Next adding these equations, we see that
\begin{align}
\begin{split}
\mathcal{J}_{mnrs} = \sum_{\theta=\alpha}^{\gamma}h_{\theta}(\boldsymbol{\phi})\mathcal{C}_{mnkl}^{\theta}\frac{\partial \epsilon_{kl}^{\theta}}{\partial \epsilon_{rs}} &= \sum_{\theta=\alpha}^{\gamma}h_{\theta}(\boldsymbol{\phi})\mathcal{C}_{mnkl}^{\theta} + h_{\alpha}h_{\beta}\left\{\mathcal{C}_{mnkl}^{\alpha} - \mathcal{C}_{mnkl}^{\beta}\right\}\frac{\partial \llbracket \epsilon_{kl}\rrbracket^{\alpha}_{\beta}}{\partial \epsilon_{rs}}\\
 &+ h_{\gamma}h_{\beta}\left\{\mathcal{C}_{mnkl}^{\beta} - \mathcal{C}_{mnkl}^{\gamma}\right\}\frac{\partial \llbracket \epsilon_{kl}\rrbracket^{\beta}_{\gamma}}{\partial \epsilon_{rs}}
 \label{EqnA26}
\end{split}
\end{align}
\renewcommand{\theequation}{C.\arabic{equation}}
\renewcommand{\thesubsection}{C.\arabic{subsection}}
\setcounter{subsection}{0}
\setcounter{equation}{0}
\section*{Appendix C}
\label{Appendix_C}
\section*{Derivation of stress and its derivatives}
Differentiating Eq. (\ref{Eqn20}) with respect to total strain yields
\begin{align}
\frac{\partial \omega_{bulk}}{\partial \epsilon_{mn}} = \sum_{\theta=\alpha}^{\gamma}h_{\theta}(\boldsymbol{\phi})\frac{\partial \omega_{bulk}^{\theta}}{\partial \epsilon_{ij}^{\theta}}\frac{\partial \epsilon_{ij}^{\theta}}{\partial \epsilon_{mn}}
\label{EqnB1}
\end{align}
Using the definition of the phase stress tensor, we can replace $\partial \omega_{bulk}^{\theta}/\partial \epsilon_{ij}$ with $\sigma_{ij}^{\theta}$. Using the relation (\ref{EqnA22}) , Eq.(\ref{EqnB1}) can be written as
\begin{align}
\frac{\partial \omega_{bulk}}{\partial \epsilon_{ij}} = \sum_{\theta=\alpha}^{\gamma}h_{\theta}\sigma_{ij}^{\theta}\frac{\partial \epsilon_{ij}^{\theta}}{\partial \epsilon_{mn}} &= \sum_{\theta=\alpha}^{\gamma}h_{\theta}(\boldsymbol{\phi})\sigma_{mn}^{\theta} 
\label{EqnB2}
\end{align}

\renewcommand{\thesubsection}{D.\arabic{subsection}}
\renewcommand{\theequation}{D.\arabic{equation}}
\setcounter{subsection}{0}
\setcounter{equation}{0}
\section*{Appendix D}
\label{Appendix_D}
\section*{Derivation of driving force and its derivatives}
Differentiating Eq. (\ref{Eqn20}) with respect to phase-field variable $\phi_{\theta}$ yields
\begin{align}
\begin{split}
\frac{\partial \omega_{bulk}}{\partial \phi_{\theta}} &= \sum_{\sigma=\alpha}^{\gamma}\frac{\partial h_{\sigma}}{\partial \phi_{\theta}}\omega^{\sigma} + \sum_{\sigma=\alpha}^{\gamma}h_{\sigma}(\boldsymbol{\phi})\frac{\partial \omega_{\sigma}}{\partial \epsilon_{ij}^{\sigma}}\frac{\partial \epsilon_{ij}^{\sigma}}{\partial \phi_{\theta}}\\
&= \sum_{\sigma=\alpha}^{\gamma}\frac{\partial h_{\sigma}}{\partial \phi_{\theta}}\omega^{\sigma} + \sum_{\sigma=\alpha}^{\gamma}h_{\sigma}(\boldsymbol{\phi})\sigma_{ij}^{\sigma}\frac{\partial \epsilon_{ij}^{\sigma}}{\partial \phi_{\theta}}
\label{EqnB1}
\end{split}
\end{align}
For $\theta=\alpha$, substituting Eqs. (\ref{EqnA01}), (\ref{EqnA02}) and (\ref{EqnA8a}) in Eq. (\ref{EqnB1}) yields
\begin{align}
\begin{split}
\frac{\partial \omega_{bulk}}{\partial \phi_{\alpha}} &= \frac{\partial h_{\beta}}{\partial \phi_{\alpha}}\left(\omega^{\beta} - \omega^{\alpha}\right) + \frac{\partial h_{\gamma}}{\partial \phi_{\alpha}}\left(\omega^{\gamma} - \omega^{\alpha}\right)\\
&-\frac{\partial h_{\alpha}}{\partial \phi_{\alpha}}\left\{\sum_{\theta=\alpha}h_{\theta}\sigma_{ij}^{\theta}\right\}\llbracket \epsilon_{ij}\rrbracket^{\alpha}_{\beta} + \frac{\partial h_{\gamma}}{\partial \phi_{\alpha}}\left\{\sum_{\theta=\alpha}^{\gamma}h_{\theta}\sigma_{ij}^{\theta}\right\}\llbracket \epsilon_{ij} \rrbracket^{\beta}_{\gamma}
\end{split}
\label{EqnB2}
\end{align}
Similarly, one can derive the bulk driving force for phase-field variables $\phi_{\beta}$ and $\phi_{\gamma}$ by using Eqs. (\ref{EqnA8b}) and (\ref{EqnA8c})
\begin{align}
\begin{split}
\frac{\partial \omega_{bulk}}{\partial \phi_{\beta}} &= \frac{\partial h_{\alpha}}{\partial \phi_{\beta}}\left(\omega^{\alpha} - \omega^{\beta}\right) + \frac{\partial h_{\gamma}}{\partial \phi_{\beta}}\left(\omega^{\gamma} - \omega^{\beta}\right)\\
&-\frac{\partial h_{\alpha}}{\partial \phi_{\beta}}\left\{\sum_{\theta=\alpha}h_{\theta}\sigma_{ij}^{\theta}\right\}\llbracket \epsilon_{ij}\rrbracket^{\alpha}_{\beta} + \frac{\partial h_{\gamma}}{\partial \phi_{\beta}}\left\{\sum_{\theta=\alpha}^{\gamma}h_{\theta}\sigma_{ij}^{\theta}\right\}\llbracket \epsilon_{ij} \rrbracket^{\beta}_{\gamma}\label{EqnB3}
\end{split}
\\
\begin{split}
\frac{\partial \omega_{bulk}}{\partial \phi_{\gamma}} &= \frac{\partial h_{\alpha}}{\partial \phi_{\gamma}}\left(\omega^{\alpha} - \omega^{\gamma}\right) + \frac{\partial h_{\beta}}{\partial \phi_{\gamma}}\left(\omega^{\beta} - \omega^{\gamma}\right)\\
&-\frac{\partial h_{\alpha}}{\partial \phi_{\gamma}}\left\{\sum_{\theta=\alpha}h_{\theta}\sigma_{ij}^{\theta}\right\}\llbracket \epsilon_{ij}\rrbracket^{\alpha}_{\beta} + \frac{\partial h_{\gamma}}{\partial \phi_{\gamma}}\left\{\sum_{\theta=\alpha}^{\gamma}h_{\theta}\sigma_{ij}^{\theta}\right\}\llbracket \epsilon_{ij} \rrbracket^{\beta}_{\gamma}\label{EqnC4}
\end{split}
\end{align}
\renewcommand{\thesubsection}{E.\arabic{subsection}}
\renewcommand{\theequation}{E.\arabic{equation}}
\setcounter{equation}{0}
\setcounter{subsection}{0}
\section*{Appendix E}
\section*{Non-dimensionalization}
Eqs. (\ref{Eqn26})-(\ref{Eqn29}) were solved in the MOOSE (Multiphysics Object-Oriented Simulation Environment) finite-element framework \cite{Gaston2009}. To ensure good convergence, we formed a non-dimensional form of these equations. In this section, we provide the dimensionless form of the governing equations.

We will denote dimensionless quantities using the symbol, $\overline{\left(\cdot\right)}$. Let $l_{c}$ and $t_{c}$ denote characteristic length and time scales. Then, the non-dimensional position and time may be written as: $\overline{\boldsymbol{x}} = \boldsymbol{x}/l_{c}$ and $\overline{t} = t/t_{c}$. The dimensionless form of the displacement field is defined as: $\overline{\boldsymbol{u}}= \boldsymbol{u}/l_{c}$. Similarly, we define the dimensionless form of the set of diffusion potentials as: $\overline{\boldsymbol{\tilde{\mu}}} = \boldsymbol{\tilde{\mu}}/RT$, where R is gas constant and T is simulation temperature. After change of variables and using the relation $c_{k} = X_{k}/V_{m}$, it can be shown that the dimensionless form of  Eqs. (\ref{Eqn26})-(\ref{Eqn29}) may be written as
\begin{align}
\sum_{\theta=1}^{p}h_{\theta}(\boldsymbol{\phi})X_{k}^{\theta}(\overline{\tilde{\boldsymbol{\mu}}}) - X_{k} &= 0,\quad \label{Eqn32} \forall\, k=1\hdots (n-1),\\
\overline{\text{div}}\left[\sum_{\theta=1}^{p}h_{\theta}(\boldsymbol{\phi})\overline{\sigma_{ij}^{\theta}}\right] &= 0,\label{Eqn33}\\
\frac{\partial \phi_{\theta}}{\partial \overline{t}} + \overline{L_{\phi}}\left[ \frac{\partial g\left(\boldsymbol{\phi}\right)}{\partial \phi_\theta} - \overline{\kappa}\overline{\Delta}\phi_{\theta} + \overline{\lambda}_{1}\frac{\partial \overline{\omega_{chem}}}{\partial \phi_{\theta}} + \overline{\lambda}_{2}\frac{\partial \overline{\omega_{mech}}}{\partial \phi_{\theta}} \right] &=0\quad  \label{Eqn34}\forall\, \theta=1\hdots p,\\
\frac{\partial X_{k}}{\partial \overline{t}} - \overline{\nabla}\left[\sum_{j=1}^{n-1}\overline{L_{kj}^{n}}\left(\overline{\boldsymbol{\tilde{\mu}},} \phi\right)\overline{\nabla} \overline{\tilde{\mu}_{j}}\left(\overline{\boldsymbol{x}},\overline{t}\right)\right] &= 0\quad \forall\, k=1\hdots (n-1),
\label{Eqn35}
\end{align}
where  
\begin{align}
\overline{\boldsymbol{\sigma}} &= \boldsymbol{\sigma}/\mu_{el},\\
\overline{L}_{\phi} &= t_{c}L_{\phi}m,\\
\overline{\kappa} &= \kappa/\left(l_{c}^{2}m\right),\\
\overline{\lambda_{1}}&= RT/(mV_{m}),\\
\overline{\lambda_{2}} &= \mu_{el}/m,\\
\overline{L_{kj}^{n}} &= L_{kj}^{n}t_{c}RT/l_{c}^{2},\\
\frac{\partial \overline{\omega_{chem}}}{\partial \phi_{\theta}} &= \left(\frac{1}{RT}\right)\left\{\sum_{\sigma=1}^{p}\frac{\partial h_{\sigma}}{\partial \phi_{\theta}}\omega_{chem}^{\sigma} \right\},\\
\frac{\partial \overline{\omega_{mech}}}{\partial \phi_{\theta}} &= \left(\frac{1}{\mu_{el}}\right)\left\{\sum_{\sigma=1}^{p}\frac{\partial h_{\sigma}}{\partial \phi_{\theta}}\omega_{elastic}^{\sigma} + \sum_{\sigma=1}^{p}h_{\sigma}\frac{\partial \omega_{elastic}^{\sigma}}{\partial \phi_{\theta}}  \right\},\\
\omega_{elastic}^{\theta} &= (1/2)\mathcal{C}_{ijkl}^{\theta}\left[\epsilon_{kl}^{\theta} - \epsilon_{kl}^{\star\theta}\right]\left[\epsilon_{ij}^{\theta} - \epsilon_{ij}^{\star\theta}\right]
\end{align}
\renewcommand{\thesubsection}{F.\arabic{subsection}}
\renewcommand{\theequation}{F.\arabic{equation}}
\renewcommand{\thefigure}{F.\arabic{figure}}
\setcounter{figure}{0}
\setcounter{equation}{0}
\setcounter{subsection}{0}
\section*{Appendix F}
\section*{Analytical solutions}
To test the simulation accuracy, we have compared our simulated results with analytically obtained solution. However, it bears emphasis that these analytical solutions require prior knowledge of the domain size, and thus of the interface positions. Therefore, we have first performed numerical simulations to calculate the position of these interfaces. Once these positions were calculated, they were used as input in the analytical solutions to make comparisons with simulated solutions. Moreover, unless stated otherwise, we have assumed zero flux boundary conditions at all boundaries. For the two planar simulations, in order to compare with analytical solutions we have taken all fields to be periodic in the top and bottom boundaries. Moreover, to reduce the computational costs by taking advantage of the domain symmetry, we have used symmetry boundary conditions at the left and bottom boundaries for two non-planar simulations (see cases III and IV).

In this section, we provide the analytical solutions to the four set of three-phase simulations performed in this paper. It must be emphasized that to analytically solve the mechanical equilibrium equations, the instantaneous positions of the two two-phase interphases are required. These prerequisite positions are therefore numerically obtained based on the phase-field results and then compared against analytical solutions. 
\subsection{Solution for the planar Ni-Al case}

\begin{figure}[!ht]
\center
\includegraphics[width=0.8\linewidth]{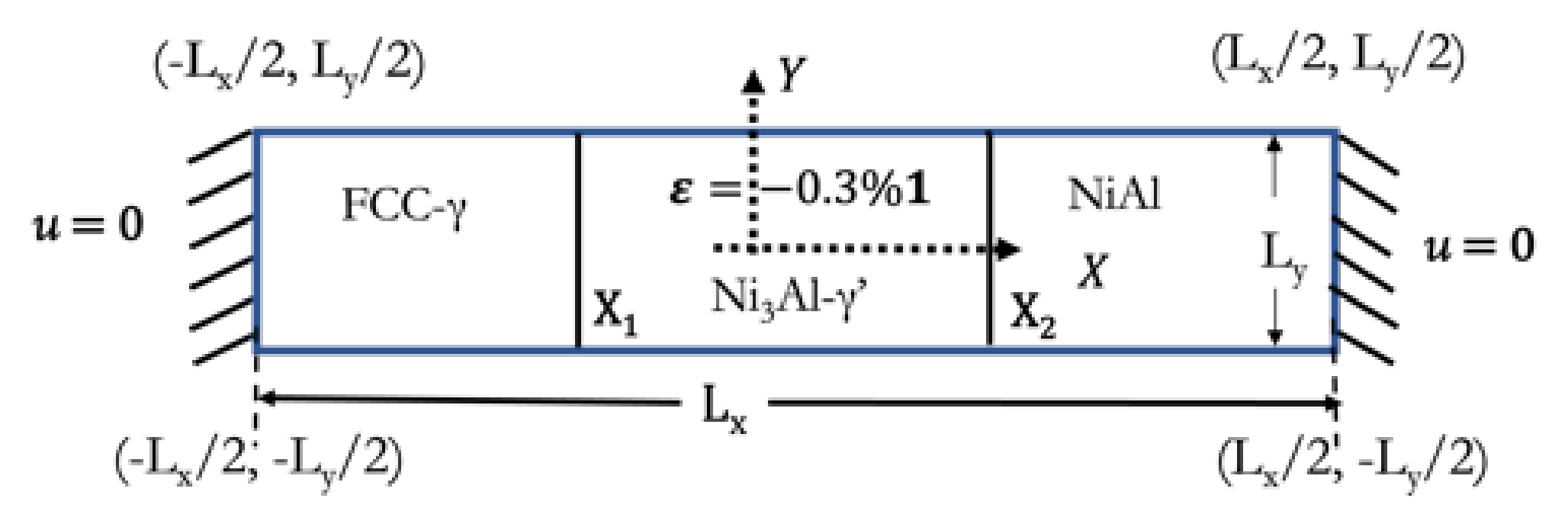}
\caption{A schematic showing the phases, eigenstrains and mechanical  boundary conditions for the planar Ni-Al case.}
\label{planar_NiAl_schematic}
\end{figure}

Fig. \ref{planar_NiAl_schematic} shows the system geometry and boundary conditions for the planar Ni-Al case. For the sake of generality, we will refer to the leftmost (fcc$-\gamma$), center (Ni$_3$Al-$\gamma^{\prime}$) and rightmost (NiAl) phases as $\alpha$, $\beta$ and $\gamma$, respectively. 
 Let $x_{1}(t)$ and $x_{2}(t)$ represent the positions of the $\alpha/\beta$ and $\beta/\gamma$ interphases at time $t$. As  mentioned before, we numerically determine these positions at any given instant from the phase-field simulations and thus the accuracy of the solution depends on the interface positions. Moreover, assuming plane stress conditions and neglecting externally applied body forces, the mechanical equilibrium equations in Cartesian frame within the bulk regions of a phase $\theta=\{\alpha,\beta,\gamma\}$ reduces to:
\begin{align}
\frac{\partial \sigma_{x}^{\theta}}{\partial x} + \frac{\partial \sigma_{xy}^{\theta}}{\partial y} &= 0, \label{EqnD1}\\
\frac{\partial \sigma_{xy}^{\theta}}{\partial y} + \frac{\partial \sigma_{yy}^{\theta}}{\partial y} &= 0.\label{EqnD2}
\end{align}
As depicted in Fig. \ref{planar_NiAl_schematic}, we have assumed that origin of the Cartesian frame lies at the center of the domain. As shown in Table 3, we have assumed the elastic constants to be isotropic but spatially heterogeneous. Thus, the stress-strain relation within the bulk phases may be written as \cite{Mura1991}:
\begin{align}
\sigma_{ij}^{\theta} = \lambda^{\theta}\delta_{ij}\left(\epsilon_{kk}^{\theta} - \epsilon_{kk}^{\star,\theta}\right) + 2\mu^{\theta}(\epsilon_{ij}^{\theta} - \epsilon_{ij}^{\star, \theta}),
\label{EqnD3}
\end{align}
where $\lambda^{\theta}$, $\mu^{\theta}$ and $\boldsymbol{\epsilon}^{\star,\theta}$ are Lame's constant, shear modulus and eigenstrain of phase $\theta$, respectively. As show in Fig. \ref{planar_NiAl_schematic}, the mechanical displacements at the left and right boundaries may be written as:
\begin{align}
u_{x}(x=\pm L_x/2,y,t) &= 0,\label{EqnD4}\\
u_{y}(x=\pm L_x/2,y,t) &= 0.\label{EqnD5}
\end{align}
On the other hand, the mechanical displacements are assumed to be periodic along the y-direction. Due to these boundary conditions, only the x-component of displacement, $u_{x}(\boldsymbol{x},t)$, and the normal strain along x-direction, $\epsilon_{x}^{\theta}=du_{x}^{\theta}/dx$, are nonzero in the bulk regions. 

Using Eq. (\ref{EqnD3}), it follows that the nonzero mechanical stresses within the bulk phases are:
\begin{align}
\sigma_{x}(\boldsymbol{x},t) &= 
\begin{cases}
\left(\lambda^{\alpha} + 2\mu^{\alpha}\right)\epsilon_{x}^{\alpha}&\quad -L_x/2 < x < x_{1},\\
\left(\lambda^{\beta} + 2\mu^{\beta}\right)\epsilon_{x}^{\beta} - 2(\lambda^{\beta} + \mu^{\beta})\epsilon^{\star}&\quad x_{1} <x < x_{2},\\
\left(\lambda^{\gamma} + 2\mu^{\gamma}\right)\epsilon_{x}^{\gamma}&\quad x_{2}< x < L_{x}/2,
\end{cases}
\label{EqnD6}
\\
\sigma_{y}(\boldsymbol{x},t) &= 
\begin{cases}
\lambda^{\alpha} \epsilon_{x}^{\alpha}&\quad -L_x/2 < x < x_{1},\\
\lambda^{\beta} \epsilon_{x}^{\beta} - 2(\lambda^{\beta} + \mu^{\beta})\epsilon^{\star}&\quad x_{1} <x < x_{2},\\
\lambda^{\gamma} \epsilon_{x}^{\gamma}&\quad x_{2}< x < L_{x}/2.
\end{cases}
\label{EqnD7}
\end{align}

It should be noticed from Eqs. (\ref{EqnD6})-(\ref{EqnD7}) that the $\beta$ stress components are different compared to the $\alpha$ (FCC) and $\gamma$ (NiAl) phases because we have assumed a two-dimensional eigenstrain $\boldsymbol{\epsilon}^{\star}  = \begin{psmallmatrix}\epsilon^{\star} & 0\\ 0 & \epsilon^{\star}\end{psmallmatrix}$ only within the Ni$_3$Al phase. Next, by substituting Eq. (\ref{EqnD6}) in Eq. (\ref{EqnD1}) and using the strain-displacement relations, we see that
\begin{align}
\left(\lambda^{\theta} + 2\mu^{\theta}\right)\frac{d^{2}u_{x}^{\theta}}{dx^{2}} = 0.
\label{EqnD8}
\end{align}
By integrating Eq. (\ref{EqnD8}), it follows that the $x$-component of displacement field must vary linearly with distance within the bulk phases. More precisely, 
\begin{align}
u_{x}(\boldsymbol{x},t) &= 
\begin{cases}
A^{\alpha} x + B^{\alpha}&\quad -L_x/2 < x < x_{1},\\
A^{\beta} x + B^{\beta} &\quad x_{1} <x < x_{2},\\
A^{\gamma} x + B^{\gamma}&\quad x_{2}< x \leq L_{x}/2,
\label{EqnD9}
\end{cases}
\end{align}
where $A^{\alpha}$, $B^{\alpha}$, $A^{\beta}$, $B^{\beta}$, $A^{\gamma}$ and $B^{\gamma}$ are unknown constants. Moreover, these six constants can be determined using the two imposed boundary conditions (Eqs. \ref{EqnD4} \& \ref{EqnD5}) and four interfacial conditions. Two of these interfacial conditions arise due to the continuity of $x$-component of displacement at the two interfaces, $\llbracket u_{x}\rrbracket = 0$, and the remaining two are a result of continuity of normal stresses along $x$,  $\llbracket \sigma_{x}\rrbracket = 0$. Specifically, 
\begin{align}
\left.u_{x}^{\alpha}\right|_{x_1} = \left.u_{x}^{\beta}\right|_{x_1}\label{EqnD10}\\
\left.u_{x}^{\beta}\right|_{x_2} = \left.u_{x}^{\gamma}\right|_{x_2}\label{EqnD11}\\
\left.\sigma_{x}^{\alpha}\right|_{x_1} = \left.\sigma_{x}^{\beta}\right|_{x_1}\label{EqnD12}\\
\left.\sigma_{x}^{\beta}\right|_{x_2} = \left.\sigma_{x}^{\gamma}\right|_{x_2}\label{EqnD13}\
\end{align}
Next, substituting the expressions in Eq. (\ref{EqnD9}) in Eqs. (\ref{EqnD4})  and  (\ref{EqnD5}) yields
\begin{align}
-A^{\alpha}L_{x}/2 + B^{\alpha} = 0\label{EqnD14}\\
A^{\gamma}L_{x}/2 + B^{\gamma} = 0\label{EqnD15}
\end{align}
Then using Eq. (\ref{EqnD9}), Eqs. (\ref{EqnD10})-(\ref{EqnD13}) may be written as
\begin{align}
A^{\alpha}x_{1} + B^{\alpha} - \left(A^{\beta}x_{1} + B^{\beta}\right) &= 0,\label{EqnD16}\\
A^{\beta}x_{2} + B^{\beta} - \left(A^{\gamma}x_{2} + B^{\gamma}\right) &= 0,\label{EqnD17}\\
(\lambda^{\alpha} + 2\mu^{\alpha})A^{\alpha} - (\lambda^{\beta} + 2\mu^{\beta})A^{\beta} + 2\left(\lambda^{\beta} + \mu^{\beta}\right)\epsilon^{\star} &= 0,\label{EqnD18}\\
(\lambda^{\beta} + 2\mu^{\beta})A^{\beta} - 2\left(\lambda^{\beta} + \mu^{\beta}\right)\epsilon^{\star} -(\lambda^{\gamma} + 2\mu^{\gamma})A^{\gamma}  &= 0\label{EqnD19}
\end{align}
Eqs. (\ref{EqnD14})-(\ref{EqnD19}) form a set of six equations that can be solved to determine the six unknowns. This was performed using the Python library for symbolic mathematics, SymPy \cite{Sympy2017}. A python script for solving these equations is available (see the python script \texttt{threephase\_planar\_analytical.py}).

\subsection{Solution for the planar Ni-Al-Cr case}
\begin{figure}[!ht]
\center
\includegraphics[width=0.8\linewidth]{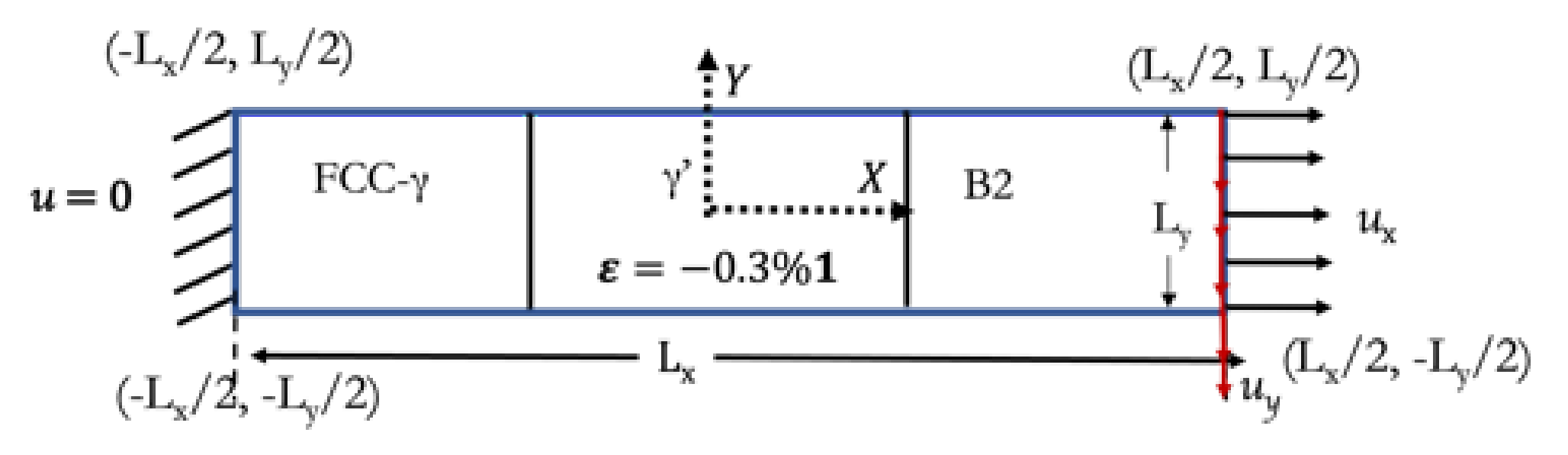}
\caption{A schematic showing the phases, eigenstrains and mechanical  boundary conditions for the planar Ni-Al-Cr case.}
\label{planar_NiAlCr_schematic}
\end{figure}

Fig.\ref{planar_NiAlCr_schematic} shows the system geometry and boundary condition for the planar Ni-Al-Cr case.

In contrast to the previous case, the leftmost (fcc-$\gamma$) and center ($\gamma^{\prime}$) phases are elastically anisotropic (see Table \ref{Table_ec}). Consequently, the stress-strain relations within these phases may be written as \cite{Mura1991}:
\begin{align}
\sigma_{ij}^{\theta} = \lambda^{\theta}\delta_{ij}\left(\epsilon_{kk}^{\theta} - \epsilon_{kk}^{\star,\theta}\right) + 2\mu^{\theta}(\epsilon_{ij}^{\theta} - \epsilon_{ij}^{\star,\theta}) + \mu^{\prime\theta}\delta_{ijkl}\left(\epsilon_{ij}^{\theta} - \epsilon_{ij}^{\star,\theta}\right),
\label{EqnD20}
\end{align}
where $\lambda^{\theta} = C_{12}^{\theta}$, $\mu^{\theta}= C_{44}^{\theta}$, $\mu^{\prime\theta}= C_{11}^{\theta} - C_{12}^{\theta} - 2C_{44}^{\theta}$ and $\delta_{ijkl}$ is zero except for $\delta_{1111} = \delta_{2222} = 1$.

As shown in Fig. \ref{planar_NiAlCr_schematic}, the imposed mechanical boundary conditions at the left and right boundaries yields:
\begin{align}
u_{x} (x=-L_x/2,y,t) &=0,\label{EqnD21}\\
u_{y}(x=-L_x/2,y,t) &=0,\label{EqnD22}\\\
u_{x}(x=L_{x}/2,y,t) &= u_{x}^{R},\label{EqnD23}\\\
u_{y}(x=L_{x/2},y,t) & = u_{y}^{R},\label{EqnD24}\
\end{align}
where $u_{x}^{R}$ and $u_{y}^{R}$ are the $x$ and $y$ components of the imposed mechanical displacement at the right boundary. Consequently, unlike the previous case, both $x$ and $y$ components of the displacement field, i.e., $u_{x}(\boldsymbol{x},t)$ and $u_{y}(\boldsymbol{x},t)$, are nonzero within the bulk phases. Precisely,
\begin{align}
u_{x}(\boldsymbol{x},t) &= 
\begin{cases}
A_{x}^{\alpha} x + B_{x}^{\alpha}&\quad -L_x/2 < x < x_{1},\\
A_{x}^{\beta} x + B_{x}^{\beta} &\quad x_{1} <x < x_{2},\\
A_{x}^{\gamma} x + B_{x}^{\gamma}&\quad x_{2}< x \leq L_{x}/2,
\label{EqnD25}
\end{cases}
\\
u_{y}(\boldsymbol{x},t) &= 
\begin{cases}
A_{y}^{\alpha} x + B_{y}^{\alpha}&\quad -L_x/2 < x < x_{1},\\
A_{y}^{\beta} x + B_{y}^{\beta} &\quad x_{1} <x < x_{2},\\
A_{y}^{\gamma} x + B_{y}^{\gamma}&\quad x_{2}< x \leq L_{x}/2,
\label{EqnD26}
\end{cases}
\end{align}
where $\{A_{x}^{\theta=\alpha,\beta,\gamma}\}$, $\{A_{y}^{\theta=\alpha,\beta,\gamma}\}$, $\{B_{x}^{\theta=\alpha,\beta,\gamma}\}$ and $\{B_{y}^{\theta=\alpha,\beta,\gamma}\}$ are the 12 unknown constants.

Next, to determine the unknown constants, we first solve the $x$-component of displacement field. This requires calculating the six unknowns: $\{A_{x}^{\theta=\alpha,\beta,\gamma}\}$ and $\{B_{x}^{\theta=\alpha,\beta,\gamma}\}$. After substituting the expressions in Eq. (\ref{EqnD25}) in Eqs. (\ref{EqnD21}) \& (\ref{EqnD23}), we get
\begin{align}
-A_{x}^{\alpha}L_{x}/2 + B_{x}^{\alpha} &= 0\label{EqnD27}\\
u_{x}^{R} - \left(A_{x}^{\gamma}L_{x}/2 + B_{x}^{\gamma}\right) &= 0\label{EqnD28}
\end{align}

The remaining four unknowns can be determined by solving continuity of $x$-component of displacement field and normal stress along $x$. Thus, using Eqs. (\ref{EqnD3}), (\ref{EqnD20}) and (\ref{EqnD25}) in Eqs. (\ref{EqnD10})-(\ref{EqnD13}), it follows that:
\begin{align}
A_{x}^{\alpha}x_{1} + B_{x}^{\alpha}- \left(A_{x}^{\beta}x_{1} + B^{\beta}_{x}\right) = 0,\label{EqnD29}\\
A_{x}^{\beta}x_{2} + B_{x}^{\beta} - \left(A_{x}^{\gamma}x_{2} + B^{\gamma}_{x}\right) = 0,\label{EqnD30}\\
(\lambda^{\alpha} + 2\mu^{\alpha})A_{x}^{\alpha} - (\lambda^{\beta} + 2\mu^{\beta})A_{x}^{\beta} + \mu^{\prime\alpha}A_x^{\alpha} - \mu^{\prime\beta}A_{x}^{\beta} + \zeta^{\beta}\epsilon^{\star}= 0,\label{EqnD31}\\\
(\lambda^{\beta} + 2\mu^{\beta})A_{x}^{\beta} - (\lambda^{\gamma} + 2\mu^{\gamma})A_{x}^{\gamma} + \mu^{\prime\beta}A_x^{\beta} - \mu^{\prime\gamma}A_{x}^{\gamma} - \zeta^{\beta}\epsilon^{\star}= 0,\label{EqnD32}\
\end{align}
where $\lambda^{\theta=\alpha,\beta} = C_{12}^{\theta}$, $\mu^{\theta=\alpha,\beta} = C_{44}^{\theta}$,  $\mu^{\prime\theta=\alpha,\beta} = C_{11}^{\theta} - C_{12}^{\theta} - 2C_{44}^{\theta}$ and $\zeta^{\beta} = 2(\lambda^{\beta} + \mu^{\beta}) + \mu^{\prime\beta}$. By solving Eqs. (\ref{EqnD27})-(\ref{EqnD32}) we can obtain six of the $12$ unknown constants. This is achieved symbolically using SymPy \cite{Sympy2017} and the python script, \texttt{threephase\_aniso\_planar\_analytical.py}, is provided with this paper.

Following this, the remaining six constants can be obtained by solving the $y$-component of displacement field. Specifically, we need another set of six equations to determine the unknown constants: $\{A_{y}^{\theta=\alpha,\beta,\gamma}\}$ and $\{B_{y}^{\theta=\alpha,\beta,\gamma}\}$. To this end, substituting Eq. (\ref{EqnD26}) in Eqs. (\ref{EqnD22}) \& (\ref{EqnD24}) yields the first two of these equations:
\begin{align}
-A_{y}^{\alpha}L_{x}/2 + B_{y}^{\alpha} &= 0\label{EqnD33}\\
u_{y}^{R} -\left(A_{y}^{\gamma}L_{x}/2 + B_{y}^{\gamma}\right) &= 0\label{EqnD34}
\end{align}

Since the y-component of displacement field must be continuous at the two interfaces, it follows that
\begin{align}
\left.u_{y}^{\alpha}\right|_{x_1} &= \left.u_{y}^{\beta}\right|_{x_1} \implies A_{y}^{\alpha}x_{1} + B_{y}^{\alpha} - \left(A_{y}^{\beta} x_{1}+ B_{y}^{\beta}\right) = 0,\label{EqnD35}\\
\left.u_{y}^{\beta}\right|_{x_2} &= \left.u_{y}^{\gamma}\right|_{x_2} \implies A_{y}^{\beta}x_{2} + B_{y}^{\beta} - \left(A_{y}^{\gamma}x_{2} + B_{y}^{\gamma}\right) = 0.\label{EqnD36}
\end{align}  

Additionally, the shear stress must be continuous at the two interfaces. This yields
\begin{align}
\left.\sigma_{xy}^{\alpha}\right|_{x_1} = \left.\sigma_{xy}^{\beta}\right|_{x_1}\label{EqnD37}\\
\left.\sigma_{xy}^{\beta}\right|_{x_2} = \left.\sigma_{xy}^{\gamma}\right|_{x_2}\label{EqnD38}
\end{align}
Using constitutive Eqs. (\ref{EqnD3}) and (\ref{EqnD20}) in Eqs. (\ref{EqnD37}) \& (\ref{EqnD38}) yields:
\begin{align}
2\mu_{\alpha} A_{y}^{\alpha} - 2\mu_{\beta}A_{y}^{\beta} &= 0\label{EqnD39},\\
2\mu_{\beta} A_{y}^{\alpha} - 2\mu_{\gamma}A_{y}^{\gamma} &= 0\label{EqnD40}.
\end{align}
Thus, by solving Eqs. (\ref{EqnD33})-(\ref{EqnD40}) the remaining six unknowns: $\{A_{y}^{\theta=\alpha,\beta,\gamma}\}$ and $\{B_{y}^{\theta=\alpha,\beta,\gamma}\}$ can be determined. These equations were  also solved symbolically. The python script, \texttt{threephase\_aniso\_shear\_components.py}, is provided with this paper. 

\subsection{Solution for the non-planar Ni-Al case}
\begin{figure}[!ht]
\center
\includegraphics[width=0.55\linewidth]{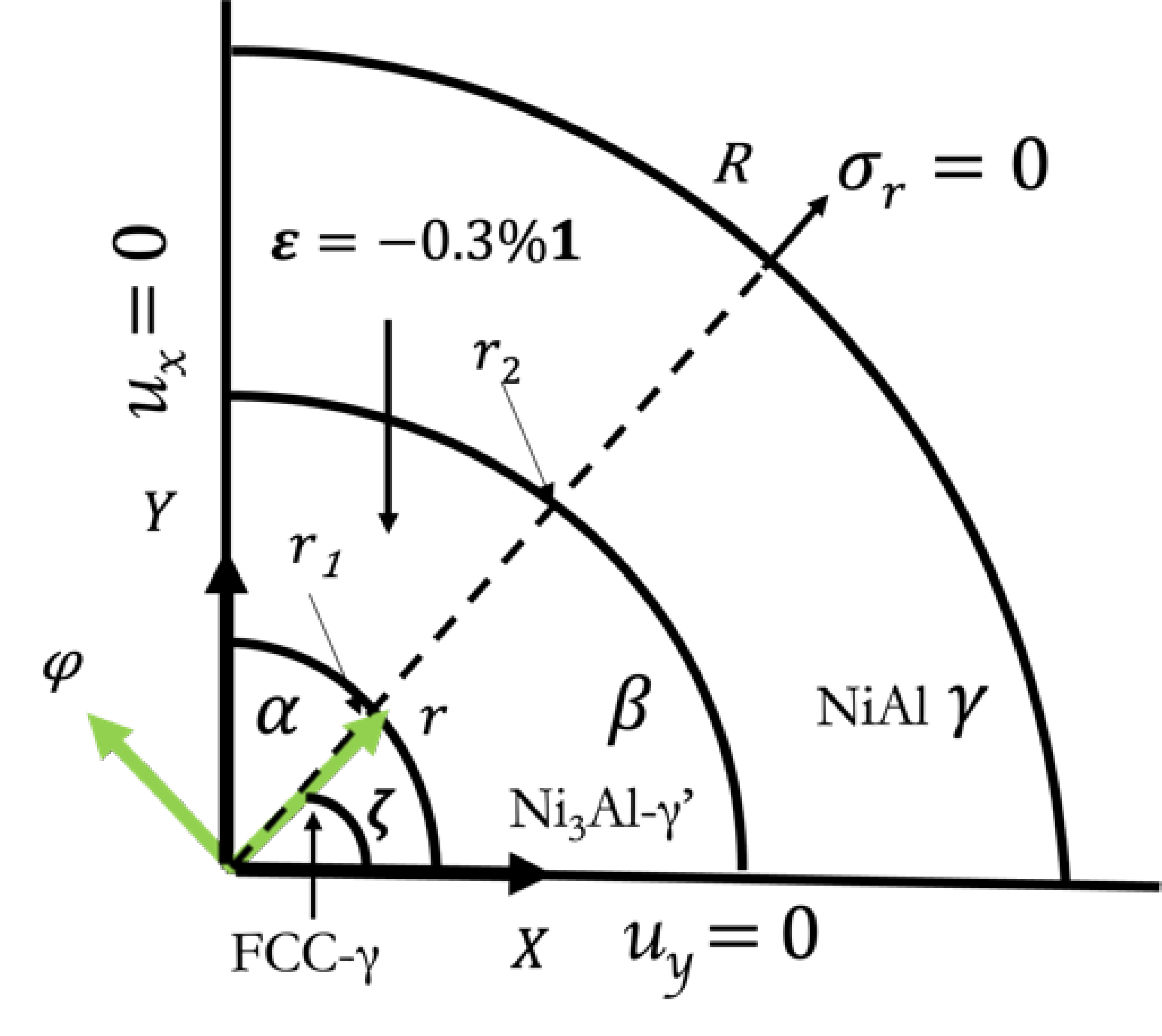}
\caption{A schematic showing the phases, eigenstrains and mechanical  boundary conditions for the concentric interface Ni-Al case.}
\label{nonplanar_NiAl_schematic}
\end{figure}

Fig. \ref{nonplanar_NiAl_schematic} shows the system geometry and boundary conditions for the three-phase Ni-Al case with concentric interfaces. As shown in Fig. \ref{nonplanar_NiAl_schematic}, the innermost (fcc-$\gamma$), center (Ni$_3$Al) and outermost (NiAl) phases are hereafter referred to as $\alpha$, $\beta$ and $\gamma$, respectively. Moreover, due to the concentric ring geometry of the system, we analytically solve the mechanical equilibrium equations in polar coordinates,  $(r,\phi)$, even though the simulation was performed in a Cartesian frame, $(x,y)$. It should be noted that to compare the analytically obtained solution against the simulated solution we transform the simulated elastic fields from the Cartesian frame to polar coordinates. 
For instance, the displacement field in polar coordinates may be calculated from Cartesian frame using
\begin{align}
\begin{Bmatrix}
u_{r}\\
u_{t}
\end{Bmatrix}
=
\begin{bmatrix}
\cos\zeta & \sin\zeta\\
-\sin\zeta & \cos\zeta
\end{bmatrix}
\begin{Bmatrix}
u_{x}\\
u_{y}
\end{Bmatrix},
\label{EqnCT}
\end{align}
where $\zeta= \tan^{-1}(y/x)$ is the angle of rotation between the two frames (Fig. \ref{nonplanar_NiAl_schematic}). For this case, the displacement field within the bulk phase $\theta$ takes the form
\begin{align}
\boldsymbol{u}^{\theta}(r,t) = u_{r}^{\theta}(r,t)\boldsymbol{e}_{r} + u_{\phi}^{\theta}(r,t)\boldsymbol{e}_{\phi}.
\end{align}
As shown in Fig. \ref{nonplanar_NiAl_schematic}, due to the imposed boundary conditions, the radial displacement is zero at the origin and the radial stress at the outer surface is zero. This yields
\begin{align}
u_{r}^{\alpha}(r=0,t)  = 0\label{EqnD42}\\
\sigma_{r}^{\gamma}(r=R,t) = 0\label{EqnD43}
\end{align}
Note that the superscripts on the mechanical fields identify the phases in the system. Because of these imposed boundary conditions, it can be assumed that the $\phi$ component of displacement field is zero throughout the system, i.e., $u_{\phi}^{\alpha} = u_{\phi}^{\beta} = u_{\phi}^{\gamma} = 0$. Consequently, the strain-displacement relation in polar coordinates within the bulk domains simplifies to
\begin{align}
\epsilon_{r}^{\theta}(r) &= du_{r}^{\theta}(r)/dr,\label{EqnD44} \\
\epsilon_{\phi}^{\theta}(r) &= u_{r}^{\theta}(r)/r,\label{EqnD45}\\
\epsilon_{r\phi}^{\theta}(r) &= 0 
\label{EqnD46}
\end{align}
Further, assuming plane stress conditions, the mechanical equilibrium equations within the bulk domains in polar coordinates simplifies to 
\begin{align}
\frac{\partial \sigma_{r}^{\theta}}{\partial r} + \frac{\sigma_{r}^{\theta} -\sigma_{\phi}^{\theta}}{r} = 0
\label{EqnD47}
\end{align}

Because we have assumed isotropic elastic properties, it follows from Eqs. (\ref{EqnD3}) \& (\ref{EqnD44})-(\ref{EqnD46})  that the nonzero stresses in polar coordinates are
\begin{align}
\sigma_{r}(r,t) &= 
\begin{cases}
\left(\lambda^{\alpha} + 2\mu^{\alpha}\right)\epsilon_{r}^{\alpha} + \lambda^{\alpha}\epsilon_{\phi}^{\alpha} &\quad 0 < r < r_{1},\\
\left(\lambda^{\beta} + 2\mu^{\beta}\right)\epsilon_{r}^{\beta} +  \lambda^{\beta}\epsilon_{\phi}^{\beta} - 2(\lambda^{\beta} + \mu^{\beta})\epsilon^{\star}&\quad r_{1} <r < r_{2},\\
\left(\lambda^{\gamma} + 2\mu^{\gamma}\right)\epsilon_{r}^{\gamma} + \lambda^{\gamma}\epsilon_{\phi}^{\gamma} &\quad r_{2}< r < R,
\end{cases}
\label{EqnD48}
\\
\sigma_{\phi}(r,t) &= 
\begin{cases}
\left(\lambda^{\alpha} + 2\mu^{\alpha}\right) \epsilon_{\phi}^{\alpha} + \lambda^{\alpha}\epsilon_{r}^{\alpha}&\quad 0 < r < r_{1},\\
\left(\lambda^{\beta} + 2\mu^{\beta}\right)\epsilon_{\phi}^{\beta} + \lambda^{\beta}\epsilon_{r}^{\beta} - 2(\lambda^{\beta} + \mu^{\beta})\epsilon^{\star}&\quad r_{1} <r < r_{2},\\
\left(\lambda^{\gamma} + 2\mu^{\gamma}\right)\epsilon_{\phi}^{\gamma} +  \lambda^{\gamma}\epsilon_{r}^{\gamma}&\quad r_{2}< r < R.
\end{cases}
\label{EqnD49}
\end{align}
Here $r_{1}(t)$ and $r_{2}(t)$ represent the numerically obtained interface positions at the $\alpha/\beta$ and $\beta/\gamma$ interfaces at time $t$ (see Fig. \ref{nonplanar_NiAl_schematic}).
Next, by substituting Eqs. (\ref{EqnD48}) \& (\ref{EqnD49}) in Eq. (\ref{EqnD47}) it can be shown that in a bulk phase $\theta$ the mechanical equilibrium equation reduces to
\begin{align}
\left(\lambda^{\theta} + 2\mu^{\theta}\right)\left[\frac{d^{2}u_{r}^{\theta}}{dr^{2}} + \frac{1}{r}\frac{du_{r}^{\theta}}{dr} - \frac{u_{r}^{\theta}}{r^{2}} \right] = 0 \Leftrightarrow \frac{d}{dr}\left[\frac{1}{r}\frac{d}{dr}\left(u^{\theta}_{r}r\right)\right] = 0.
\label{EqnD50}
\end{align}
Integrating Eq. (\ref{EqnD50}) yields the radial displacement within the bulk phases yields
\begin{align}
u_{r}(r)=
\begin{cases}
u_{r}^{\alpha} := A^{\alpha}r  + B^{\alpha}/r&\quad 0 < r < r_{1},\\
u_{r}^{\beta} := A^{\beta}r + B^{\beta}/r &\quad r_{1} < r < r_{2},\\
u_{r}^{\gamma} := A^{\gamma}r + B^{\gamma}/r&\quad r_{2}< r \leq R,
\end{cases}
\label{EqnD51}
\end{align}
Now, the problem reduces to finding the solution to the six unknown constants: $\{A^{\theta=\alpha,\beta,\gamma}\}$ and $\{B^{\theta=\alpha,\beta,\gamma}\}$. Since the displacement field must be bounded as $r\rightarrow 0$, the constant $B^{\alpha}$ must be zero. This ensures that the radial displacement is zero at the origin (see Eq. (\ref{EqnD42})). 

Using the strain-displacement relations, i.e., Eqs.(\ref{EqnD44})-(\ref{EqnD46}), and Eq.(\ref{EqnD51}), it can be shown that the nonzero strains within the bulk phases are
\begin{align}
\epsilon_{r}(r)=
\begin{cases}
\epsilon_{r}^{\alpha} := A^{\alpha} &\quad 0 < r < r_{1},\\
\epsilon_{r}^{\beta} := A^{\beta} - B^{\beta}/r^2 &\quad r_{1} < r < r_{2},\\
\epsilon_{r}^{\gamma} := A^{\gamma} - B^{\gamma}/r^2&\quad r_{2}< r \leq R,
\end{cases}
\label{EqnD52}
\\
\epsilon_{\phi}(r)=
\begin{cases}
\epsilon_{r}^{\alpha} := A^{\alpha} &\quad 0 < r < r_{1},\\
\epsilon_{r}^{\beta} := A^{\beta} + B^{\beta}/r^2 &\quad r_{1} < r < r_{2},\\
\epsilon_{r}^{\gamma} := A^{\gamma} + B^{\gamma}/r^2&\quad r_{2}< r \leq R,
\end{cases}
\label{EqnD53}
\end{align}
Note that we have set $B^{\alpha}$ to be zero in Eqs.(\ref{EqnD52}) \& (\ref{EqnD53}). To determine the remaining five unknowns, we need five equations. The first equation is a consequence of  boundary condition at the outer surface, i.e., Eq. (\ref{EqnD43}). Thus, substituting Eqs. (\ref{EqnD52}) and (\ref{EqnD53}) in Eq. (\ref{EqnD48}) and setting $r=R$ yields
\begin{align}
\left(\lambda^{\gamma} + 2\mu^{\gamma}\right)\left[A^{\gamma} -B^{\gamma}/R^{2} \right]  + \lambda^{\gamma}\left[A^{\gamma} + B^{\gamma}/R^{2}\right] = 0
\label{EqnD54}
\end{align}
The four remaining equations are obtained as a consequence of the interfacial conditions, specifically the continuity of radial displacement and radial stress. The continuity of displacement field yields:
\begin{align}
\left.u_{r}^{\alpha}\right|_{r_1} &= \left.u_{r}^{\beta}\right|_{r_1} \label{EqnD55}\\
\left.u_{r}^{\beta}\right|_{r_2} &= \left.u_{r}^{\gamma}\right|_{r_2}  \label{EqnD56}
\end{align}
Similarly, stress continuity implies:
\begin{align}
\left.\sigma_{r}^{\alpha}\right|_{r_1} &= \left.\sigma_{r}^{\beta}\right|_{r_1} \label{EqnD57}\ \\
\left.\sigma_{r}^{\beta}\right|_{r_2} &= \left.\sigma_{r}^{\gamma}\right|_{r_2}  \label{EqnD58}
\end{align}
Substituting Eq. (\ref{EqnD51}) in Eqs. (\ref{EqnD55}) and (\ref{EqnD56}) yields two of the required equations
\begin{align}
\left(A_{\alpha} - A_{\beta}\right)r_{1} - B_{\beta}/r_{1} &= 0\label{EqnD59}\\
\left(A_{\beta} - A_{\gamma}\right)r_{2} + \left(B_{\beta} - B_{\gamma}\right)/r_{1} &= 0\label{EqnD60}
\end{align}
Similarly, using  Eqs. (\ref{EqnD48}),  (\ref{EqnD51}) \& (\ref{EqnD52})  in Eqs. (\ref{EqnD57}) \&  (\ref{EqnD58}) yields the remaining two equations:
\begin{align}
\begin{split}
&\left[(\lambda^{\alpha} + 2\mu^{\alpha})A^{\alpha}+\lambda^{\alpha}A^{\alpha}\right] -\left[(\lambda^{\beta} + 2\mu^{\beta})\left(A^{\beta} - B^{\beta}/r_{1}^{2}\right)\right] \\
&- \lambda^{\beta}\left[A^{\beta} + B^{\beta}/r_{1}^{2}\right] + 2(\lambda^{\beta} + \mu^{\beta})\epsilon^{\star} = 0
 \end{split}
 \label{EqnD61}
 \end{align}
 \begin{align}
\begin{split}
&\left[(\lambda^{\beta} + 2\mu^{\beta})\left(A^{\beta} - B^{\beta}/r_{2}^{2}\right)\right] + \lambda^{\beta}\left[A^{\beta} + B^{\beta}/r_{2}^{2}\right] - 2(\lambda^{\beta} + \mu^{\beta})\epsilon^{\star} \\
&-\left[(\lambda^{\gamma} + 2\mu^{\gamma})\left(A^{\gamma} - B^{\gamma}/r_{2}^{2}\right)\right] - \lambda^{\gamma}\left[A^{\gamma} + B^{\gamma}/r_{2}^{2}\right]= 0
\end{split}
\label{EqnD62}
\end{align}
By solving Eq. (\ref{EqnD54}) and Eqs. (\ref{EqnD59})-(\ref{EqnD62}) yields the five unknown constants. A python script,  \texttt{threephase\_nonplanar\_analytical.py}, to solve these equations symbolically is provided with this paper. 

\subsection{Solution for the non-planar Ni-Al-Cr case}
\begin{figure}[!ht]
\center
\includegraphics[width=0.55\linewidth]{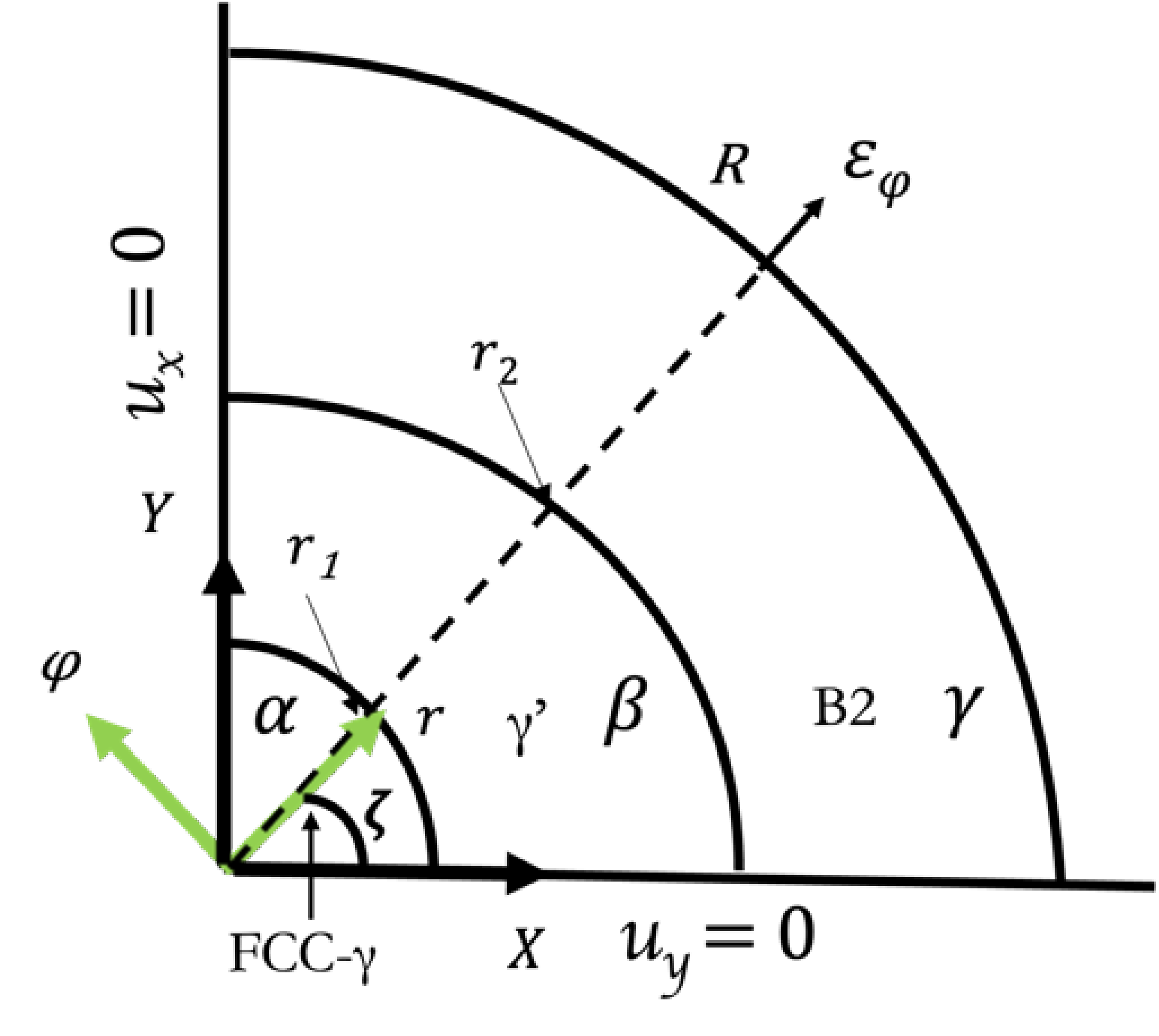}
\caption{A schematic showing the phases, eigenstrains and mechanical  boundary conditions for the concentric interface Ni-Al-Cr case.}
\label{nonplanar_NiAlCr_schematic}
\end{figure}

Fig. \ref{nonplanar_NiAlCr_schematic} shows the simulation domain and boundary conditions for the concentric ring Ni-Al-Cr case. Despite the similarities, there are two important differences that affects the analytical solution. First, we have assumed that there are no eigenstrains in the system; and second, we have imposed a hoop strain at the outer boundary. The outer boundary condition may be written as:
\begin{align}
\epsilon_{\phi}^{\gamma}(r=R,t) = \epsilon_{R}^{g},
\label{EqnD63}
\end{align}
where $\epsilon_{R}^{g}$ is the assumed hoop strain.
In the simulation, this hoop strain is imposed by assuming that the Cartesian displacements at the outer boundary are:
\begin{align}
\begin{split}
u_{x}(r=R,t) = \epsilon_{R}^{g}x\\
u_{y}(r=R,t) = \epsilon_{R}^{g}y
\end{split}
\label{ExtraF}
\end{align}
Using Eqs. (\ref{EqnCT}), (\ref{EqnD45}) and (\ref{ExtraF}), it can be shown that the hoop strain at the outer boundary is equal to $ \epsilon_{R}^{g}$. Moreover, due to fact that the geometry of the system is similar to the previous case, it can be assumed that the displacement fields within the bulk regions are given by Eq. (\ref{EqnD51}). Furthermore, since the boundary conditions at the left and bottom boundaries are identical to the previous case, it follows that the radial displacement at the origin must be zero, and consequently $B^{\alpha}=0$. Therefore, we need to solve for only the five unknown constants in Eq. (\ref{EqnD51}).

The first of these conditions is obtained by solving the outer boundary condition. Thus, substituting Eq. (\ref{EqnD51}) in Eq. (\ref{EqnD53}) and using Eq. (\ref{EqnD63}) yields
\begin{align}
A^{\gamma} + B^{\gamma}/R^2 = \epsilon_{r}^{g}
\label{EqnD64}
\end{align}
Similar to the previous case, the remaining four equations arise from the interfacial conditions. Moreover, the equations resulting from continuity of displacement field are identical to the previous case, i.e., Eqs. (\ref{EqnD59}) \& (\ref{EqnD60}). The remaining two equations are obtained assuming that the radial stress is continuous at the two interfaces. Specifically,
\begin{align}
\begin{split}
&\left[(\lambda^{\alpha} + 2\mu^{\alpha})A^{\alpha}+\lambda^{\alpha}A^{\alpha}\right] -\left[(\lambda^{\beta} + 2\mu^{\beta})\left(A^{\beta} - B^{\beta}/r_{1}^{2}\right)\right] \\
&- \lambda^{\beta}\left[A^{\beta} + B^{\beta}/r_{1}^{2}\right]  = 0
 \end{split}
 \label{EqnD65}
 \end{align}
 \begin{align}
\begin{split}
&\left[(\lambda^{\beta} + 2\mu^{\beta})\left(A^{\beta} - B^{\beta}/r_{2}^{2}\right)\right] + \lambda^{\beta}\left[A^{\beta} + B^{\beta}/r_{2}^{2}\right] \\
&-\left[(\lambda^{\gamma} + 2\mu^{\gamma})\left(A^{\gamma} - B^{\gamma}/r_{2}^{2}\right)\right] - \lambda^{\gamma}\left[A^{\gamma} + B^{\gamma}/r_{2}^{2}\right]= 0
\end{split}
\label{EqnD66}
\end{align}

Thus, by solving Eqs. (\ref{EqnD64}), (\ref{EqnD65}),  (\ref{EqnD66}), (\ref{EqnD59}) and (\ref{EqnD60}) we can obtain the five unknowns in Eq. (\ref{EqnD51}. This was achieved using the python package SymPy \cite{Sympy2017}. The python script, \texttt{threephase\_iso\_nonplanar}-\\
\texttt{\_applied\_strain.py}, is also available with this paper.

\section*{Data Availability}
The processed data required to reproduce the figures are available from the corresponding author on request. The simulation software required to reproduce the results is available to download from 
\href{https://github.com/souravmat-git/gibbs}{https://github.com/souravmat-git/gibbs.} The MOOSE input files required to run the simulations are available to download from the folder \href{https://github.com/souravmat-git/gibbs/tree/master/tested/stresssed_multiphase}{stressed\_multiphase.} The MATLAB scripts required to reproduce the precomputed input thermodynamic and kinetic properties are available to download from the folder \href{https://data.mendeley.com/datasets/4wzgh8nwzr/1}{Precomputed properties} \cite{SC_mendley}. Finally, the python scripts required to symbolically calculate the constants in the analytical solutions are available to download from the folder \href{https://github.com/souravmat-git/gibbs/tree/master/symbolic_python}{symbolic\_python.}
\clearpage
\bibliography{paper2}

\begin{thebibliography}{61}
\providecommand{\natexlab}[1]{#1}
\providecommand{\url}[1]{\texttt{#1}}
\expandafter\ifx\csname urlstyle\endcsname\relax
  \providecommand{\doi}[1]{doi: #1}\else
  \providecommand{\doi}{doi: \begingroup \urlstyle{rm}\Url}\fi

\bibitem[Fried and Gurtin(1999)]{Fried1999}
Eliot Fried and Morton~E Gurtin.
\newblock Coherent solid-state phase transitions with atomic diffusion: a
  thermomechanical treatment.
\newblock \emph{Journal of statistical physics}, 95\penalty0 (5):\penalty0
  1361--1427, 1999.

\bibitem[Gurtin and Voorhees(1993)]{Gurtin1993}
Morton~E. Gurtin and Peter~W. Voorhees.
\newblock The continuum mechanics of coherent two-phase elastic solids with
  mass transport.
\newblock \emph{Proceedings of the Royal Society. A, Mathematical and physical
  sciences}, 440\penalty0 (1909):\penalty0 323--343, 1993.

\bibitem[Provatas and Elder(2011)]{NProvatas2011}
N.~Provatas and K.~Elder.
\newblock \emph{Phase-Field Methods in Materials Science and Engineering}.
\newblock Wiley, 2011.

\bibitem[Chen and Wang(1996)]{Chen1996}
Long-Qing Chen and Yunzhi Wang.
\newblock The continuum field approach to modeling microstructural evolution.
\newblock \emph{JOM}, 48\penalty0 (12):\penalty0 13--18, Dec 1996.

\bibitem[Chen(2002)]{LQChen2002}
Long-Qing Chen.
\newblock Phase-field models for microstructure evolution.
\newblock \emph{Annual Review of Materials Research}, 32\penalty0 (1):\penalty0
  113--140, 2002.

\bibitem[Moelans et~al.(2008{\natexlab{a}})Moelans, Blanpain, and
  Wollants]{Moelans2008}
Nele Moelans, Bart Blanpain, and Patrick Wollants.
\newblock An introduction to phase-field modeling of microstructure evolution.
\newblock \emph{Calphad}, 32\penalty0 (2):\penalty0 268 -- 294,
  2008{\natexlab{a}}.

\bibitem[Steinbach(2009)]{Steinbach2009}
Ingo Steinbach.
\newblock Phase-field models in materials science.
\newblock \emph{Modelling and Simulation in Materials Science and Engineering},
  17\penalty0 (7):\penalty0 073001, jul 2009.

\bibitem[Nestler and Choudhury(2011)]{Nestler2011}
Britta Nestler and Abhik Choudhury.
\newblock Phase-field modeling of multi-component systems.
\newblock \emph{Current Opinion in Solid State and Materials Science},
  15\penalty0 (3):\penalty0 93--105, 2011.
\newblock Applications of Phase Field Modeling in Materials Science and
  Engineering.

\bibitem[Moelans(2011)]{Moelans2011}
Nele Moelans.
\newblock A quantitative and thermodynamically consistent phase-field
  interpolation function for multi-phase systems.
\newblock \emph{Acta Materialia}, 59\penalty0 (3):\penalty0 1077--1086, 2011.

\bibitem[Plapp(2011)]{Plapp2011}
Mathis Plapp.
\newblock Unified derivation of phase-field models for alloy solidification
  from a grand-potential functional.
\newblock \emph{Phys. Rev. E}, 84\penalty0 (3):\penalty0 031601--1--15, 2011.

\bibitem[Karma and Tourret(2016)]{Karma2016}
Alain Karma and Damien Tourret.
\newblock Atomistic to continuum modeling of solidification microstructures.
\newblock \emph{Current Opinion in Solid State and Materials Science},
  20\penalty0 (1):\penalty0 25--36, 2016.
\newblock Recent Advances in Solidification Microstructure- Experiments and
  Computational Analysis.

\bibitem[Kim et~al.(1998)Kim, Kim, and Suzuki]{Kim1998}
Seong~Gyoon Kim, Won~Tae Kim, and Toshio Suzuki.
\newblock Interfacial compositions of solid and liquid in a phase-field model
  with finite interface thickness for isothermal solidification in binary
  alloys.
\newblock \emph{Phys. Rev. E}, 58\penalty0 (3):\penalty0 3316--3323, Sep 1998.

\bibitem[Kim et~al.(1999)Kim, Kim, and Suzuki]{Kim1999}
Seong~Gyoon Kim, Won~Tae Kim, and Toshio Suzuki.
\newblock Phase-field model for binary alloys.
\newblock \emph{Phys. Rev. E}, 60\penalty0 (6):\penalty0 7186--7197, 1999.

\bibitem[Plapp(2015)]{Plapp2015}
M~Plapp.
\newblock \emph{Phase-field models}.
\newblock Handbook of Crystal Growth, 2nd edition, 2015.

\bibitem[Durga et~al.(2013)Durga, Wollants, and Moelans]{DurgaA2013}
A~Durga, P~Wollants, and N~Moelans.
\newblock Evaluation of interfacial excess contributions in different
  phase-field models for elastically inhomogeneous systems.
\newblock \emph{Modelling and simulation in materials science and engineering},
  21\penalty0 (5):\penalty0 55018, 2013.

\bibitem[Schneider et~al.(2015)Schneider, Tschukin, Choudhury, Selzer,
  B{\"o}hlke, and Nestler]{Schneider2015}
Daniel Schneider, Oleg Tschukin, Abhik Choudhury, Michael Selzer, Thomas
  B{\"o}hlke, and Britta Nestler.
\newblock Phase-field elasticity model based on mechanical jump conditions.
\newblock \emph{Computational Mechanics}, 55\penalty0 (5):\penalty0 887--901,
  May 2015.

\bibitem[Choudhury and Nestler(2012)]{Choudhary2012}
Abhik Choudhury and Britta Nestler.
\newblock Grand-potential formulation for multicomponent phase transformations
  combined with thin-interface asymptotics of the double-obstacle potential.
\newblock \emph{Phys. Rev. E}, 85\penalty0 (2):\penalty0 021602--1--16, 2012.

\bibitem[Chatterjee and Moelans(2021{\natexlab{a}})]{Chatterjee2021}
Sourav Chatterjee and Nele Moelans.
\newblock A grand-potential based phase-field approach for simulating growth of
  intermetallic phases in multicomponent alloy systems.
\newblock \emph{Acta Materialia}, 206:\penalty0 116630, 2021{\natexlab{a}}.

\bibitem[Eiken et~al.(2006)Eiken, B\"ottger, and Steinbach]{Eiken2006}
J.~Eiken, B.~B\"ottger, and I.~Steinbach.
\newblock Multiphase-field approach for multicomponent alloys with
  extrapolation scheme for numerical application.
\newblock \emph{Phys. Rev. E}, 73\penalty0 (6):\penalty0 066122, Jun 2006.

\bibitem[Kim et~al.(2021)Kim, Kim, Cha, Lee, Lee, Park, and Oh]{Kim2021}
Seong~Gyoon Kim, Won~Tae Kim, Pil-Ryung Cha, Byeong-Joo Lee, Jae~Sang Lee,
  Jiwon Park, and Chang-Seok Oh.
\newblock Phase-field model with relaxation of the partition coefficient.
\newblock \emph{Computational Materials Science}, 188:\penalty0 110184, 2021.

\bibitem[Böttger et~al.(2015)Böttger, Eiken, and Apel]{Bottger2015}
B.~Böttger, J.~Eiken, and M.~Apel.
\newblock Multi-ternary extrapolation scheme for efficient coupling of
  thermodynamic data to a multi-phase-field model.
\newblock \emph{Computational Materials Science}, 108:\penalty0 283 -- 292,
  2015.

\bibitem[Jiang et~al.(2019)Jiang, Zhang, Zhang, Yin, and Qu]{Jiang2019}
Xue Jiang, Ruijie Zhang, Cong Zhang, Haiqing Yin, and Xuanhui Qu.
\newblock Fast prediction of the quasi phase equilibrium in phase field model
  for multicomponent alloys based on machine learning method.
\newblock \emph{Calphad}, 66:\penalty0 101644, 2019.

\bibitem[Zhu et~al.(2004)Zhu, Wang, Ardell, Zhou, Liu, and Chen]{Zhu2004}
J.Z. Zhu, T.~Wang, A.J. Ardell, S.H. Zhou, Z.K. Liu, and L.Q. Chen.
\newblock Three-dimensional phase-field simulations of coarsening kinetics of
  {$\gamma^{\prime}$} particles in binary ni–al alloys.
\newblock \emph{Acta Materialia}, 52\penalty0 (9):\penalty0 2837--2845, 2004.

\bibitem[Ammar et~al.(2009)Ammar, Appolaire, Cailletaud, and Forest]{Ammar2009}
Kais Ammar, Benoît Appolaire, Georges Cailletaud, and Samuel Forest.
\newblock Combining phase field approach and homogenization methods for
  modelling phase transformation in elastoplastic media.
\newblock \emph{European Journal of Computational Mechanics}, 18\penalty0
  (5-6):\penalty0 485--523, 2009.

\bibitem[Steinbach and Apel(2007)]{Steinbach2007Acta}
I.~Steinbach and M.~Apel.
\newblock The influence of lattice strain on pearlite formation in fe–c.
\newblock \emph{Acta Materialia}, 55\penalty0 (14):\penalty0 4817--4822, 2007.

\bibitem[Durga et~al.(2015)Durga, Wollants, and Moelans]{Durga2015}
A.~Durga, P.~Wollants, and N.~Moelans.
\newblock A quantitative phase-field model for two-phase elastically
  inhomogeneous systems.
\newblock \emph{Computational Materials Science}, 99:\penalty0 81--95, 2015.

\bibitem[Mushongera et~al.(2015)Mushongera, Fleck, Kundin, Wang, and
  Emmerich]{Mushongera2015}
L.T. Mushongera, M.~Fleck, J.~Kundin, Y.~Wang, and H.~Emmerich.
\newblock Effect of re on directional $\gamma^{\prime}$-coarsening in
  commercial single crystal ni-base superalloys: A phase field study.
\newblock \emph{Acta Materialia}, 93:\penalty0 60--72, 2015.

\bibitem[Tschukin et~al.(2019)Tschukin, Schneider, and Nestler]{Tschukin2017a}
O.~Tschukin, D.~Schneider, and B.~Nestler.
\newblock An elasto-chemical phase-field model for isotropic solids.
\newblock \emph{European Journal of Mechanics - A/Solids}, 73:\penalty0
  181--191, 2019.

\bibitem[Simon et~al.(2020)Simon, Aagesen, Motta, and Tonks]{Simon2020}
Pierre-Clément~A. Simon, Larry~K. Aagesen, Arthur~T. Motta, and Michael~R.
  Tonks.
\newblock The effects of introducing elasticity using different interpolation
  schemes to the grand potential phase field model.
\newblock \emph{Computational Materials Science}, 183:\penalty0 109790, 2020.

\bibitem[Chatterjee et~al.(2021)Chatterjee, Schwen, and Moelans]{SC2}
Sourav Chatterjee, Daniel Schwen, and Nele Moelans.
\newblock An efficient and quantitative phase-field model for elastically
  heterogeneous solids based on a partial rank-one homogenization scheme.
\newblock \emph{Under Review in International Journal of Solids and
  Structures}, 2021.

\bibitem[Mosler et~al.(2014)Mosler, Shchyglo, and {Montazer
  Hojjat}]{Mosler2014}
J.~Mosler, O.~Shchyglo, and H.~{Montazer Hojjat}.
\newblock A novel homogenization method for phase field approaches based on
  partial rank-one relaxation.
\newblock \emph{Journal of the Mechanics and Physics of Solids}, 68:\penalty0
  251 -- 266, 2014.

\bibitem[Svendsen et~al.(2018)Svendsen, Shanthraj, and Raabe]{Svendsen2018}
Bob Svendsen, Pratheek Shanthraj, and Dierk Raabe.
\newblock Finite-deformation phase-field chemomechanics for multiphase,
  multicomponent solids.
\newblock \emph{Journal of the Mechanics and Physics of Solids}, 112:\penalty0
  619--636, 2018.

\bibitem[Steinbach and Apel(2006)]{Steinbach2006}
I.~Steinbach and M.~Apel.
\newblock Multi phase field model for solid state transformation with elastic
  strain.
\newblock \emph{Physica D: Nonlinear Phenomena}, 217\penalty0 (2):\penalty0 153
  -- 160, 2006.

\bibitem[Durga et~al.(2020)Durga, Wollants, and Moelans]{Durga2020}
A.~Durga, P.~Wollants, and N.~Moelans.
\newblock Phase-field study of imc growth in sn–cu/cu solder joints including
  elastoplastic effects.
\newblock \emph{Acta Materialia}, 188:\penalty0 241--258, 2020.

\bibitem[Schneider et~al.(2018)Schneider, Schoof, Tschukin, Reiter, Herrmann,
  Schwab, Selzer, and Nestler]{Schneider2018}
Daniel Schneider, Ephraim Schoof, Oleg Tschukin, Andreas Reiter, Christoph
  Herrmann, Felix Schwab, Michael Selzer, and Britta Nestler.
\newblock Small strain multiphase-field model accounting for configurational
  forces and mechanical jump conditions.
\newblock \emph{Computational Mechanics}, 61\penalty0 (3):\penalty0 277--295,
  Mar 2018.

\bibitem[Tschukin(2017)]{Tschukin2017}
Oleg Tschukin.
\newblock \emph{Phase-Field Modelling of Welding and of Elasticity-Dependent
  Phase Transformations}.
\newblock PhD thesis, Karlsruher Institut für Technologie (KIT), 2017.

\bibitem[{Kubendran Amos} et~al.(2018){Kubendran Amos}, Schoof, Schneider, and
  Nestler]{Amos2018}
P.G. {Kubendran Amos}, Ephraim Schoof, Daniel Schneider, and Britta Nestler.
\newblock Chemo-elastic phase-field simulation of the cooperative growth of
  mutually-accommodating widmanstätten plates.
\newblock \emph{Journal of Alloys and Compounds}, 767:\penalty0 1141--1154,
  2018.

\bibitem[Bartels et~al.(2021)Bartels, Kurzeja, and Mosler]{Bartel2021}
Alexander Bartels, Patrick Kurzeja, and Jörn Mosler.
\newblock Cahn–hilliard phase field theory coupled to mechanics:
  Fundamentals, numerical implementation and application to topology
  optimization.
\newblock \emph{Computer Methods in Applied Mechanics and Engineering},
  383:\penalty0 113918, 2021.

\bibitem[Wheeler et~al.(1992)Wheeler, Boettinger, and McFadden]{Wheeler1992}
A.~A. Wheeler, W.~J. Boettinger, and G.~B. McFadden.
\newblock Phase-field model for isothermal phase transitions in binary alloys.
\newblock \emph{Phys. Rev. A}, 45\penalty0 (10):\penalty0 7424--7439, May 1992.

\bibitem[Sarhil et~al.(2021)Sarhil, Shchyglo, Brands, Steinbach, and
  Schr{\"o}der]{Sarhil2021}
Mohammad Sarhil, Oleg Shchyglo, Dominik Brands, Ingo Steinbach, and J{\"o}rg
  Schr{\"o}der.
\newblock Martensitic transformation in a two-dimensional polycrystalline shape
  memory alloys using a multi-phase-field elasticity model based on pairwise
  rank-one convexified energies at small strain.
\newblock \emph{PAMM}, 20\penalty0 (1):\penalty0 e202000200, 2021.

\bibitem[Folch and Plapp(2003)]{Folch2003}
R~Folch and M~Plapp.
\newblock Towards a quantitative phase-field model of two-phase solidification.
\newblock \emph{Physical Review E}, 68\penalty0 (1):\penalty0 010602, 2003.

\bibitem[Schneider et~al.(2017)Schneider, Schwab, Schoof, Reiter, Herrmann,
  Selzer, B{\"o}hlke, and Nestler]{Schneider2017}
Daniel Schneider, Felix Schwab, Ephraim Schoof, Andreas Reiter, Christoph
  Herrmann, Michael Selzer, Thomas B{\"o}hlke, and Britta Nestler.
\newblock On the stress calculation within phase-field approaches: a model for
  finite deformations.
\newblock \emph{Computational Mechanics}, 60\penalty0 (2):\penalty0 203--217,
  Aug 2017.

\bibitem[Herrmann et~al.(2018)Herrmann, Schoof, Schneider, Schwab, Reiter,
  Selzer, and Nestler]{Herrmann2018}
Christoph Herrmann, Ephraim Schoof, Daniel Schneider, Felix Schwab, Andreas
  Reiter, Michael Selzer, and Britta Nestler.
\newblock Multiphase-field model of small strain elasto-plasticity according to
  the mechanical jump conditions.
\newblock \emph{Computational Mechanics}, 62\penalty0 (6):\penalty0 1399--1412,
  Dec 2018.

\bibitem[Steinbach et~al.(2012)Steinbach, Zhang, and Plapp]{Steinbach2012}
Ingo Steinbach, Lijun Zhang, and Mathis Plapp.
\newblock Phase-field model with finite interface dissipation.
\newblock \emph{Acta Materialia}, 60\penalty0 (6-7):\penalty0 2689--2701, 2012.

\bibitem[Kazaryan et~al.(2001)Kazaryan, Wang, Dregia, and Patton]{Kazaryan2001}
A.~Kazaryan, Y.~Wang, S.~A. Dregia, and Bruce~R. Patton.
\newblock Grain growth in systems with anisotropic boundary mobility:
  Analytical model and computer simulation.
\newblock \emph{Phys. Rev. B}, 63:\penalty0 184102, Apr 2001.

\bibitem[Nestler et~al.(2005)Nestler, Garcke, and Stinner]{Nestler2005}
Britta Nestler, Harald Garcke, and Bj{\"o}rn Stinner.
\newblock Multicomponent alloy solidification: phase-field modeling and
  simulations.
\newblock \emph{Physical Review E}, 71\penalty0 (4):\penalty0 041609, 2005.

\bibitem[Moelans et~al.(2008{\natexlab{b}})Moelans, Blanpain, and
  Wollants]{Moelans2008PRB}
Nele Moelans, Bart Blanpain, and Patrick Wollants.
\newblock Quantitative analysis of grain boundary properties in a generalized
  phase field model for grain growth in anisotropic systems.
\newblock \emph{Physical Review B}, 78\penalty0 (2):\penalty0 024113,
  2008{\natexlab{b}}.

\bibitem[Kiefer et~al.(2017)Kiefer, Furlan, and Mosler]{Kiefer2017}
B.~Kiefer, T.~Furlan, and J.~Mosler.
\newblock A numerical convergence study regarding homogenization assumptions in
  phase field modeling.
\newblock \emph{International Journal for Numerical Methods in Engineering},
  112\penalty0 (9):\penalty0 1097--1128, 2017.

\bibitem[Hansen-D{\"o}rr et~al.(2021)Hansen-D{\"o}rr, Brummund, and
  K{\"a}stner]{Hansen2021}
Arne~Claus Hansen-D{\"o}rr, J{\"o}rg Brummund, and Markus K{\"a}stner.
\newblock Phase-field modeling of fracture in heterogeneous materials: jump
  conditions, convergence and crack propagation.
\newblock \emph{Archive of Applied Mechanics}, 91\penalty0 (2):\penalty0
  579--596, 2021.

\bibitem[Hötzer et~al.(2015)Hötzer, Jainta, Steinmetz, Nestler, Dennstedt,
  Genau, Bauer, Köstler, and Rüde]{Hotzer2015}
Johannes Hötzer, Marcus Jainta, Philipp Steinmetz, Britta Nestler, Anne
  Dennstedt, Amber Genau, Martin Bauer, Harald Köstler, and Ulrich Rüde.
\newblock Large scale phase-field simulations of directional ternary eutectic
  solidification.
\newblock \emph{Acta Materialia}, 93:\penalty0 194 -- 204, 2015.

\bibitem[Cogswell(2015)]{Cogswell2015}
Daniel~A. Cogswell.
\newblock Quantitative phase-field modeling of dendritic electrodeposition.
\newblock \emph{Phys. Rev. E}, 92\penalty0 (1):\penalty0 011301, Jul 2015.

\bibitem[Aagesen et~al.(2018)Aagesen, Gao, Schwen, and Ahmed]{Aagensen2018}
Larry~K. Aagesen, Yipeng Gao, Daniel Schwen, and Karim Ahmed.
\newblock Grand-potential-based phase-field model for multiple phases, grains,
  and chemical components.
\newblock \emph{Phys. Rev. E}, 98\penalty0 (2):\penalty0 023309, Aug 2018.

\bibitem[Voorhees and Johnson(2004)]{Voorhees2004}
{P. W.} Voorhees and {William C.} Johnson.
\newblock \emph{The thermodynamics of elastically stressed crystals}, pages
  1--201.
\newblock Number~C in Solid State Physics - Advances in Research and
  Applications. Academic Press Inc, c edition, 2004.

\bibitem[Kim(2007)]{Kim2007}
Seong~Gyoon Kim.
\newblock A phase-field model with antitrapping current for multicomponent
  alloys with arbitrary thermodynamic properties.
\newblock \emph{Acta Materialia}, 55\penalty0 (13):\penalty0 4391--4399, 2007.

\bibitem[Gaston et~al.(2009)Gaston, Newman, Hansen, and
  Lebrun-Grandié]{Gaston2009}
Derek Gaston, Chris Newman, Glen Hansen, and Damien Lebrun-Grandié.
\newblock Moose: A parallel computational framework for coupled systems of
  nonlinear equations.
\newblock \emph{Nuclear Engineering and Design}, 239\penalty0 (10):\penalty0
  1768 -- 1778, 2009.

\bibitem[Socrate and Parks(1993)]{Socrate1992}
S.~Socrate and D.M. Parks.
\newblock Numerical determination of the elastic driving force for directional
  coarsening in ni-superalloys.
\newblock \emph{Acta Metallurgica et Materialia}, 41\penalty0 (7):\penalty0
  2185--2209, 1993.

\bibitem[Miracle(1993)]{Miracle1993}
D.B. Miracle.
\newblock Overview no. 104 the physical and mechanical properties of nial.
\newblock \emph{Acta Metallurgica et Materialia}, 41\penalty0 (3):\penalty0
  649--684, 1993.

\bibitem[Ardell(2014)]{Ardell2014}
Alan Ardell.
\newblock The effects of elastic interactions on precipitate microstructural
  evolution in elastically inhomogeneous nickel-base alloys.
\newblock \emph{Philosophical Magazine}, 94, 04 2014.

\bibitem[Mura(1991)]{Mura1991}
Toshio Mura.
\newblock \emph{Micromechanics of defects in solids}.
\newblock Kluwer academic, Dordrecht, 1991.

\bibitem[Meurer et~al.(2017)Meurer, Smith, Paprocki, {\v{C}}ert{\'\i}k,
  Kirpichev, Rocklin, Kumar, Ivanov, Moore, Singh, et~al.]{Sympy2017}
Aaron Meurer, Christopher~P Smith, Mateusz Paprocki, Ond{\v{r}}ej
  {\v{C}}ert{\'\i}k, Sergey~B Kirpichev, Matthew Rocklin, AMiT Kumar, Sergiu
  Ivanov, Jason~K Moore, Sartaj Singh, et~al.
\newblock Sympy: symbolic computing in python.
\newblock \emph{PeerJ Computer Science}, 3:\penalty0 e103, 2017.

\bibitem[Chatterjee and Moelans(2021{\natexlab{b}})]{SC_mendley}
Sourav Chatterjee and Nele Moelans.
\newblock A grand-potential based phase-field approach for simulating growth of
  intermetallic phases in multicomponent alloy systems.
\newblock Mendeley Data, v1, 2021{\natexlab{b}}.

\end{thebibliography}

\end{document}